\documentclass[12pt]{article}
\usepackage{amsmath,amssymb,amsthm,amsfonts,bm,mathrsfs,mathtools,float,algorithm,algorithmic,caption,subcaption,setspace}
\usepackage{avant,array,geometry,fontenc,inputenc,natbib,hyperref,multirow,enumerate,rotating,booktabs,color,latexsym,times,stmaryrd,xr}

\newtheorem{theorem}[]{Theorem}

\newtheorem{corollary}[]{Corollary}
\newtheorem{proposition}[]{Proposition}
\newtheorem{definition}[]{Definition}

\def\E{\mathbb E}
\def\N{\mathbb N}
\def\P{\mathbb P}
\def\R{\mathbb R}

\newcommand{\Bcal}{{\mathcal B}}
\newcommand{\Ccal}{{\mathcal C}}

\newcommand{\Gcal}{{\mathcal G}}
\newcommand{\Hcal}{{\mathcal H}}
\newcommand{\Ical}{{\mathcal I}}

\newcommand{\Ncal}{{\mathcal N}}

\newcommand{\Xcal}{{\mathcal X}}

\newcommand{\trans}{^{\mbox{\tiny {\sf T}}}}
\DeclarePairedDelimiter{\ceil}{\lceil}{\rceil}

\newtheorem{defi}[]{Definition}

\begin{document}

\title{\Large{\textbf{Orthogonalized Kernel Debiased Machine Learning \\ 
for Multimodal Data Analysis}}}
%\title{\Large{\textbf{Orthogonal Statistical Inference for \\ 
%Multimodal Data Analysis}}} 
\author{
\bigskip
{\sc Xiaowu Dai and Lexin Li} \\
{\it {\normalsize University of California at Berkeley}}\\
\\
 {\normalsize To appear in  \emph{Journal of the American Statistical Association: Theory and Method }}
}
\date{}
\maketitle

\begin{abstract}
Multimodal imaging has transformed neuroscience research. While it presents unprecedented opportunities, it also imposes serious challenges. Particularly, it is difficult to combine the merits of the interpretability attributed to a simple association model with the flexibility achieved by a highly adaptive nonlinear model. In this article, we propose an orthogonalized kernel debiased machine learning approach, which is built upon the Neyman orthogonality and a form of decomposition orthogonality, for multimodal data analysis. We target the setting that naturally arises in almost all multimodal studies, where there is a primary modality of interest, plus additional auxiliary modalities. We establish the root-$N$-consistency and asymptotic normality of the estimated primary parameter, the semi-parametric estimation efficiency, and the asymptotic validity of the confidence band of the predicted primary modality effect. Our proposal enjoys, to a good extent, both model interpretability and model flexibility. It is also considerably different from the existing statistical methods for multimodal data integration, as well as the orthogonality-based methods for high-dimensional inferences. We demonstrate the efficacy of our method through both simulations and an application to a multimodal neuroimaging study of Alzheimer's disease. 
\end{abstract}
\bigskip


\newpage
\baselineskip=21pt

%%%%%%%%%%%%%%%%%%%%%%%%%%%%%%%%%%%%%%%%%%%%%%%%%%%
\section{Introduction}
\label{sec:intro}

Multimodal neuroimaging, where different types of images are acquired for a common set of experimental subjects, is becoming a norm in neuroscience research. It utilizes different physical and physiological sensitivities of imaging scanners and technologies, and measures distinct brain characteristics including brain structures, functions and chemical constituents. Multimodal neuroimaging analysis aggregates such diverse but often complementary information, consolidates knowledge across different modalities, and produces improved understanding of neurological development or disorders \citep{Uludaug2014}. Multimodal data also frequently arise in many other scientific applications, e.g., integrative genomics \citep{Richardson2016}, multimodal healthcare \citep{Cai2019}, and audio-visual speech recognition \citep{Baltrusaitis2019}. 

Our motivation is a multimodal neuroimaging study of Alzheimer's disease (AD). AD is an irreversible neurodegenerative disorder and the leading form of dementia in elderly subjects. The most notable AD imaging biomarker is the brain grey matter cortical atrophy measured by structural magnetic resonance imaging (MRI). Meanwhile, amyloid-$\beta$ and tau are two hallmark pathological proteins that are believed to be part of the driving mechanism of AD, and both can be measured by positron emission tomography (PET) using different nuclear tracers. The current model of AD pathogenesis hypothesizes a sequence of biological cascade among different AD biomarkers \citep{Jack2010}. It is of great scientific interest to study how they interact with each other and how they affect the cognitive outcome. These questions are crucial for our understanding of AD pathophysiology, and also have important therapeutic implications. 

While multimodal neuroimaging presents unprecedented opportunities, it also imposes numerous serious challenges. First, neuroimaging data are typically high-dimensional and highly correlated, with measurements of brain characteristics at hundreds of brain regions and millions of brain voxel locations, and those measurements are often spatially or temporally correlated. Besides, the associations between different imaging modalities, and between images and phenotypic outcomes, are complicated. A  linear association model, despite its wide usage, is hardly adequate to capture such complex associations. Second, it is particularly challenging to balance between model interpretability and model flexibility. \citet{breiman2001statistical} contrasted two modeling cultures: the ``data modeling culture", which adopts parametric models that are easier to interpret and to perform inference but much less flexible, versus the ``algorithmic modeling culture", also known as machine learning, which involves complex and sometimes black-box type models that are highly flexible and nonlinear but difficult to interpret and infer. Both approaches have been frequently adopted in neuroimaging analysis. Nevertheless, it is difficult to combine the merits of both. Most existing works on multimodal data integration either assume a simple parametric model for easy interpretation \citep[e.g.,][]{Sperling2019, LiLi2020factor}, or consider a flexible nonlinear model but sacrifice the interpretability or inference capability \citep[e.g.,][]{hinrichs2011predictive,WangYP2018}. Finally, rigorously quantifying statistical significance of the primary parameter of interest remains a fundamental question in scientific inquiries. There have been a large number of highly successful nonlinear modeling techniques, ranging from the more classical splines, reproducing kernels, and random forests, to more recent deep neural network models. However, it is notoriously difficult to carry out statistical inference when utilizing those flexible methods. Moreover, when it comes to inference, naively adding multiple modalities together may suffer from serious biases and produce misleading results, as we show later. 

In this article, we propose an orthogonalized kernel debiased machine learning approach, built upon the Neyman orthogonality \citep{Neyman1959, Neyman1979}, and a form of decomposition orthogonality \citep[Chapter 3]{wahba1990}, for multimodal data analysis. The principal setting we target is that there is a \emph{primary} modality of interest, plus additional \emph{auxiliary} modalities. Such a setting naturally arises in almost all multimodal studies, and is particularly useful from the perspective of scientific inquiries. For instance, in AD pathophysiology modeling \citep{Jack2010}, it is often of interest to quantify the effect of brain structural atrophy on cognition after accounting for amyloid-$\beta$ and tau accumulations. In this case, the structural atrophy can be treated as the primary modality, while amyloid-beta and tau are the auxiliary modalities. In imaging genetics studies \citep{ZhuHT2014, Zhu2017review}, brain imaging features often play the role of intermediate phenotype between the genetic variants and clinical outcome. In this case, the brain image can be taken as the primary modality, and the genetic variants as the auxiliary modality.  Under this setting, we employ a basis expansion type model along with model error to characterize the association between the primary modality and the outcome, and develop rigorous inference methods for the main parameter of interest as well as the predicted primary modality effect. Meanwhile, we employ highly flexible machine learning methods to model the complex associations both between the auxiliary modalities and the outcome, and between the primary and auxiliary modalities. A key challenge that comes with flexible machine learning modeling is that its associated regularization bias and overfitting would introduce heavy bias in the estimation of the main parameter of interest. To remove such an impact, we employ two types of orthogonality formulations based on \citet{Neyman1959, Neyman1979}, \citet{Chernozhukov2018}, and \citet{wahba1990}. We establish the $\sqrt{N}$-consistency and asymptotic normality of the estimated main parameter, the semi-parametric estimation efficiency, as well as the asymptotic validity of the confidence band of the predicted primary modality effect, where $N$ is the sample size. Our proposed framework thus enjoys, to a good extent, both model interpretability and model flexibility. 

Our proposal is considerably different from the existing statistical methods for multimodal data integration. Particularly, there have been a class of unsupervised multimodal analysis built on matrix or tensor factorization \citep{Lock2013}, or canonical correlation analysis \citep{mai2019iterative, Shu2019dcca}. By contrast, we aim at a supervised regression problem. Under the  regression setting with multimodal predictors, \citet{LiChen2019} proposed an integrative reduced-rank regression. \citet{XueQu2019} developed an estimating equations approach to accommodate block missing patterns. \citet{LiLi2020factor} developed a factor analysis-based linear regression model. These methods are supervised, but all of them still assume linear type associations, and none utilizes any nonlinear machine learning modeling. 

Relatedly, the Neyman orthogonality has played an important role in both statistics and econometrics. Early works date back to \citet{newey1990semiparametric}, \citet{robins1995semiparametric} and  \cite{van2006targeted}. Meanwhile, it has received revived interest in high-dimensional statistical inference in recent years, thanks to, most notably, \cite{Chernozhukov2018}; see also many references therein. Our proposal can be viewed as an extension of the double/debiased machine learning framework developed by \cite{Chernozhukov2018}. However, there are some fundamental differences. First and most importantly, we allow an additional model error for the primary modality, which has crucial implications in terms of model interpretation, estimation and theoretical analysis. In particular,  \cite{Chernozhukov2018} focused on a low-dimensional primary parameter involving no additional error. \citet{kozbur2020inference} extended to a nonparametric primary function through basis expansion, but required that the function can be well approximated with a vanishing approximation error. By contrast, we do not impose a vanishing error, which distinguishes our proposal from \cite{Chernozhukov2018, kozbur2020inference} and other double/debiased machine learning methods. This additional model error essentially offers improved inferential robustness. Depending on the scientific context, one may choose a simple and interpretable yet less accurate model for the primary modality, or one may choose a more accurate but perhaps less interpretable model, and our method works for both cases. On the other hand, this error imposes numerous new challenges. To address those challenges, we introduce a second form of orthogonality, similar to the perpendicularity in smoothing splines \citep{wahba1990}, to ensure the parameter identifiability. We construct a new reproducing kernel Hilbert space (RKHS) and employ residual learning to decouple and remove the impact of the model error in parameter estimation. We also develop new theoretical tools to establish the asymptotic guarantees of the estimated primary parameter under model error. Second, we establish the confidence band for the nonparametric primary regression function given the high-dimensional nonlinear nuisance function. This quantity is of key scientific interest, as it quantifies the predicted effect and the contribution of the primary modality. However, its inference is challenging, due to the nonparametric nature of the model, high dimensionality, and strong correlations between the modalities. The existing literature on high-dimensional nonparametric inference usually requires stronger conditions that are unlikely to hold in multimodal neuroimaging data. We extend the framework of \citet{chernozhukov2014anti}, and approximate the supremum of high-dimensional empirical processes by a Gaussian multiplier process to obtain the asymptotically valid confidence band. Later we further compare with a number of alternative solutions, both analytically and numerically.

The rest of the article is organized as follows.  We introduce the model framework in Section \ref{sec:model}, and develop an estimation procedure in Section \ref{sec:estimation}. We derive the orthogonal statistical inference procedure and the theoretical guarantees in Section \ref{sec:inference}. We analytically compare with the alternative methods in Section \ref{sec:comparison}. We present the simulations in Section \ref{sec:simulations}, and revisit the multimodal AD study in Section \ref{sec:realdata}. We conclude the paper with a further discussion on the innovation of our method in Section \ref{sec:conclusion}, and relegate all technical proofs to the Supplementary Appendix.

%%%%%%%%%%%%%%%%%%%%%%%%%%%%%%%%%%%%%%%%%%%%%%%%%%%
\section{Model}
\label{sec:model}

Suppose there are $M+1$ modalities of predictors. Let $X =(X_{(1)},\ldots,X_{(p)})\trans \in \Xcal^{p}$ denote the $p$-dimensional vector of the primary modality, where $\mathcal X \subset \R$ is a compact domain and $X$ follows the distribution $P$ in $\Xcal^{p}$. Let $Z_{(m)} \in \R^{p'_m}$ denote the $p'_m$-dimensional vector of the $m$th auxiliary modality, $m = 1, \ldots, M$, and let $Z = (Z_{(1)}\trans, \ldots, Z_{(M)}\trans )\trans \in \R^{p'}$ collect all auxiliary modalities, $p' = p'_1+\ldots+p'_M$.  Let $Y \in \R$ denote the response variable. We propose the following model framework, 
\vspace{-0.05in}
\begin{equation} \label{eqn:regeqn}
Y  = f_0(X) + g_0(Z) + U, 
\end{equation}
where  $U \in \R$ is the measurement error that is independent of $(X,Z)$ and $\E[U]=0$ and $\E[U^2]=\sigma^2<\infty$, $f_0$ is the regression function  capturing the effect of the primary modality on the response, and $g_0$ is the function capturing the collective effects of the auxiliary modalities.  We also note that we can extend \eqref{eqn:regeqn} from a linear model form to a generalized linear model form, so that it works for a binary or count type of response variable. 

Next, assuming that $f_0:\Xcal^p\to\R$ resides in an RKHS \citep{wahba1990}, we decompose $f_0$ as, 
\begin{equation} \label{eqn:decomoff0}
f_0(x) = \eta(x, \theta_0) + \delta_0(x),
\end{equation}
where $\eta(\cdot, \theta_0)$ is a parametric component that preserves the interpretability of $f_0(\cdot)$, and $\delta_0$ is a nonparametric component that accounts for model error. Together, they form a nonparametric model for $f_0(x)$. Despite the wide use of a simple linear model for $f_0(\cdot)$ in the literature, there has been ample evidence showing that the linear model is inadequate to capture the complex association between $X$ and $Y$ \citep[e.g.,][]{WangYP2018}. This has motivated us to consider a more flexible model for $\eta(\cdot,\theta_0)$, meanwhile taking into account the model error $\delta_0(\cdot)$ as in \eqref{eqn:decomoff0}.  

Next, we employ a basis expansion type model for $\eta(x, \theta_0)$, due to its ease of interpretation, relative flexibility, as well as computational efficiency \citep{huang2007efficient, wang2014estimation, shujie2015estimation}. Specifically, let $\{\phi_1,\ldots,\phi_s\}$ denote a collection of orthonormal and centered basis functions in $\Xcal$, satisfying that $\mathbb E[\phi_k(X_{(j)})]=0$, $j = 1, \ldots, p, k = 1, \ldots, s$, where $s$ is the number of basis functions. There is a rich library of basis functions, including polynomial basis, Fourier basis, B-splines, among others. Denote $\Bcal_s(x_{(j)}) = \text{Span}\{1,\phi_1(x_{(j)}),\ldots,\phi_s(x_{(j)})\}$ as the space spanned by these basis functions. Let the parametric component $\eta(x,\theta_0)$ be the projection of $f_0$ onto the space spanned by the tensor product of the basis functions, i.e., 
\begin{equation} \label{eqn:lineartheta}
\begin{aligned}
\eta(x,\theta_0) = \underset{f\in\otimes_{j=1}^{p}\Bcal_s(x_{(j)})}{\arg\min}\int_{\Xcal^p}[f(x)-f_0(x)]^2dP(x) = \Phi(x)\trans \theta_0,
\end{aligned}
\end{equation}
where $x=(x_{(1)},\ldots,x_{(p)})\trans\in\Xcal^p$, the basis vector $\Phi(x) = [1,\phi_1(x_{(1)}), \ldots, \phi_s(x_{(1)}), \ldots, \phi_1(x_{(p)})$, $\ldots, \phi_s(x_{(p)}),\ldots,\phi_1(x_{(1)})\cdots\phi_p(x_{(p)})]\trans \in \R^{d}$, and $d = (s+1)^p$. Model \eqref{eqn:lineartheta} is a general model that includes main effects $\phi_i(x_{(j)})$, $i=1,\ldots,s,j=1,\ldots,p$, pairwise interactions $\phi_{i_1}(x_{(j_1)})\phi_{i_2}(x_{(j_2)})$, $i_1,i_2=1,\ldots,s,j_1, j_2 = 1, \ldots, p$, as well as higher-order interactions. It includes additive model \citep{hastie1990generalized}, linear model, and functional ANOVA model \citep{LinZhang2006} as special cases. That is, when $\eta(x,\theta_0)$ is the projection of $f_0$ onto the space $\oplus_{j=1}^p\Bcal_s(x_{(j)})$ spanned by the sum of the basis, then \eqref{eqn:lineartheta} is essentially an additive model.  When $s=1$ and $\phi_s(\cdot)$ is a centered linear basis function, \eqref{eqn:lineartheta} becomes a linear model. When $\eta(x,\theta_0)$ is the projection of $f_0$ onto the space spanned by the tensor product of the basis with pairwise or higher-order interactions, \eqref{eqn:lineartheta} becomes a functional ANOVA model.

Finally, we characterize the association between the primary modality $X$ and the auxiliary modalities $Z$ as,
\vspace{-0.05in}
\begin{equation} \label{eqn:confandmediator}
\Phi(X) = r_0(Z) + V, \quad \mathbb E[V | Z] = 0,
\end{equation}
where $V \in \R^{d}$ accounts for the part of the variation in $\Phi(X)$ that cannot be explained by $Z$, and $r_0$ captures the complicated association between $Z$ and $\Phi(X)$. 

Suppose the observed data $\{(X_i,Z_i,Y_i):i=1,\ldots,N\}$ are independent copies of $(X,Z,Y)$ and satisfy the system of models \eqref{eqn:regeqn} to \eqref{eqn:confandmediator}. Our main goal is the statistical inference of $\theta_0$, which reflects the interpretable effect of the primary modality $X$ on the outcome $Y$, and of $f_0$, which reflects the predicted effect of the primary modality, and is also directly related to some causal effect and the quantification of the contribution of $X$. Meanwhile, we view $\{g_0,\delta_0,r_0\}$ as nuisance functions, and propose to use highly flexible machine learning methods, e.g., random forests, reproducing kernels, or neural networks, to model them. The machine learning methods often use regularization to avoid overfitting, especially when $X$ and $Z$ are high-dimensional and highly nonlinear. However, regularization would introduce sizable bias, and would invalidate the subsequent inference on $\theta_0$ and $f_0$. Actually, the naive estimator of $\theta_0$ by simply plugging in the machine learning estimators of $\{g_0,\delta_0,r_0\}$ would fail to be $\sqrt{N}$-consistent; see Section \ref{sec:comparison}. This has motivated us to develop an orthogonal statistical inference framework to correct the bias introduced by the flexible estimators of $\{g_0,\delta_0,r_0\}$, and to perform a valid inference for $\theta_0$ and $f_0$.

%%%%%%%%%%%%%%%%%%%%%%%%%%%%%%%%%%%%%%%%%%%%%%%%%%%
\section{Orthogonalized Kernel Debiased Machine Learning}
\label{sec:estimation}

We consider two orthogonality formulations that are essential for the construction of our estimator. We then present our estimation algorithm built on those orthogonal formulations.

\subsection{Orthogonality}
\label{sec:orthogonality}

The first is the Neyman orthogonality \citep{Neyman1959, Neyman1979,  Chernozhukov2018}, which allows the estimation of $\theta_0$ to be locally insensitive to the values of nuisance functions, and thus one can plug in noisy estimates of the nuisance functions for the inference of $\theta_0$. We consider the target parameter $\theta \in \R^d$, and the nuisance functions $r \in \Hcal_r, g \in \Hcal_g, \delta \in \Hcal_\delta$, where $\Hcal_r$ and $\Hcal_g$ are functional spaces of finite mean squared functions, and $\Hcal_\delta$ is an RKHS.

\begin{definition}[Neyman orthogonality]  \label{defi:neyman-orth}
A score function $\psi(\theta,r,g,\delta)$ is said to satisfy the Neyman orthogonality \citep{Neyman1959, Neyman1979, Chernozhukov2018} if (i) The mean $\E[\psi(\theta_0,r_0,g_0,\delta_0)] = 0$ at $(\theta_0,r_0,g_0,\delta_0)$; (ii) The pathwise derivative map, $\partial_r\{\E[\psi(\theta_0,r_0+t(r-r_0),g_0+t(g-g_0),\delta_0+t(\delta-\delta_0))]\}$, exists for all $t\in[0,1)$,  where $r$, $g$ and $\delta$ lie in a neighborhood of $r_0\in\Hcal_r$, $g_0\in\Hcal_g$ and $\delta_0\in\Hcal_\delta$, respectively; (iii) The pathwise derivative vanishes at $t=0$, in that $\partial_t\{\E[\psi(\theta_0,r_0+t(r-r_0),g_0+t(g-g_0),\delta_0+t(\delta-\delta_0))]\}\vert_{t=0} = 0$. 
\end{definition}

\begin{proposition} \label{thm:neymanscore}
Define the score function, 
\vspace{-0.05in}
\begin{equation*}
\psi(W;\theta,r,g,\delta) = [Y-\Phi(X)\trans\theta-g(Z)-\delta(X)] [r(Z)-\Phi(X)],
\end{equation*}
where $W=(X,Y,Z)$. Then under the system of models (\ref{eqn:regeqn}) to (\ref{eqn:confandmediator}), the  score $\psi(W;\theta,m,\delta,g) $ is Neyman orthogonal at $(\theta_0,r_0,g_0,\delta_0)$. 
\end{proposition}

\noindent 
We briefly comment that a similar idea to Neyman orthogonality is also used in targeted maximum likelihood estimation \citep{van2006targeted, zheng2011cross}, which constructs an estimation equation for a target parameter and requires the score function to be in the orthogonal complement of the tangent space of the nuisance parameter.

In addition to the Neyman orthogonality, we also require the functions $\Phi$ and $\delta_0$ in models \eqref{eqn:decomoff0} and \eqref{eqn:lineartheta} to satisfy a decomposition orthogonality, which is necessary for the identifiability of $\theta_0$. 

\begin{definition}[Decomposition orthogonality] 
\label{defi:decomp-orth}
Suppose that $\Phi(\cdot)$ is bounded on $\Xcal^p$. The functions $\Phi$ and $\delta_0$ are said to satisfy the decomposition orthogonality if $\E_X[\Phi(X)\delta_0(X)] = 0$. 
\end{definition}

\begin{proposition}
\label{thm:orthogonalofphi}
Under models \eqref{eqn:decomoff0} and \eqref{eqn:lineartheta}, $\theta_0$ is identifiable only if $\Phi$ and $\delta_0$ satisfy the decomposition orthogonality. Moreover, for any reproducing kernel $K(\cdot,\cdot)$ on $\Xcal^p\times \Xcal^p$, define
\begin{align*}
K_\delta(x,x') = K(x,x')-&\E_X[\Phi(X)\trans K(x,X)] \\
& \quad \times \left(\E_{X}\{\E_{X'}[\Phi(X')K(X',X)]\Phi(X)\trans\}\right)^{-1}\E_{X'}[\Phi(X') K(x',X')],
\end{align*}
where $X$ and $X'$ are i.i.d.\ copies of the primary modality. Then $K_\delta(\cdot,\cdot): \Xcal^p\times \Xcal^p \to \R$ is positive definite. Besides, for any $\widehat{\delta}(x) = \sum_{i=1}^mc_iK_\delta(x,x_i)$, with $c_i\in\R, x_i\in\Xcal^p$ and $m\geq 1$, $\Phi(X)$ and $\widehat{\delta}(X)$ satisfy the decomposition orthogonality. 
\end{proposition}

\noindent
The decomposition orthogonality in Definition \ref{defi:decomp-orth} is similar to the perpendicularity requirement in the smoothing splines literature \citep[see, e.g.,][Chapter~3]{wahba1990}, where the null space and the RKHS need to be perpendicular under certain norms in order to find a consistent estimator as the sample size diverges, while we use an $\ell_2$-norm with respect to the distribution of $X$. Hereinafter, let $\Hcal_\delta$ be the corresponding RKHS of the kernel $K_\delta(\cdot,\cdot)$. By the representer theorem \citep{wahba1990}, the $M$-estimator in RKHS $\Hcal_\delta$ can be found in a finite-dimensional subspace of $\Hcal_\delta$, i.e., it can be written as $\widehat{\delta}(x) = \sum_{i=1}^mc_iK_\delta(x,x_i)$, with $c_i\in\R, x_i\in\Xcal^p$ and $m \geq 1$. Proposition \ref{thm:orthogonalofphi} shows that $\widehat{\delta}(X)$ and $\Phi(X)$ satisfy the decomposition orthogonality, which in turn ensures the identifiability of the primary parameter $\theta_0$ we target.

\subsection{Iterative cross-fitting procedure}

We next present an estimation algorithm of $\theta_0$ based on the orthogonality formulations in Propositions \ref{thm:neymanscore} and \ref{thm:orthogonalofphi}. The algorithm consists of five main steps. In the first step, we obtain the initial estimators of $\{ \theta_0, g_0, \delta_0 \}$. In the second step, we split the data into $Q$ disjoint chunks. In the third step, we estimate $r_0$, and in the fourth step, we iteratively update the estimates of $\{ g_0, \delta_0 \}$ and $\theta_0$. In these two steps, we obtain the estimates by leaving out some chunk of data in turn. In the fifth step, we construct the final estimator of $\theta_0$, by first using only one chunk of data at a time, then averaging over all $Q$ chunks. When estimating the nuisance functions $\{r_0, g_0, \delta_0\}$, we employ some penalized learning methods, where we denote $\text{PEN}_{\Hcal_r}(r)$, $\text{PEN}_{\Hcal_g}(g)$,  $\text{PEN}_{\Hcal_\delta}(\delta)$ as the penalty functionals in the candidate functional spaces $\Hcal_r$,  $\Hcal_g$, $\Hcal_\delta$, respectively. Here, $\Hcal_\delta$ is chosen to be the corresponding RKHS of $K_\delta(\cdot,\cdot)$ in Proposition \ref{thm:orthogonalofphi}, and $\text{PEN}_{\Hcal_\delta}(\delta)$ is the penalty based on the squared RKHS-norm in $\Hcal_\delta$. The choices of $\{\Hcal_r,\Hcal_g\}$ as well as the penalty functions depend on specific data applications, and the tuning follows the usual tuning procedures in penalized learning. We first summarize the procedure in Algorithm \ref{alg:trainofrl}, then detail the main steps.

\begin{algorithm}[t!]
\caption{Orthogonalized kernel debiased machine learning algorithm} 
\begin{algorithmic}[1]
\STATE Obtain the initial estimators $\widehat{\theta}^{(0)}, \widehat{g}^{(0)}, \widehat{\delta}^{(0)}$ by (\ref{eqn:initialestimateoftheta}) using all the data. 
\STATE Split the data randomly into $Q$ non-overlapping chunks of equal size. For $q \in [Q]$, denote $I_q$ as the corresponding set of data indices of the $q$th chunk, and $I_q^c = [N] \backslash I_q$. 
\FOR{$q=1$ to $Q$} 
\STATE Obtain the estimator $\widehat{r}_0$ by (\ref{eqn:estimateofr}) using the data in $I_q^c$. 
\ENDFOR
\REPEAT
\FOR{$q=1$ to $Q$} 
\STATE Obtain the iterative estimators $\{ \widehat{g}^{(t)}_{q}, \widehat{\delta}^{(t)}_{q} \}$ by (\ref{eqn:estdeltag}) using the data in $I_q^c$. 
\STATE Obtain the iterative estimator $\widetilde{\theta}^{(t)}_q$ by (\ref{eqn:thetatildeq}) using the data in $I_q$.
\ENDFOR
\STATE Obtain the iterative estimator $\widehat{\theta}^{(t)}$  by (\ref{eqn:thetaq}). 
\UNTIL the stopping criterion is met.
\STATE Construct the final estimator $\widehat{\theta}\in\R^d$ by (\ref{eqn:TDE}) using cross-fitting.
\end{algorithmic} 
\label{alg:trainofrl}
\end{algorithm}

In the first step, we obtain the initial estimators of $\{ \theta_0, g_0, \delta_0 \}$ as, 
\begin{align} \label{eqn:initialestimateoftheta}
\begin{split}
\widehat{\theta}^{(0)} & = \underset{\theta\in\R^d}{\arg\min} \Bigg\{ \frac{1}{N}\sum_{i=1}^N \left[ Y_i-\Phi(X_i)\trans\theta - \widehat{g}^{(0)}(Z) \right]^2 \Bigg\}, \\
\widehat{g}^{(0)} & = \underset{g\in\Hcal_g}{\arg\min} \Bigg\{ \frac{1}{N}\sum_{i=1}^N \left[ Y_i-g(Z_i) \right]^2+\lambda^g_{N}\text{PEN}_{\Hcal_g}(g) \Bigg\},
\end{split}
\end{align}
and $\widehat{\delta}^{(0)} = 0$. Here, $\lambda_N^g \geq 0$ is a tuning parameter, and we use all the $N$ data samples. 

In the second step, we randomly split the sample observations into $Q \geq 2$ non-overlapping chunks of equal size $n=N/Q$. For notational simplicity, we assume $N$ is divisible by $Q$. For each $q \in [Q] = \{1,\ldots,Q\}$, we denote $I_q$ as the set of indices in $[N] = \{1,\ldots,N\}$ corresponding to the data in the $q$th chunk, and denote $I_q^c = [N] \backslash I_q$ as the indices of the complementary data. 

In the third step, we estimate the function $r_0$ by, 
\begin{equation} \label{eqn:estimateofr}
\widehat{r}_{q}  =  \underset{r\in\Hcal_r}{\arg\min} \Bigg\{ \frac{1}{n}\sum_{i\in I_q^c} \left[ \Phi(X_i)-r(Z_i) \right]^2 + \lambda_n^r\text{PEN}_{\Hcal_r}(r) \Bigg\},
\end{equation}
where $\lambda_n^r \geq 0$ is a tuning parameter. Note that we only utilize the data from $I_q^c$ in \eqref{eqn:estimateofr}. Besides, we estimate $r_0$ only once, without any iterations, for each $q \in [Q]$. 

In the fourth step, we iteratively update the estimates of $\{g_0,\delta_0\}$ and $\theta_0$. That is, 
\begin{eqnarray} 
\Big\{ \widehat{g}^{(t)}_{q}, \widehat{\delta}^{(t)}_{q} \Big\} & = & \underset{g\in\Hcal_g,\delta\in\Hcal_\delta}{\arg\min} \Bigg\{ \frac{1}{n}\sum_{i\in I_q^c}\left[ Y_i-\Phi(X_i)\widehat{\theta}^{(t-1)} - \delta(X_i)-g(Z_i) \right]^2 \nonumber \\
&& \quad\quad\quad\quad\quad\quad\quad\quad\quad\quad\; \;\;+ \; \lambda^g_{n}\text{PEN}_{\Hcal_g}(g) + \lambda^\delta_{n}\text{PEN}_{\Hcal_\delta}(\delta) \Bigg\}, \label{eqn:estdeltag} \\
\widetilde{\theta}^{(t)}_q & = & \Bigg\{ \frac{1}{n}\sum_{i\in I_q}\left[ \Phi(X_i)-\widehat{r}_{q}(Z_i) \right] \Phi(X_i)\trans \Bigg\}^{-1} \nonumber \\
&& \quad\quad\quad \times \; \frac{1}{n}\sum_{i\in I_q} \left[ \Phi(X_i)-\widehat{r}_{q}(Z_i) \right] \left[ Y_i-\widehat{g}^{(t)}_{q}(Z_i)-\widehat{\delta}^{(t)}_{q}(X_i) \right], \label{eqn:thetatildeq} \\
\widehat{\theta}^{(t)} & = & \frac{1}{Q} \sum_{q=1}^{Q} \widetilde{\theta}_{q}^{(t)}, \label{eqn:thetaq}
\end{eqnarray}
where $\lambda_n^g, \lambda_n^{\delta} \geq 0$ are the tuning parameters. The estimation in (\ref{eqn:estdeltag}) employs residual learning, since it is based on the residual $[Y-\Phi(X)\widehat{\theta}_q^{(t-1)}]$. The resulting estimator $\widehat{\delta}_q^{(t)}$ satisfies the decomposition orthogonality relative to $\Phi$ in Proposition \ref{thm:orthogonalofphi}. Besides, it involves only the complementary data in $I_q^c$. The estimation in (\ref{eqn:thetatildeq}) employs the Neyman orthogonality formulation in Proposition \ref{thm:neymanscore}, and involves only the data in $I_q$. The estimation in (\ref{eqn:thetaq}) averages $\widetilde{\theta}^{(t)}_{q}$ from (\ref{eqn:thetatildeq}) across all $q = 1, \ldots Q$. Moreover, (\ref{eqn:thetatildeq}) and (\ref{eqn:thetaq}) together utilize the idea of centralized training with decentralized execution \citep{Lowe2017}, which greatly facilitates the convergence of the algorithm. We stop the iterations when some stopping criterion is met, e.g., when the difference between two consecutive estimates of $\theta_0$ is smaller than a threshold value. We also remark that, this step is essentially a Gauss-Seidel iterative algorithm that has been widely used in statistics \citep{buja1989linear}. In our simulations, we find the algorithm converges fast, usually after only 3 to 5 iterations. We denote the final estimators for $\{g_0,\delta_0\}$ as $\{\widehat{g}_q, \widehat{\delta}_q\}, q \in [Q]$. 

In the final step, we construct our orthogonal estimator for $\theta_0$ using cross-fitting, 
\begin{align} \label{eqn:TDE}
\begin{split}
\widehat{\theta} & =  \Bigg\{ \frac{1}{Q}\sum_{q=1}^Q\frac{1}{n}\sum_{i\in I_q}\left[ \Phi(X_i)-\widehat{r}_{q}(Z_i) \right] \Phi(X_i)\trans \Bigg\}^{-1}\\
& \quad\quad\quad  \times \frac{1}{Q}\sum_{q=1}^Q\frac{1}{n}\sum_{i\in I_q}\left[ \Phi(X_i)-\widehat{r}_{q}(Z_i) \right] \left[ Y_i-\widehat{g}_{q}(Z_i)-\widehat{\delta}_{q}(X_i) \right].
\end{split}
\end{align}
That is, for each $q \in [Q]$, we use the chunk of data that is left out when estimating $\{r_0, g_0, \delta_0\}$ earlier, then average over all $Q$ chunks. Cross-fitting has been commonly used in high-dimensional inferences in recent years; see, e.g., \citet{Chernozhukov2018,  newey2018cross}. By swapping the roles of each chunk and the complementary chunks $Q$ times, it ensures good statistical properties while regaining the efficiency of making use of all available data observations. Later, we show the  estimator $\widehat{\theta}$ in \eqref{eqn:TDE} is actually semi-parametric efficient.

%%%%%%%%%%%%%%%%%%%%%%%%%%%%%%%%%%%%%%%%%%%%%%%%%%%
\section{Statistical Inference}
\label{sec:inference}

We aim at two key inference questions: inference for the primary parameter of interest $\theta_0$, and inference for the primary regression function $f_0(\cdot)$. Both are crucial for scientific inquires. The former directly quantifies the relevance of the variables of the primary modality to the outcome. The latter captures the predicted effect and the contribution of the primary modality, and also has some causal interpretation under additional conditions.

\subsection{Inference of the primary parameter $\theta_0$}
\label{sec:parameterinference}

We begin with the study of the asymptotic behavior of the estimator $\widehat{\theta}$ in (\ref{eqn:TDE}) as the sample size $N$ tends to infinity. We establish the $\sqrt{N}$-convergence that $\|\widehat{\theta}-\theta_0\|_{\ell_2}=O_p(N^{-1/2})$, as well as the asymptotic normality that $\sqrt{N}(\widehat{\theta}-\theta_0)$ approaches a  normal distribution. We note that this $\sqrt{N}$-convergence result is highly nontrivial, because the estimator $\widehat{\theta}$ in (\ref{eqn:TDE}) involves the nuisance estimators $\{ \widehat{r}_q, \widehat{g}_q, \widehat{\delta}_q \}$. When $\{r_0,g_0,\delta_0\}$ are estimated nonparametrically, the convergence rates of the estimators $\{ \widehat{r}_q, \widehat{g}_q, \widehat{\delta}_q \}$ are generally slower than $O_p(N^{-1/2})$ \citep{Vandevaart1998}. Later in Section \ref{sec:comparison}, we show that many popular alternative methods cannot achieve the $\sqrt{N}$-consistency.

We first present a set of regularity conditions. 
\vspace{-0.05in}
\begin{enumerate}
\item[(C1)] The basis vector $\Phi(\cdot)$ in (\ref{eqn:lineartheta}) satisfies that $\E[ \|\Phi(X)\|^2_{\ell_2} ] < \infty$.
\item[(C2)] The error term $V\in\R^d$ in (\ref{eqn:confandmediator}) satisfies that $\E(VV\trans)$ is invertible and $\E(V\trans V)<\infty$. 
\item[(C3)] The estimators $\widehat{r}_q$ as constructed in (\ref{eqn:estimateofr}), and $\{ \widehat{g}_q, \widehat{\delta}_q \}$ as constructed in (\ref{eqn:estdeltag}) at the algorithmic convergence satisfy that $\E[ \|\widehat{r}_q(Z)-r_0(Z)\|_{\ell_2}^2 ] = o(N^{-1/2})$, $\E\{ [ \widehat{g}_q(Z)-g_0(Z) ]^2 \} = o(N^{-1/2})$, and $\E\{ [ \widehat{\delta}_q(X)-\delta_0(X) ]^2 \} = o(N^{-1/2})$, for $q \in [Q]$ and $Q$ is finite. 
\vspace{-0.05in}
\end{enumerate}

\noindent
Condition (C1) is mild and holds for most practical choices of the basis functions. For example, (C1) holds with the continuous basis over the compact domain $\Xcal^{p}$. Condition (C2) is a fairly standard regularity condition, and is needed for the asymptotic normality of parameter estimation in moment-based problems \citep{Chernozhukov2018}. Condition (C3) is different from requiring the estimators $\{ \widehat{r}_q, \widehat{g}_q, \widehat{\delta}_q \}$ to be $\sqrt{N}$-consistent, which is difficult to satisfy for many nonparametric estimators. Instead, (C3) holds for a wide range of popular machine learning methods; for instance, it holds for the $\ell_1$-penalized linear regression in a variety of sparse models \citep{bickel2009simultaneous, buhlmann2011statistics}, a class of random forests \citep{biau2012analysis}, a class of neural networks \citep{chen1999improved}, and numerous kernel methods in RKHS \citep{wahba1990, Vandevaart1998}, among others. Moreover, we note that (C3) is generally less restrictive than the Donsker conditions, which are commonly assumed in semi-parametric statistical analysis \citep{kosorok2007introduction}. The Donsker conditions require the functional spaces $\{\Hcal_r,\Hcal_g,\Hcal_\delta\}$ to have a bounded complexity, or more specifically, a bounded entropy integral. However, for multimodal data analysis where the dimension of the auxiliary modalities $Z$ increases with the sample size, such a requirement fails even in the linear model setting with the parameter space specified by the Euclidean ball of unit radius \citep{Raskutti2011}. By contrast, (C3) holds in this example. 

Under (C1) to (C3), we obtain the main theoretical result for our estimator $\widehat{\theta}$. 

\begin{theorem}
\label{thm:mainresulttri}
Suppose the system of models (\ref{eqn:regeqn}) to (\ref{eqn:confandmediator}), and the regularity conditions (C1) to (C3) hold. The orthogonalized kernel debiased machine learning estimator $\widehat{\theta}$ in (\ref{eqn:TDE}) satisfies that, 
\vspace{-0.01in}
\begin{equation*}
 \widehat{\theta}-\theta_0 = [\E(VV\trans)]^{-1} \left( \frac{1}{N}\sum_{i=1}^N V_i U_i \right) + o_p(N^{-1/2}). 
\end{equation*}
where $\{(U_i,V_i):i=1,\ldots,N\}$ are independent copies of the error terms $(U,V)$ in (\ref{eqn:regeqn}) and (\ref{eqn:confandmediator}).
\end{theorem}

\noindent
The proof of this theorem is given in Appendix \ref{sec:pfthmparameter}. We make two remarks. First, a direct implication of Theorem \ref{thm:mainresulttri} is the asymptotic normality of $\widehat{\theta}$, i.e., 
\begin{equation} \label{eqn:asympnormalityoftheta}
\sqrt{N}(\widehat{\theta}-\theta_0)\overset{d}{\to} \mathcal N\left( 0,\sigma^2 [\E(VV\trans)]^{-1} \right).
\end{equation}
Second, the asymptotic normality in (\ref{eqn:asympnormalityoftheta}) further implies that we can construct the confidence interval for the primary parameter of interest $\theta_0$ as, 
\begin{equation*}
\text{CI}(\theta_0) = \widehat{\theta}\pm F_{\mathcal N}^{-1}(1-\alpha/2) \sqrt{ \sigma^2 (\E[VV\trans])^{-1}/N},
\end{equation*}
where $F_{\mathcal N}(\cdot)$ denotes the cumulative distribution function of the standard normal distribution. When the variance term $\sigma^2\E[VV\trans]$ in \eqref{eqn:asympnormalityoftheta} is unknown, we use a plug-in estimator, 
\begin{equation*}
\widehat{\Sigma}(\widehat{\theta}) = \widehat{J}^{-1} \Bigg\{ \frac{1}{nQ}\sum_{q=1}^Q \sum_{i\in I_q}
\Big[ Y_i-\Phi(X_i)\trans\widehat{\theta}-\widehat{g}_q(Z_i)-\widehat{\delta}_q(X_i) \Big]^2 [\widehat{r}_q(Z_i)-\Phi(X_i)][\widehat{r}_q(Z_i)-\Phi(X_i)]\trans \Bigg\} \widehat{J}^{-1},
\end{equation*} 
where $\widehat{J} = (nQ)^{-1} \sum_{q=1}^Q \sum_{i\in I_q}[\Phi(X_i)-\widehat{r}_{q}(Z_i)]\Phi(X_i)\trans$. The next corollary shows that this plug-in estimator is consistent, and its proof is given in Appendix \ref{sec:proofofcovarianceestimation}.

\begin{corollary} \label{thm:covarestimation}
Suppose the conditions of Theorem \ref{thm:mainresulttri} hold. If $U$ in  (\ref{eqn:regeqn}) and the elements of $V$ in (\ref{eqn:confandmediator}) have bounded fourth moment, then the plug-in estimator $\widehat{\Sigma}(\widehat{\theta})$ is consistent, in that 
\begin{equation*}
\widehat{\Sigma}(\widehat{\theta}) \overset{p}{\to} \sigma^2 \left( \E[VV\trans] \right)^{-1}.
\end{equation*} 
\end{corollary}

Next, we discuss the efficiency of the estimator $\widehat{\theta}$. We first note that the estimation problem for $\theta_0$ under the system of models (\ref{eqn:regeqn}) to (\ref{eqn:confandmediator}) is semi-parametric. This is because the parameter of interest $\theta_0 \in \R^d$ is finite-dimensional as specified in (\ref{eqn:lineartheta}), while the parameter space of models (\ref{eqn:regeqn}) and (\ref{eqn:decomoff0}) contains high-dimensional, or infinite-dimensional functional spaces as $\{g_0,\delta_0\}\in\Hcal_g\otimes\Hcal_\delta$. We also allow the dimensions of $g_0$ and $\delta_0$ to grow with the sample size $N$. The next theorem shows that $\widehat{\theta}$ in (\ref{eqn:TDE}) is semi-parametric efficient \citep{kosorok2007introduction}, in that it achieves the highest possible efficiency, if the measurement error $U$ follows a normal distribution. The proof of this theorem is given in Appendix \ref{sec:pfofsemi-parametric}, along with a brief review of the background on semi-parametric estimation efficiency. 

\begin{theorem}
\label{thm:semvariance}
Suppose the conditions of Theorem \ref{thm:mainresulttri} hold. If the measurement error $U$ in (\ref{eqn:regeqn}) follows a normal distribution, then the estimator $\widehat{\theta}$ in (\ref{eqn:TDE}) is semi-parametric efficient. 
\end{theorem}

\subsection{Inference of the primary function $f_0$}
\label{sec:uqoff0}

We next consider inference of the primary regression function $f_0(\cdot)$, which is of particular interest for several reasons. First of all, it quantifies the predicted effect of the primary modality $X$ on the outcome $Y$. In addition, it also captures the amount of contribution of the primary modality, in terms of the percentage of variation explained, given all other modalities in the model. Finally, under some additional assumptions, $f_0$ is directly related to the notions of the partial dependence of $Y$ on $X$, as well as the total effect of $X$ on $Y$ in a causal inference sense. 

Given the orthogonal estimator $\widehat{\theta}$ in (\ref{eqn:TDE}), a natural estimator for $f_0$ is $\widehat{f}(x) = \Phi(x)\trans\widehat{\theta}$.  We seek the confidence band for $f_0$. A confidence band $\Ccal_N$ is a set of confidence intervals, $\Ccal_N = \big\{ \Ccal_N(x) = [c_L(x),c_U(x)] \ \big| \ x\in\Xcal^p \big\}$.  Consider the empirical process $\sup_{x\in\Xcal^p}\sqrt{N}[\widehat{f}(x)-f_0(x)]$, whose distribution can be approximated by a Gaussian multiplier process, 
\begin{equation*}
\widehat{\mathbb H}_N(x) = \sqrt{N}\Phi(x)\trans\Bigg\{ \frac{1}{nQ}\sum_{q=1}^Q \sum_{i\in I_q}[\Phi(X_i)-\widehat{r}_{q}(Z_i)]\Phi(X_i)\trans \Bigg\}^{-1} 
\frac{1}{nQ}\sum_{q=1}^Q \sum_{i\in I_q}[\Phi(X_i)-\widehat{r}_{q}(Z_i)]\widehat{\sigma}(\widehat{\theta})\xi_{i},
\end{equation*}
where the estimator $\widehat{\sigma}^2(\widehat{\theta}) = (nQ)^{-1} \sum_{q=1}^Q \sum_{i\in I_q}[Y_i-\Phi(X_i)\trans\widehat{\theta}-\widehat{g}_{q}(Z_i)-\widehat{\delta}_{q}(X_i)]^2$, and $\xi = (\xi_1,\ldots,\xi_N)\trans\in\R^N$ are independent $\mathcal N(0,1)$ random variables. Let $\widehat{c}_N(\alpha/2)$ be the $(1-\alpha/2)$th quantile of $\sup_{x\in\Xcal^p}\widehat{\mathbb H}_N(x)$. We construct the $100\times(1-\alpha)\%$ confidence band for $f_0$ as, 
\vspace{-0.01in}
\begin{equation} \label{eqn:cioff0}
\Ccal_N= \left.\left\{\Ccal_N(x) = \left[\widehat{f}(x) - \frac{\widehat{c}_N(\alpha/2)}{\sqrt{N}}, \widehat{f}(x) + \frac{\widehat{c}_N(\alpha/2)}{\sqrt{N}}\right]\ \right|\ x\in\Xcal^p\right\}. 
\end{equation}

To establish the asymptotic validity of \eqref{eqn:cioff0}, we first present a modified version of the regularity condition (C3), and an additional condition regarding the function $f_0$.

\vspace{-0.05in}
\begin{enumerate}
\item[(C3$'$)] The estimators $\widehat{r}_q$ as constructed in (\ref{eqn:estimateofr}), and $\{ \widehat{g}_q, \widehat{\delta}_q \}$ as constructed in (\ref{eqn:estdeltag}) at the algorithmic convergence satisfy that $\E[\|\widehat{r}_q(Z)-r_0(Z)\|_{\ell_2}^2] = O(N^{-1/2-c_r})$, $\E[(\widehat{g}_q(Z)-g_0(Z))^2] = O(N^{-1/2-c_g})$, and $\E[(\widehat{\delta}_q(X)-\delta_0(X))^2] = O(N^{-1/2-c_\delta})$, for some constants $c_r,c_g,c_\delta\in(0,1/2]$, $q \in [Q]$, and $Q$ is finite. 

\item[(C4)] The function $f_0: \Xcal^p\to\R$ resides in the $k$th-order Sobolev space, $k>p$, in that $f_0$ and the derivatives $f_0^{(\nu)}$ are absolutely continuous for any vector of nonnegative integers $\nu \in \N_0^p$ with $\|\nu\|_{\ell_1}\leq k-1$, and $\E\{[f_0^{(\nu)}(X)]^2\}<\infty$ for any $\nu\in\N_0^p$ with $\|\nu\|_{\ell_1}=k$.
\vspace{-0.05in}
\end{enumerate}

\noindent 
Condition (C3$'$) is slightly stronger than (C3), which is necessary to obtain the asymptotic validity of the confidence band $\Ccal_N$ in \eqref{eqn:cioff0}. Nevertheless, (C3$'$) continues to hold for a wide range of commonly-used machine learning methods, including all the aforementioned ones where (C3) holds. Condition (C4) is a standard regularity condition in the literature on nonparametric estimations \citep{wahba1990,Vandevaart1998}. 

The next theorem shows that the confidence band $\Ccal_N$ in \eqref{eqn:cioff0} is asymptotically valid, in the sense that the coverage holds uniformly for all $x\in\Xcal^p$ under a fixed $f_0$,
\begin{equation*} \label{eqn:asympthonest}
\underset{N\to\infty}{\lim\inf}\ \P\big[ f_0(x) \in \Ccal_N(x),\text{ for all }x\in\Xcal^p \big] \geq 1-\alpha.
\end{equation*}

\begin{theorem}
\label{thm:infonpred}
Suppose the system of models (\ref{eqn:regeqn}) to (\ref{eqn:confandmediator}), and the regularity conditions (C1), (C2), (C3$'$) and (C4) hold. Let $s$ be the number of bases for each function component in (\ref{eqn:lineartheta}), and $c_{\min} = \min\{c_r,c_g,c_\delta\}>0$. Suppose the measurement error $U$ in (\ref{eqn:regeqn}) follows a normal distribution, and the number of basis functions $s= \ceil{N^{(1+2c)/2k}}$ for a constant $c \in \big( 0, (k-p)/2(k+p) \big]$. Then, there exist a constant $C>0$, such that the coverage of the confidence band $\Ccal_N$ in \eqref{eqn:cioff0} satisfies, 
\begin{equation*}
\P\big[ f_0(x) \in \Ccal_N(x), \text{ for all }x\in\Xcal^p \big] \geq 1-\alpha-CN^{-c}, \; \textrm{ for any } \; 0 < \alpha < 1.
\end{equation*}
Consequently, the confidence band $\Ccal_N$ in \eqref{eqn:cioff0} is asymptotically valid.
\end{theorem}

\noindent
The proof of this theorem is given in Appendix \ref{sec:pfofasympthonestci}, and is built upon the framework of using the Gaussian multiplier process to approximate the distribution of the supremum of empirical processes \citep{chernozhukov2014anti}. We first note that, for the inference of $f_0$, we require the number of basis functions $s$ to diverge with the sample size, but for the inference of $\theta_0$, we do \emph{not} require a diverging $s$. When $s$ diverges, the error term $V \in \R^{(s+1)^p}$ in \eqref{eqn:confandmediator} has a diverging dimension too. Nevertheless, Theorem \ref{thm:infonpred} continues to hold. We next compare Theorem \ref{thm:infonpred} with \citet{lu2020kernel} and \citet{kozbur2020inference}. \citet{lu2020kernel} studied the inference of nonparametric additive models, but required there only exists a weak dependency between the covariates, e.g., between $X$ and $Z$, in that the difference between the joint distribution and the product of marginal distributions is small under a certain norm. Multimodal data, however, are typically highly correlated \citep{Uludaug2014}, and as such, the requirement of \citet{lu2020kernel} may not always hold. By contrast, we allow a strong dependency between $X$ and $Z$, and employ \eqref{eqn:confandmediator} to model potentially complex dependency between $X$ and $Z$. \citet{kozbur2020inference} considered a nonparametric primary function $f_0$ through basis expansion, but required the approximation error to vanish at a rate faster than $\sqrt{N}$, which can be rather restrictive. By contrast, we do not require a vanishing approximation error for our method. This has a crucial implication, because it essentially allows one to use a simple and interpretable model to characterize the parametric component of $f_0$, e.g., a linear model, which itself can be inaccurate and may induce a non-negligible approximation error. Finally, we briefly comment that, to establish an honest confidence band with a uniform coverage for all $f_0 \in \Hcal_f$ and data-generating functions, one needs to fully characterize $\Hcal_f$ and to extend the classical Smirnov-Bickel-Rosenblatt condition \citep{gine2009exponential} to the multimodal setting. We leave a full investigation as future research.

In addition to the predicted effect, the function $f_0$ also captures the amount of contribution of the primary modality given other modalities. Recall that in the classical linear regression model, the coefficient of determination $R^2$ measures the percentage of total variation in the response that has been explained by the predictors. We next show that $f_0$ is directly related to $R^2$, then derive the confidence interval for the $R^2$ measure. Consider the population version of $R^2$, 
\begin{equation}
\label{eqn:ciofr2}
R^2 = 1-\frac{\E(\text{RSS}) }{\E(\text{TSS})}, \;\; \text{ where } \; \E(\text{RSS}) = \E\left[ \{Y- f_0(X)\}^2 \right], \; \E(\text{TSS}) = \E\left[ (Y-\bar{Y})^2 \right], 
\end{equation}
$\bar{Y} = N^{-1}\sum_{i=1}^NY_i$, and $\text{RSS}$ and $\text{TSS}$ denote the residual sum of squares and total sum of squares, respectively. Define $\widehat{f}_{(1)}(x) = \widehat{f}(x)- N^{-1/2}\widehat{c}_N(\alpha/2)$, and $\widehat{f}_{(2)}(x) = \widehat{f}(x)+ N^{-1/2}\widehat{c}_N(\alpha/2)$. Then denote $R^2_{(1)} =  1-\sum_{i=1}^N[Y_i- \widehat{f}_{(1)}(X_i)]^2/\sum_{i=1}^N(Y_i-\bar{Y})^2$, and $R^2_{(2)} = 1-\sum_{i=1}^N[Y_i- \widehat{f}_{(2)}(X_i)]^2/\sum_{i=1}^N(Y_i-\bar{Y})^2$. We construct  the $100\times(1-\alpha)\%$ confidence interval for $R^2$ as,
\begin{equation*} \label{eqn:constructofciforr2}
\text{CI}(R^2) = \left( \min(R^2_{(1)},R^2_{(2)}), \; \max(R^2_{(1)},R^2_{(2)}) \right). 
\end{equation*}
The next corollary, following directly from Theorem \ref{thm:infonpred}, shows this is a valid confidence interval.
\begin{corollary}
\label{cor:ciforr2}
Suppose the conditions of Theorem \ref{thm:infonpred} hold. The confidence interval $\text{CI}(R^2)$ is valid, in that $\underset{N\to\infty}{\lim\inf}\ \P\left[ R^2\in \text{CI}(R^2) \right] \geq 1-\alpha$.
\end{corollary}

Finally, we note that $f_0$, under some additional conditions, has a causal interpretation, and is directly related to the notions of partial dependence and total effect. Consequently, our proposed orthogonal inference procedure for $f_0$ may be useful for inferring causal effect. 

Specifically, following \citet{friedman2001greedy}, the partial dependence of the response $Y$ on the primary modality $X=x_0\in\Xcal^p$ is defined as, 
\begin{equation} \label{eqn:pdp}
\E_{Z}\left[ \E_{U}(Y) \right] = \E_{Z}\left[ f_0(x_0) + g_0(z) \right] = f_0(x_0)+c,\quad c\in\R,
\end{equation}
where $(X,Z,Y)$ follows model (\ref{eqn:regeqn}). That is, the partial dependence is the expectation of $Y$ over the marginal distribution of all modalities other than $X$. It is different from the conditional expectation, $\E_{Z}[\E_{U}(Y) | X=x_0] = \E_{Z|X=x_0}[f_0(x_0) + g_0(z)]$, where the expectation is taken over the conditional distribution of $Z$ given $X=x_0$. By \eqref{eqn:pdp}, we see that the partial dependence is equal to $f_0(x_0)$ up to an additive constant $c$. This property does not hold for the conditional expectation. 

Next, following \citet{pearl2009causality} and \citet{zhao2019causal}, the partial dependence measure in (\ref{eqn:pdp}) coincides with the back-door adjustment formula for identifying the causal effect of $X$ on $Y$ given the observational data. More specifically, view (\ref{eqn:regeqn}) as a structural equation model, where each of the $(M+1)$ modalities $\{X,Z_{(1)},\ldots,Z_{(M)}\}$ corresponds to one of the $(M+1)$ nodes in a directed acyclic graph \citep{pearl2009causality}. Let a path be a consecutive sequence of edges of the directed graph, and a back-door path be a path that contains an arrow into $X$. If the following back-door criteria are satisfied, such that none of $\{Z_{(1)},\ldots,Z_{(M)}\}$ is a descendant of $X$, and $\{Z_{(1)},\ldots,Z_{(M)}\}$ blocks all back-door paths between $X$ and $Y$, then the partial dependence measure in (\ref{eqn:pdp}), or equivalently $f_0(\cdot)$, can be interpreted as the total effect of the primary modality $X$ affecting the outcome $Y$.

%%%%%%%%%%%%%%%%%%%%%%%%%%%%%%%%%%%%%%%%%%%%%%%%%%%
\section{Comparison with Alternative Methods}
\label{sec:comparison}

We next analytically compare our method with a number of important alternative solutions, and carefully evaluate the asymptotic behavior of each estimator.

\subsection{Uni-modality regression}
\label{sec:unimodelreg}

A common solution in practice is to focus on a single data modality and exclude all other modalities from the analysis. This approach is simple, and shares a similar spirit as the marginal regression \citep{fan2008sure}. We term it as the \emph{uni-modality regression}. Specifically, it regresses the outcome on the primary modality, and estimate the primary parameter $\theta_0$ by,
\begin{equation*} \label{eqn:estimatorur}
\widehat{\theta}_{\text{UR}} = \underset{\theta\in\R^d}{\arg\min} \left\{ \frac{1}{N}\sum_{i=1}^N \left[ Y_i-\Phi(X_i)\trans\theta \right]^2 \right\}.
\end{equation*}

Proposition \ref{prop:unimodelregression} characterizes the asymptotic behavior of the uni-modality estimator $\widehat{\theta}_{\text{UR}}$. 

\begin{proposition} \label{prop:unimodelregression}
Suppose the system of models (\ref{eqn:regeqn}) to (\ref{eqn:confandmediator}) hold. Suppose $\E[\Phi(X)\Phi(X)\trans]$ is invertible. Then the uni-modality regression estimator $\widehat{\theta}_{\text{UR}}$ satisfies that, 
\begin{equation*}
\widehat{\theta}_{\text{UR}} -\theta_0 = \left\{ \E[ \Phi(X)\Phi(X)\trans ] \right\}^{-1} \left\{ \frac{1}{N}\sum_{i=1}^N \Phi(X_i) \left[ \delta_0(X_i)+g_0(Z_i)+U_i \right] \right\}+ o_p(N^{-1/2}).
\end{equation*}
\end{proposition}

\noindent
The proof of this proposition is given in Appendix \ref{sec:pfofunimodal}. We next compare the behavior of $\widehat{\theta}_{\text{UR}}$ with our orthogonal estimator $\widehat{\theta}$ in (\ref{eqn:TDE}) in terms of the asymptotic bias and variance, respectively. 

In terms of the bias, we note that $\widehat{\theta}_{\text{UR}}$ may suffer from a severe bias, because 
\begin{equation*} \label{eqn:biasofur}
\E(\widehat{\theta}_{\text{UR}}) - \theta_0 = \left\{ \E[\Phi(X)\Phi(X)\trans] \right\}^{-1} \E\{ \Phi(X) [ \delta_0(X)+g_0(Z) ] \} + o(N^{-1/2}),
\end{equation*}
which can be arbitrarily large, due to both the model error $\delta_0$ in \eqref{eqn:decomoff0}, and the effect of the auxiliary modality reflected by $g_0$ in (\ref{eqn:regeqn}). In multimodal analysis, however, both $\delta_0$ and $g_0$ can be substantial. Because of this bias, we have $\sqrt{N}(\widehat{\theta}_{\text{UR}} -\theta_0) = O_p(\sqrt{N})$, which diverges as $N$ tends to infinity. Consequently, $\widehat{\theta}_{\text{UR}}$ is unsuitable for statistical inference tasks. By contrast, the proposed orthogonal estimator $\widehat{\theta}$ is asymptotically unbiased. 
 
In terms of the variance, we note that $\widehat{\theta}_{\text{UR}}$ achieves a variance that is no larger than that of $\widehat{\theta}$. Specifically, the asymptotic variance of $\widehat{\theta}_{\text{UR}}$ is $\text{Var}(\widehat{\theta}_{\text{UR}}) = N^{-1}\sigma^2 \{\E[\Phi(X)\Phi(X)\trans]\}^{-1}$. Compared to the asymptotic variance of our orthogonal estimator $\widehat{\theta}$ as given in (\ref{eqn:asympnormalityoftheta}), we have, 
\begin{equation*} \label{eqn:varthetislarger}
\text{Var}(\widehat{\theta}) - \text{Var}( \widehat{\theta}_{\text{UR}} ) \geq 0, \text{ as } N\to\infty,
\end{equation*} 
in the sense that the difference of the two covariance matrices is semi-positive definite. The two asymptotic variances are equal only when $r_0 = 0$ in (\ref{eqn:confandmediator}), i.e., when the primary and auxiliary modalities are completely independent of each other. The inflated variance of $\widehat{\theta}$ compared to that of $\widehat{\theta}_{\text{UR}}$ is due to the intrinsic correlation between $X$ and $Z$ that is modeled by $r_0$. It can be viewed as a generalization of the well-known variance inflation phenomenon in the classical linear regression model due to the collinearity. For instance, consider the linear model $Y = X\theta_0+ Z\trans\beta_0+U$, with $\E(X)=\E(Y)=0$. The variance of the least squared estimator becomes $\E(U^2) / [\E(X^2) (1-\kappa)]$ after incorporating the auxiliary modality $Z$, where $\kappa = \E(XZ\trans) [\E(ZZ\trans)]^{-1} \E(ZX) / \E(X^2)$ characterizes the correlation between $X$ and $Z$. This variance increases compared to the case when there is no $Z$ in the model.  On the other hand, we also note that, the orthogonal estimator $\widehat{\theta}$ actually attains the smallest possible variance when $Z$ is incorporated, as shown in Theorem \ref{thm:semvariance}.

\subsection{Debiased uni-modality regression}

We next consider a debiased version of the uni-modality regression. Numerous debiasing strategies have been successfully developed in high-dimensional regression modeling in recent years \citep[see, e.g.,][among others]{Zhang2014,Vandegeer2014,cai2017}. The debiased estimator is obtained in two stages. First, the model error $\delta_0$ is estimated based on the uni-modality regression estimator $\widehat{\theta}_{\text{UR}}$ and some machine learning method as in (\ref{eqn:estdeltag}), 
\vspace{-0.05in}
\begin{equation*}
\widehat{\delta}_{\text{DUR}}   = \underset{\delta\in\Hcal_\delta}{\arg\min} \left\{ \frac{1}{N}\sum_{i=1}^N \left[ Y_i-\Phi(X_i)\widehat{\theta}_{\text{UR}}- \delta(X_i) \right]^2+\lambda^\delta_{N}\text{PEN}_{\Hcal_\delta}(\delta) \right\}, 
\end{equation*}
where $\lambda_N^\delta\geq 0$ is a tuning parameter. Then the debiased estimator of $\theta_0$ is obtained by explicitly taking the model error into account, 
\begin{equation*} \label{eqn:defofthetacheck}
\widehat{\theta}_{\text{DUR}} = \underset{\theta\in\R^d}{\arg\min} \left\{ \frac{1}{N}\sum_{i=1}^N \left[ Y_i-\Phi(X_i)\trans\theta-\widehat{\delta}_{\text{DUR}}(X_i) \right]^2 \right\}.
\end{equation*}

Proposition \ref{thm:twostep} characterizes the asymptotic behavior of the debiased uni-modality estimator $\widehat{\theta}_{\text{DUR}}$. 

\begin{proposition} \label{thm:twostep}
Suppose the conditions of Proposition \ref{prop:unimodelregression} hold. Suppose the regularity condition (C1) holds. Then the debiased uni-modality regression estimator $\widehat{\theta}_{\text{DUR}}$ satisfies that, 
\vspace{-0.01in}
\begin{align*}
\begin{split}
\widehat{\theta}_{\text{DUR}} -\theta_0 = \{ \E[\Phi(X)\Phi(X)\trans] \}^{-1} \left\{ \frac{1}{N}\sum_{i=1}^N \Phi(X_i)[g_0(Z_i)+U_i] \right\}\\
+ O_p[(\E\{ [\widehat{\delta}_{\text{DUR}}(X)-\delta_0(X)]^2\} )^{1/2}]+o_p(N^{-1/2}).
\end{split}
\end{align*}
\end{proposition}

\noindent
The proof of this proposition is given in Appendix \ref{sec:pfofpropsingledebiased}. We make two observations regarding the asymptotic bias of $\widehat{\theta}_{\text{DUR}}$. First, $\widehat{\theta}_{\text{DUR}}$ indeed achieves a reduced bias compared to the uni-modality estimator $\widehat{\theta}_{\text{UR}}$. 
This is because under the regularity condition (C3),  the bias of $\widehat{\theta}_{\text{DUR}}$ is
\begin{equation*} \label{eqn:biasofudr}
\E(\widehat{\theta}_{\text{DUR}}) - \theta_0 = \{ \E[\Phi(X)\Phi(X)\trans] \}^{-1} \E[\Phi(X)g_0(Z)] + o(N^{-1/4}).
\end{equation*}
Comparing this bias with that of $\widehat{\theta}_{\text{UR}}$, we see that $\widehat{\theta}_{\text{DUR}}$ removes the bias term due to the model error $\delta_0$ as $N\to\infty$,  but $\widehat{\theta}_{\text{UR}}$ does not. On the other hand, $\widehat{\theta}_{\text{DUR}}$ is still an inconsistent and biased estimator of $\theta_0$, because $\widehat{\theta}_{\text{DUR}}$ does not remove the bias due to the effect of the auxiliary modality $g_0$. Consequently,  $\widehat{\theta}_{\text{DUR}}$ is unsuitable for statistical inference neither.

\subsection{Simple joint regression}
\label{sec:sjr}

Another common solution in multimodal analysis is to incorporate multiple data modalities in a simple additive fashion into a single regression model. This strategy is intuitive, and we term it as the \emph{simple joint regression}. Specifically, it obtains the joint estimator for $\{\theta_0,g_0\}$ as, 
\begin{equation*} \label{eqn:sjr}
\{\widehat{\theta}_{\text{SJR}},\widehat{g}_{\text{SJR}}\} = \underset{\theta\in\R^d,g\in\Hcal_g}{\arg\min} \left\{ \frac{1}{N}\sum_{i=1}^N \left[ Y_i-\Phi(X_i)\trans\theta - g(Z) \right]^2+\lambda^g_{N}\text{PEN}_{\Hcal_g}(g) \right\},
\end{equation*}
where $\lambda_N^g\geq 0$ is a tuning parameter, and $\widehat{g}_{\text{SJR}}$ is obtained by a machine learning method as in (\ref{eqn:estdeltag}). 

Proposition \ref{thm:sjr} characterizes the asymptotic behavior of the simple joint estimator $\widehat{\theta}_{\text{SJR}}$. 

\begin{proposition} \label{thm:sjr}
Suppose the conditions of Proposition \ref{prop:unimodelregression} hold. Suppose the regularity condition (C1) holds. Then the simple joint regression estimator $\widehat{\theta}_{\text{SJR}}$ satisfies that, 
\begin{equation*}
\begin{aligned}
\widehat{\theta}_{\text{SJR}} - \theta_0 = \{ \E[\Phi(X)\Phi(X)\trans] \}^{-1} \left\{ \frac{1}{N}\sum_{i=1}^N\Phi(X_i)[\delta_0(X_i)+U_i] \right\} \\
+ O_p( (\E\{ [\widehat{g}_{\text{SJR}}(Z)- g_0(Z)]^2\} )^{1/2}) + o_p(N^{-1/2}).
\end{aligned}
\end{equation*}
\end{proposition}

\noindent
The proof of this proposition is given in Appendix \ref{sec:pfofsjr}. We again study the asymptotic behavior of $\widehat{\theta}_{\text{SJR}}$. Under the regularity condition (C3), the asymptotic bias of $\widehat{\theta}_{\text{SJR}}$ is, 
\begin{equation*} \label{eqn:biasofsjr}
\E(\widehat{\theta}_{\text{SJR}}) - \theta_0 = \{\E[\Phi(X)\Phi(X)\trans]\}^{-1} \E[\Phi(X)\delta_0(X)] + o(N^{-1/4}),
\end{equation*}
which is not vanishing due to the non-zero model error $\delta_0$. The mean squared error of $\widehat{\theta}_{\text{SJR}}$ is,
\begin{equation*}
\E\big[ (\widehat{\theta}_{\text{SJR}}-\theta_0)^2 \big] = O(\E\{ [\widehat{g}_{\text{SJR}}(Z)- g_0(Z)]^2+\delta^2_0(X) \}),
\end{equation*}
which does not converge at the rate of $N^{-1}$ if $\widehat{g}_{\text{SJR}}$ is estimated using  machine learning methods, or if $\delta_0$ is not negligible. Consequently, $\widehat{\theta}_{\text{SJR}}$ is generally an inefficient and biased estimator of $\theta_0$.

\subsection{Double/debiased machine learning}
\label{sec:dml}

The seminal work of \cite{Chernozhukov2018} developed the framework of double/debiased machine learning (DML), which lays the foundation for the inference of the primary parameter of interest in the presence of high-dimensional nuisance parameters. Our proposal extends the DML framework to incorporate the additional model error $\delta_0$. More specifically, DML randomly splits the data into $Q$ disjoint chunks, and estimates $g_0$ by
\begin{equation*}
\widehat{g}_{\text{DML},q} = \underset{g\in\Hcal_g}{\arg\min} \left\{ \frac{1}{n}\sum_{i\in I_q^c}[Y_i-g(Z_i)]^2+\lambda^g_{n}\text{PEN}_{\Hcal_g}(g) \right\}. 
\end{equation*}
where $\lambda^g_{n} \geq 0$ is a tuning parameter. It then estimates $\theta_0$ by 
\begin{equation*} \label{eqn:defofdml}
\widehat{\theta}_{\text{DML}} = \left\{\frac{1}{nQ}\sum_{q=1}^Q \sum_{i\in I_q}[\Phi(X_i)-\widehat{r}_{q}(Z_i)]\Phi(X_i)\trans\right\}^{-1} 
\frac{1}{nQ}\sum_{q=1}^Q\sum_{i\in I_q}[\Phi(X_i)-\widehat{r}_{q}(Z_i)][Y_i-\widehat{g}_{\text{DML},q} (Z_i)].
\end{equation*}

Proposition \ref{prop:propertyofDML} characterizes the asymptotic behavior of DML estimator $\widehat{\theta}_{\text{DML}}$.

\begin{proposition} \label{prop:propertyofDML}
Suppose the conditions of Proposition \ref{prop:unimodelregression} hold. Suppose the regularity conditions (C1) to (C3) hold. Then the DML estimator $\widehat{\theta}_{\text{DML}}$ satisfies that, 
\begin{equation*}
\widehat{\theta}_{\text{DML}} -\theta_0 = (\E[VV\trans])^{-1} \left( \frac{1}{N}\sum_{i=1}^N V_iU_i \right) + O_p(\{\E[\delta^2_0(X)]\}^{1/2})+o_p(N^{-1/2}).
\end{equation*}
\end{proposition}

\noindent
The proof is given in Appendix \ref{sec:pfofdml}. The mean squared error of $\widehat{\theta}_{\text{DML}}$ is, 
\begin{equation*}
\E\big[ (\widehat{\theta}_{\text{DML}} -\theta_0)^2 \big] = \frac{1}{N}\sigma^2(\E[VV\trans])^{-1}+ O(\E[\delta^2_0(X)])+ o(N^{-1}).
\end{equation*}
Compared to our estimator $\widehat{\theta}$, whose mean squared error is $N^{-1} \sigma^2(\E[VV\trans])^{-1}+ o(N^{-1})$, $\widehat{\theta}_{\text{DML}}$ has an inflated mean squared error at the order of $\E[\delta^2_0(X)]$. Consequently, it cannot achieve the $\sqrt{N}$-consistency if the model error $\delta_0$ is not negligible.

%%%%%%%%%%%%%%%%%%%%%%%%%%%%%%%%%%%%%%%%%%%%%%%%%%%
\section{Simulations}
\label{sec:simulations}

We next study the finite-sample performance of the proposed orthogonalized kernel debiased machine learning (OKDML) method. We first evaluate the performance of inferring $\theta_0$ in an additive model setting. We also numerically compare with the alternative methods of uni-modality regression (UR), debiased uni-modality regression (DUR), simple joint regression (SJR), and double machine learning (DML) that ignores $\delta_0$. We next evaluate the performance of inferring $f_0$ in a high-dimensional additive setting. We also study the sensitivity of using different machine learning methods for nuisance function estimation when inferring $\theta_0$, and report the results in Section \ref{sec:sim-sensitivity} of the Appendix. In all these examples, the model error $\delta_0$ is estimated in the RKHS constructed as in Proposition \ref{thm:orthogonalofphi}. We use the Mat\'{e}rn kernel $K(x,x') = (1+\sqrt{5}\|x-x'\|+5\|x-x'\|^2/3)\exp(-\sqrt{5}\|x-x'\|)$, where the corresponding RKHS contains twice differentiable functions. The tuning parameter $\lambda_n^\delta$ in (\ref{eqn:estdeltag}) is selected by generalized cross-validation \citep{wahba1990}. We set $Q=2$ in Algorithm \ref{alg:trainofrl}.

\subsection{Empirical performance of inference on $\theta_0$}
\label{sec:eg1}

We begin with an additive model, $Y_i = f_0(X_i) + g_{01}(Z_{i1})+ g_{02}(Z_{i2})+ g_{03}(Z_{i3}) + U_i$, where 
\begin{align*}
f_0(x) & =  5x - [\cos(2\pi x)+\sin(2\pi x)], \\
g_{01}(z_1) & =  6\big[0.1\sin(2\pi z_1)+0.2\cos(2\pi z_1)+0.3\sin^2(2\pi z_1) + 0.4\cos^3(2\pi z_1)+0.5\sin^3(2\pi z_1))\big],\\
g_{02}(z_2) & =  3(2z_2-1)^2, \quad g_{03}(z_3)=\frac{4\sin(2\pi z_3)}{2-\sin(2\pi z_3)}.
\end{align*} 
We generate random variables $E_1, \ldots, E_5$ independently from Uniform$[0,1]$, and set the primary and auxiliary modalities as $X = (E_1+\rho E_5)/(1+\rho) \in \Xcal = [0,1]$, and $Z_j=(E_{j+1}+\rho E_5)/(1+\rho)$, for some $\rho > 0$ and $j=1,2,3$. The correlation between any two variables in $X$ and $Z$ is thus $\rho^2/(1+\rho^2)$. We generate i.i.d.\ copies $(X_i,Z_{i1},Z_{i2},Z_{i3})$ of $(X,Z_{1},Z_{2},Z_{3})$, and generate the error $U_i$ from $\Ncal(0,\sigma^2)$. We set the sample size $N=500$. We set $\eta(x,\theta_0) = \theta_0x$, and apply the random forests averaged over $500$ trees to estimate the nuisance functions $\{r_0,g_0\}$.

\begin{figure}[t!]
\centering
\begin{tabular}{ccc}
\includegraphics[width=0.32\textwidth,height=1.525in]{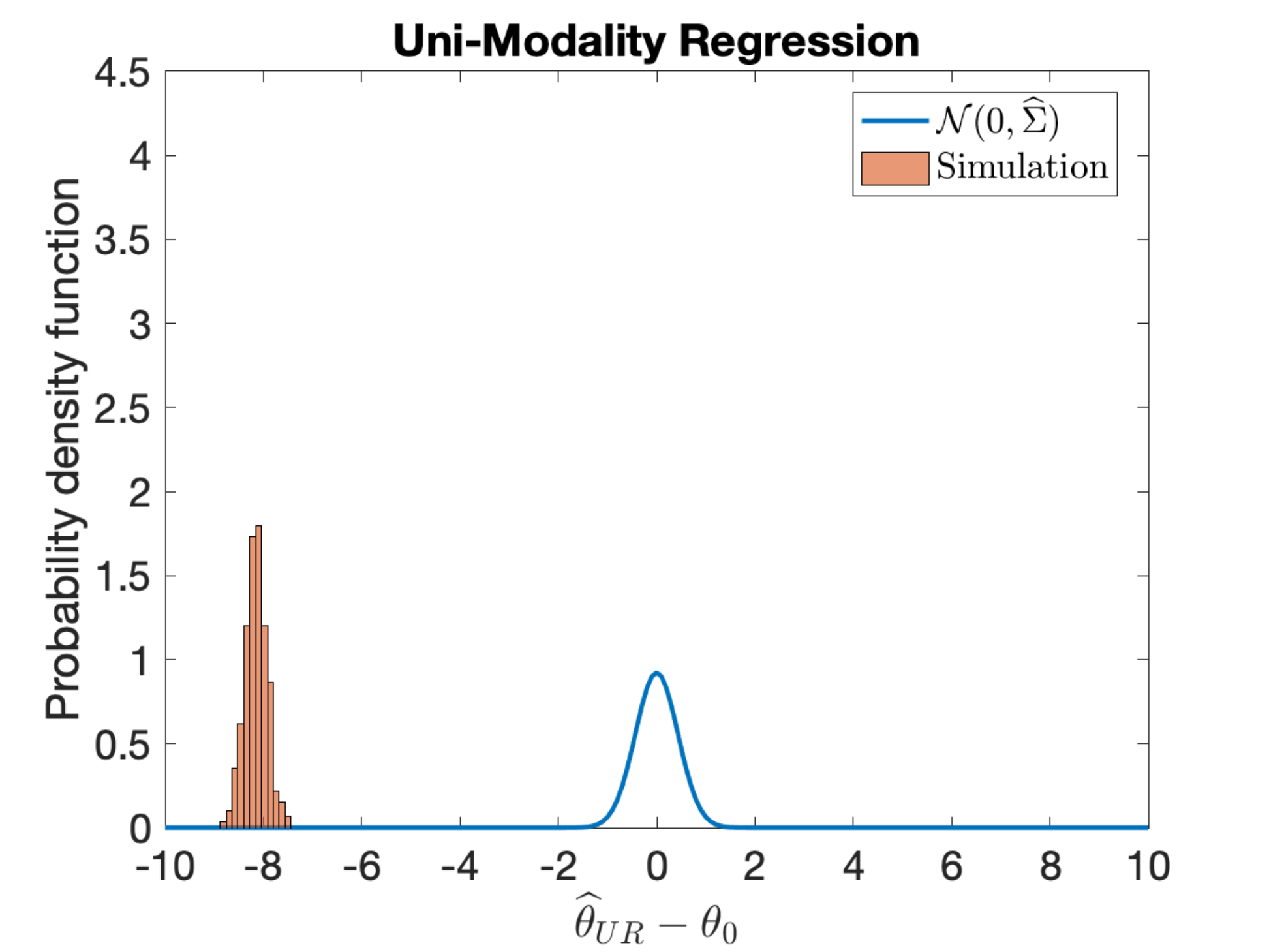} &
\includegraphics[width=0.32\textwidth,height=1.525in]{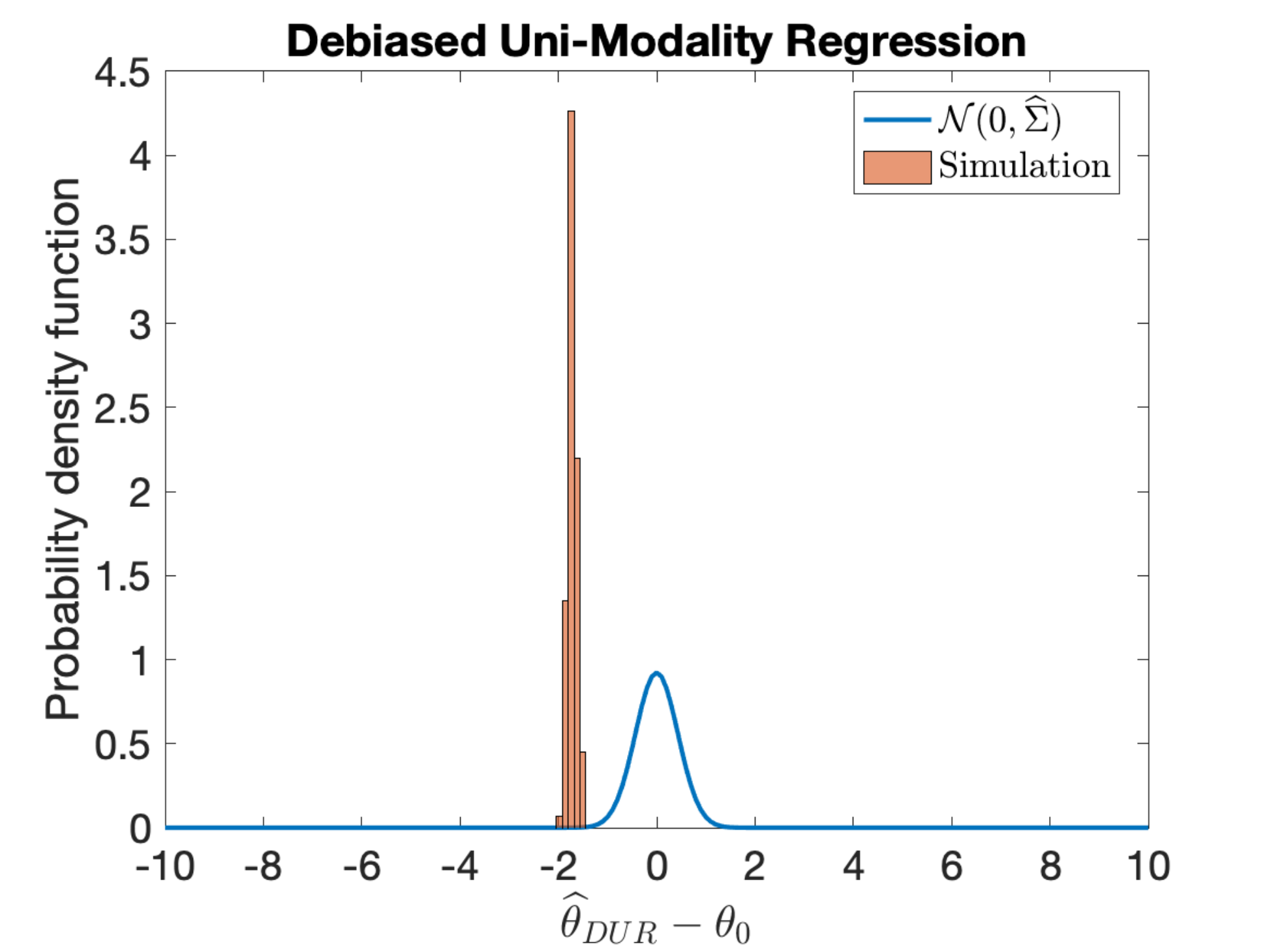} \\
\includegraphics[width=0.32\textwidth,height=1.525in]{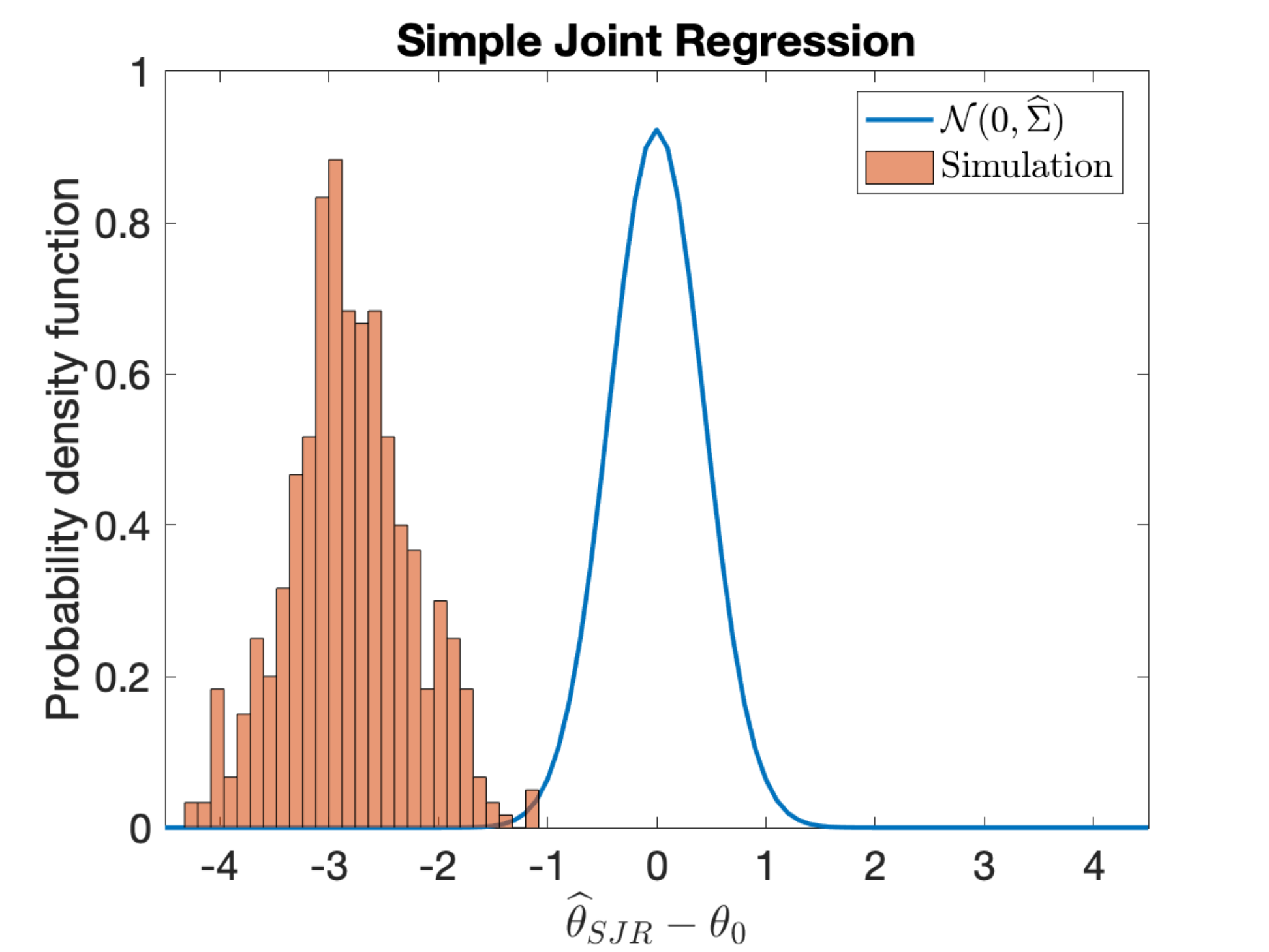} &
\includegraphics[width=0.32\textwidth,height=1.525in]{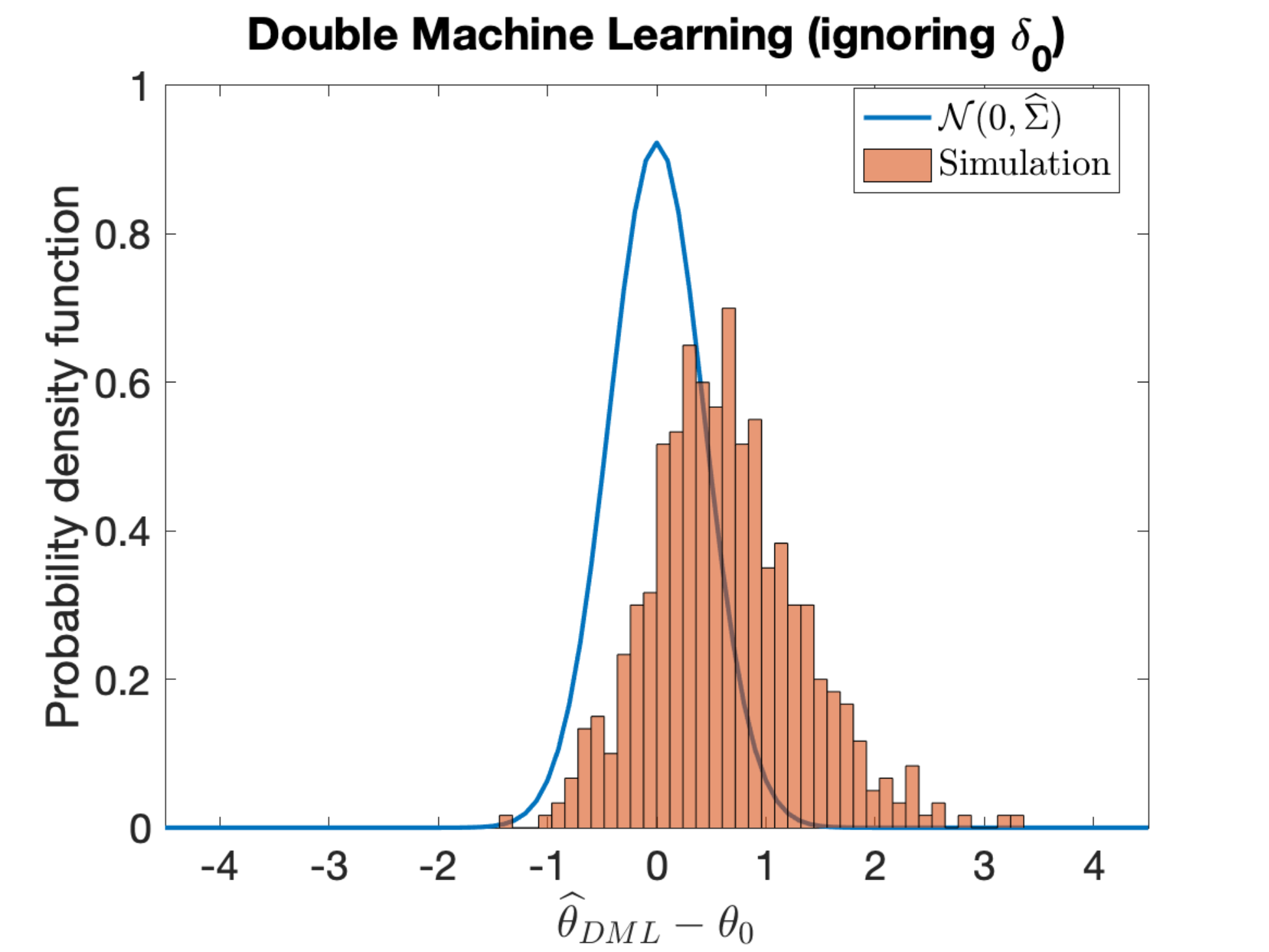} &
\includegraphics[width=0.32\textwidth,height=1.525in]{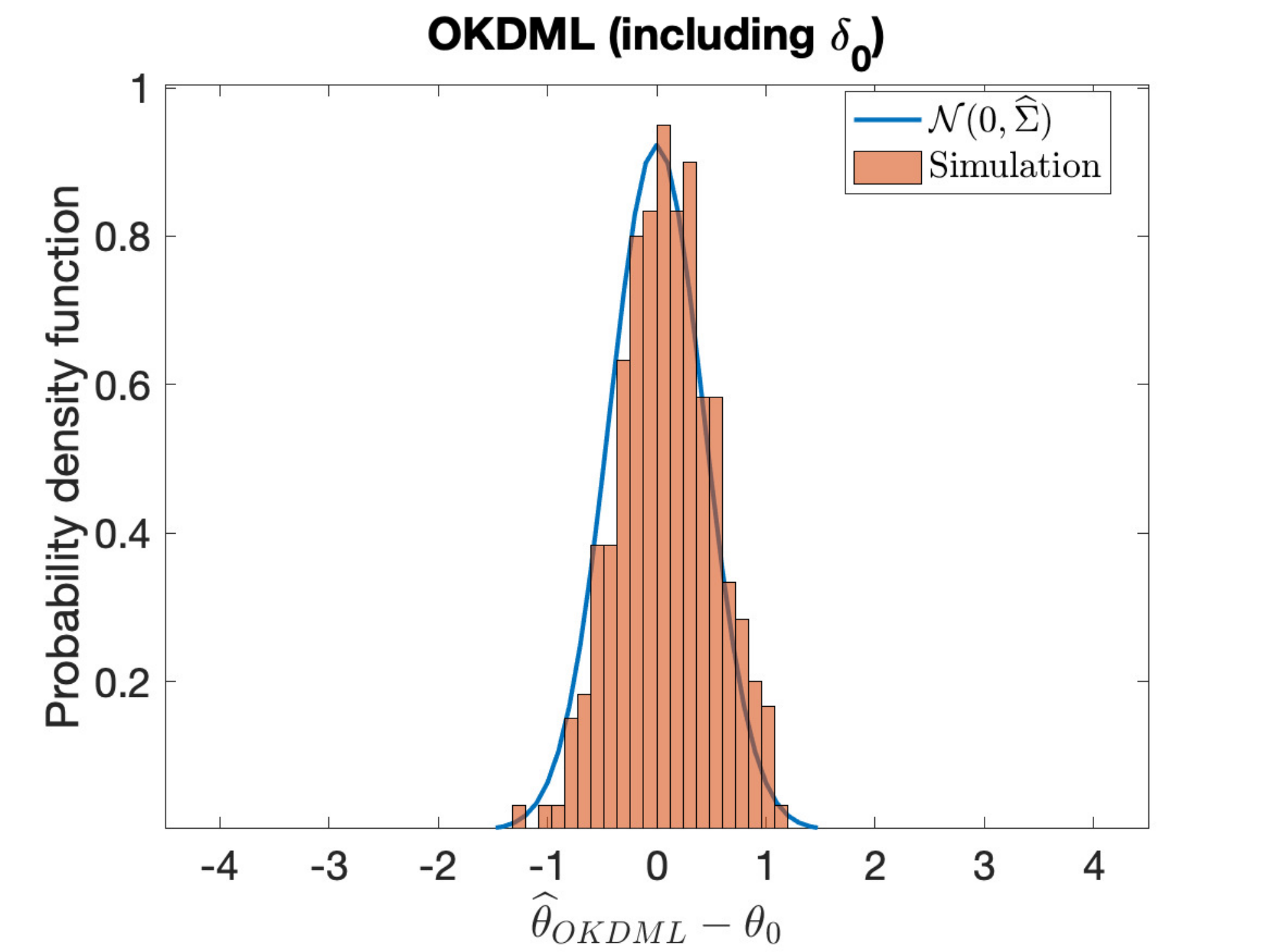} 
\end{tabular}
\caption{Empirical distribution of the estimator of $\theta_0$ based on 500 data replications. The bell-shape curve denotes the oracle normal distribution.}
\label{fig:eg1-hist} 
\end{figure}

\begin{figure}[b!]
\centering
\includegraphics[width=\textwidth]{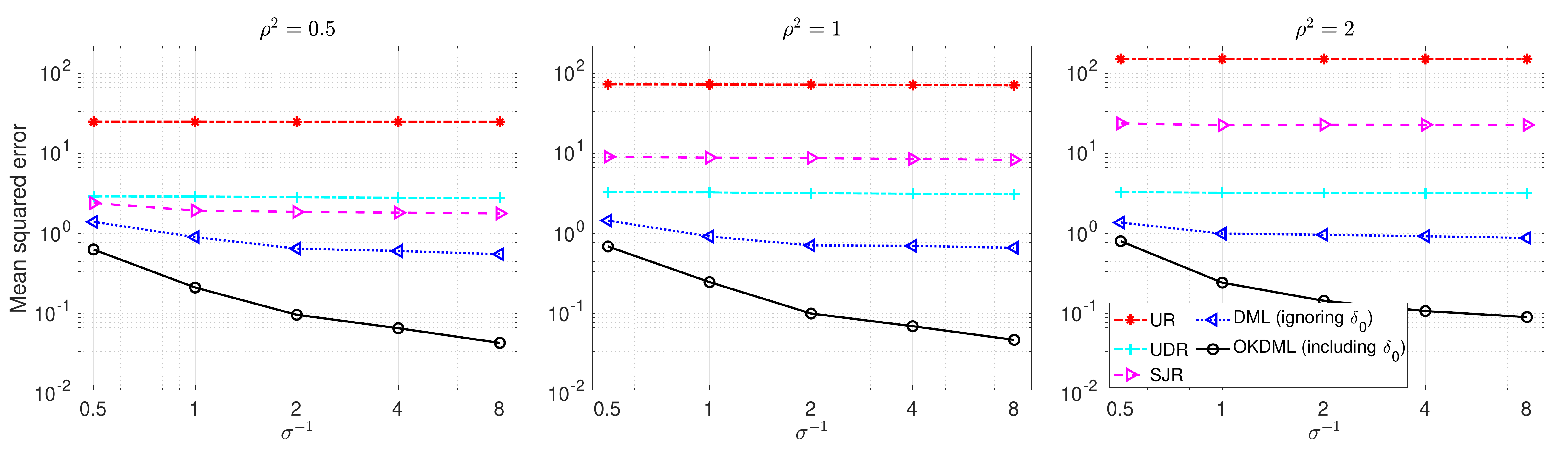}
\caption{Mean squared error of the estimator of $\theta_0$ with varying noise level $\sigma$ and correlation level $\rho$. Both axes are in the log scale.}
\label{fig:eg1-noise-corr} 
\end{figure}

Figure \ref{fig:eg1-hist} shows the histograms of the competing estimators, $\widehat{\theta}_{\text{UR}}, \widehat{\theta}_{\text{DUR}}, \widehat{\theta}_{\text{SJR}}, \widehat{\theta}_{\text{DML}}$, and our proposed OKDML estimator $\widehat{\theta}_{\text{OKDML}}$, under $\rho=1$ and $\sigma=1$, based on 500 data replications. 
It is clearly seen that all four competing estimators are biased, whereas the histogram of the OKDML estimator $\widehat{\theta}_{\text{OKDML}}$ matches that of the normal distribution. Figure \ref{fig:eg1-noise-corr} further reports the empirical mean squared error of different estimators under various combinations of the noise level $\sigma$ and the correlation level $\rho$. When $\sigma^{-1}$ increases, the signal-to-noise ratio increases. However, the mean squared errors of the four competing methods do not decrease much due to the estimation bias, whereas the mean squared error of our OKDML estimator continuously decreases.

\subsection{Empirical performance of inference on $f_0$}

We next consider a high-dimensional additive model, $Y_i = f_0(X_i) + \sum_{j=1}^{600}g_{0j}(Z_{ij}) + U_i$, where $f_0(x),g_{01}(z_1),g_{02}(z_2),g_{03}(z_3)$ are the same as the first example, and 
\begin{equation*}
 g_{0j}(z_j)=z_j, \; \textrm{ for } \  j\in\{4,\ldots, 100\}, \quad  g_{0j}(z_j)=0, \; \textrm{ for } \ j\in\{101,\ldots, 600 \}.
\end{equation*}
We generate random variables $E_1,\ldots,E_{602}$ independently from Uniform$[0,1]$, and set the primary and auxiliary modalities as $X = (E_1+\rho E_{602})/(1+\rho)$, and $Z_j=(E_{j+1}+\rho E_{602})/(1+\rho)$, for $\rho=1$ and $j=1,\ldots,600$. We generate i.i.d.\ copies $(X_i,Z_{i1},Z_{i2},\ldots,,Z_{i600})$ of $(X,Z_{1},Z_{2},\ldots,Z_{600})$, and generate the error $U_i$ from $\Ncal(0,\sigma^2)$ with $\sigma\in\{0.25,0.5,1\}$. We set the sample size  $N=500$. 

We construct both the confidence band (\ref{eqn:cioff0}) for the primary effect $f_0(x)$, and the confidence interval (\ref{eqn:ciofr2}) for the coefficient of determination $R^2$. We use polynomial basis functions with $s=5$ following Theorem \ref{thm:infonpred}, while we estimate $\delta_0$ in a similar way as in the first example. We employ the Lasso to estimate the nuisance functions $\{r_0,g_0\}$ due to the high-dimensionality of this example, and tune the Lasso parameter using tenfold cross-validation. We compute the quantile estimator $\widehat{c}_N(\alpha/2)$ in (\ref{eqn:cioff0}) by bootstrap with $500$ replications.

\begin{figure}[b!]
\centering
\includegraphics[width=\textwidth]{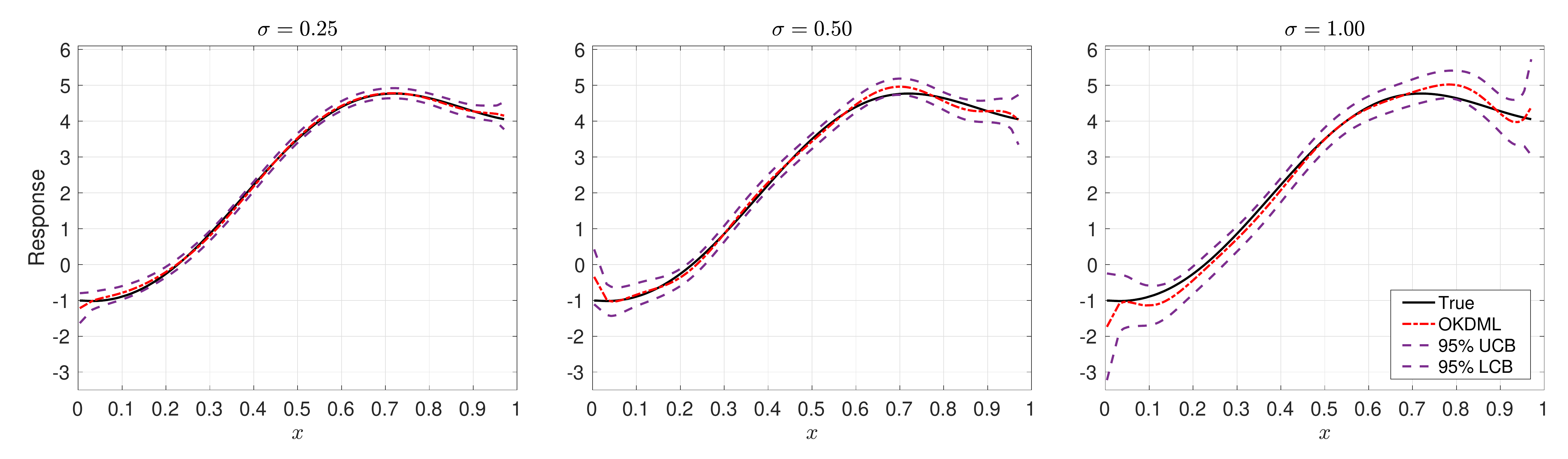}
\caption{The true and estimated primary function $f_0(x)$, with the $95\%$ upper and lower confidence bounds, of the OKDML method, under varying noise level $\sigma$.}
\label{fig:eg3} 
\end{figure}

Figure \ref{fig:eg3} shows the true and estimated primary function $f_0(x)$, along with the $95\%$ upper and lower confidence bounds, of the proposed orthogonal method with the varying noise level $\sigma$. We also compute the empirical coverage probability of the confidence band $\Ccal_N$ at the significance level $95\%$, by discretizing the interval $\Xcal=[0,1]$ into $1000$ grids, then calculating the percentage that the confidence band covers the truth on the $1000$ grid points in $500$ data replications. The resulting coverage probability is $0.968, 0.958$ and $0.946$, when $\sigma = 0.25, 0.50$ and $1.00$, respectively. Moreover, we compute the empirical coverage probability of $ \text{CI}(R^2)$ as the percentage that the confidence interval covers the true $R^2$. The resulting coverage probability is $0.990, 0.972$ and $0.964$, when $\sigma = 0.25, 0.50$ and $1.00$, respectively. It is seen from both the estimated function and the coverage probability that our proposed method works well.

%%%%%%%%%%%%%%%%%%%%%%%%%%%%%%%%%%%%%%%%%%%%%%%%%%%
\section{Multimodal Neuroimaging Study for Alzheimer's Disease}
\label{sec:realdata}

We revisit the motivating example of multimodal neuroimaging analysis for Alzheimer's disease. The data is part of the Berkeley Aging Cohort Study, and consists of 697 subjects. For each subject, the imaging data includes the anatomical MRI scan, which measures brain cortical thickness and is summarized as a 68-dimensional vector that corresponds to 68 predefined brain regions-of-interest (ROIs), and the PET scan, which measures tau deposition and is summarized as a 70-dimensional vector that corresponds to 70 ROIs. In addition, the subject's age, gender, education, and a scalar measure of the total amyloid-$\beta$ accumulation are collected. The response is a composite cognition score that combines assessments of episodic memory, timed executive function, and global cognition. We study two scientific questions given this data, first, the effect of brain atrophy on cognition after controlling for demographic variables and amyloid-$\beta$, tau depositions, and second, the cascade of AD biomarkers as suggested by \citet{Jack2010}.

\begin{table}[b!]
\centering
\caption{Multimodal study of AD: the identified significant brain regions.}
\begin{tabular}{lcccccc}
\toprule
& &  Estimate  & & SD  & &  $p$-value \\ [0.2ex] \midrule
Entorhinal cortex, left              & & $ 3.214$   & &  $0.709$ & &   $6.957\times 10^{-6}$  \\ [0.2ex]
Entorhinal cortex, right                & &   $2.853 $ & &  $0.671$ & &   $2.454\times 10^{-5}$   \\ [0.2ex]
Superior temporal cortex, left          & & $10.42$   & &  $2.444$ & &   $2.321\times 10^{-5}$  \\ [0.2ex]
Superior temporal cortex, right    & &   $5.061$ & &  $1.451$ & &  $5.213\times 10^{-4}$ \\ [0.2ex]
Parahippocampal gyrus, left   & & $1.076$   & &  $0.362$ & &   $3.112\times 10^{-3}$  \\ [0.2ex]
Parahippocampal gyrus, right  & & $1.366$   & &  $0.474$ & &   $4.098\times 10^{-3}$  \\ [0.2ex]
\bottomrule
\end{tabular}
\label{table:app1}
\end{table}

\begin{figure}[t!]
\centering
\begin{tabular}{ccc}
\includegraphics[width=0.32\textwidth,height=1.5in]{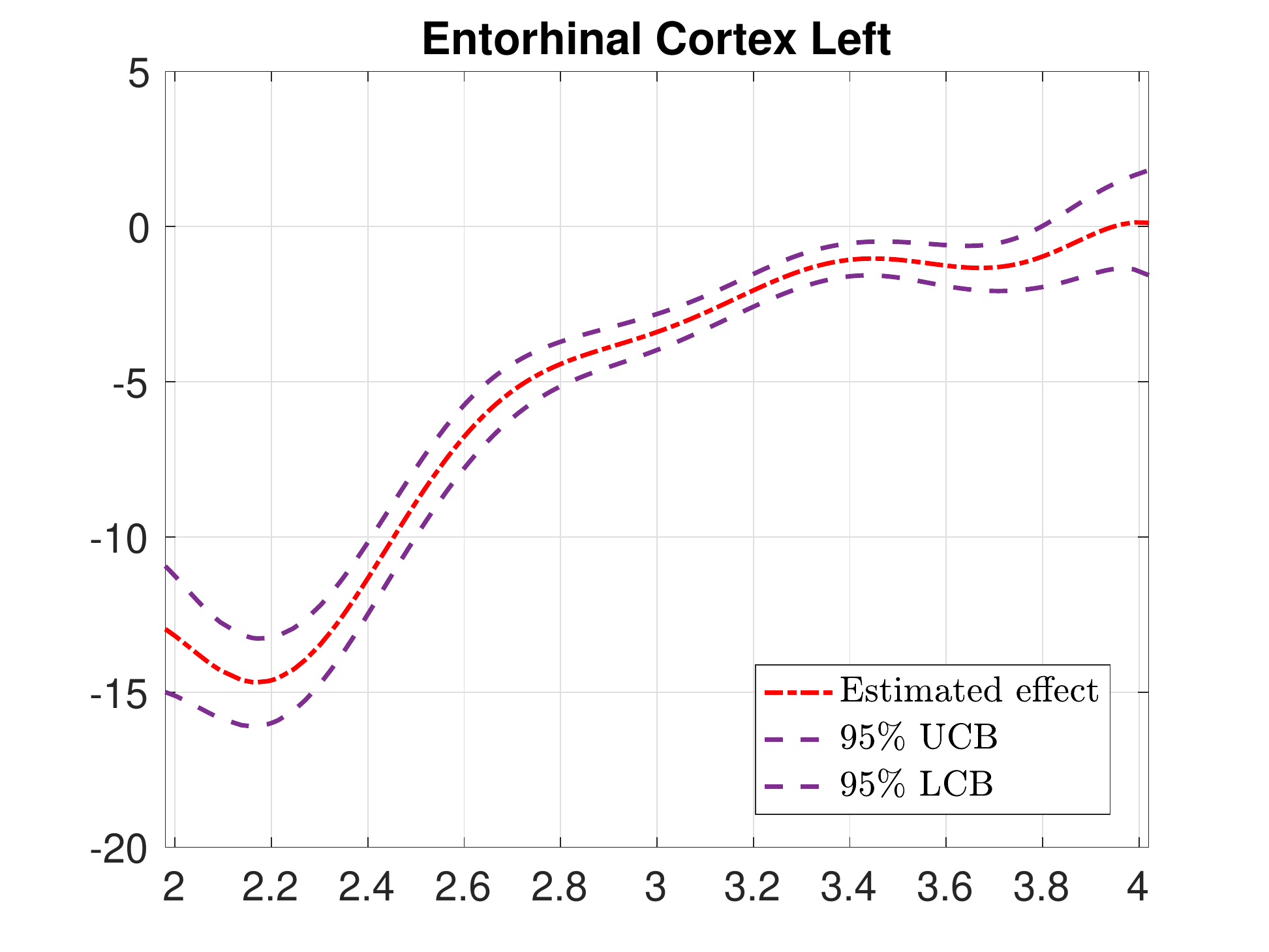} &
\includegraphics[width=0.32\textwidth,height=1.5in]{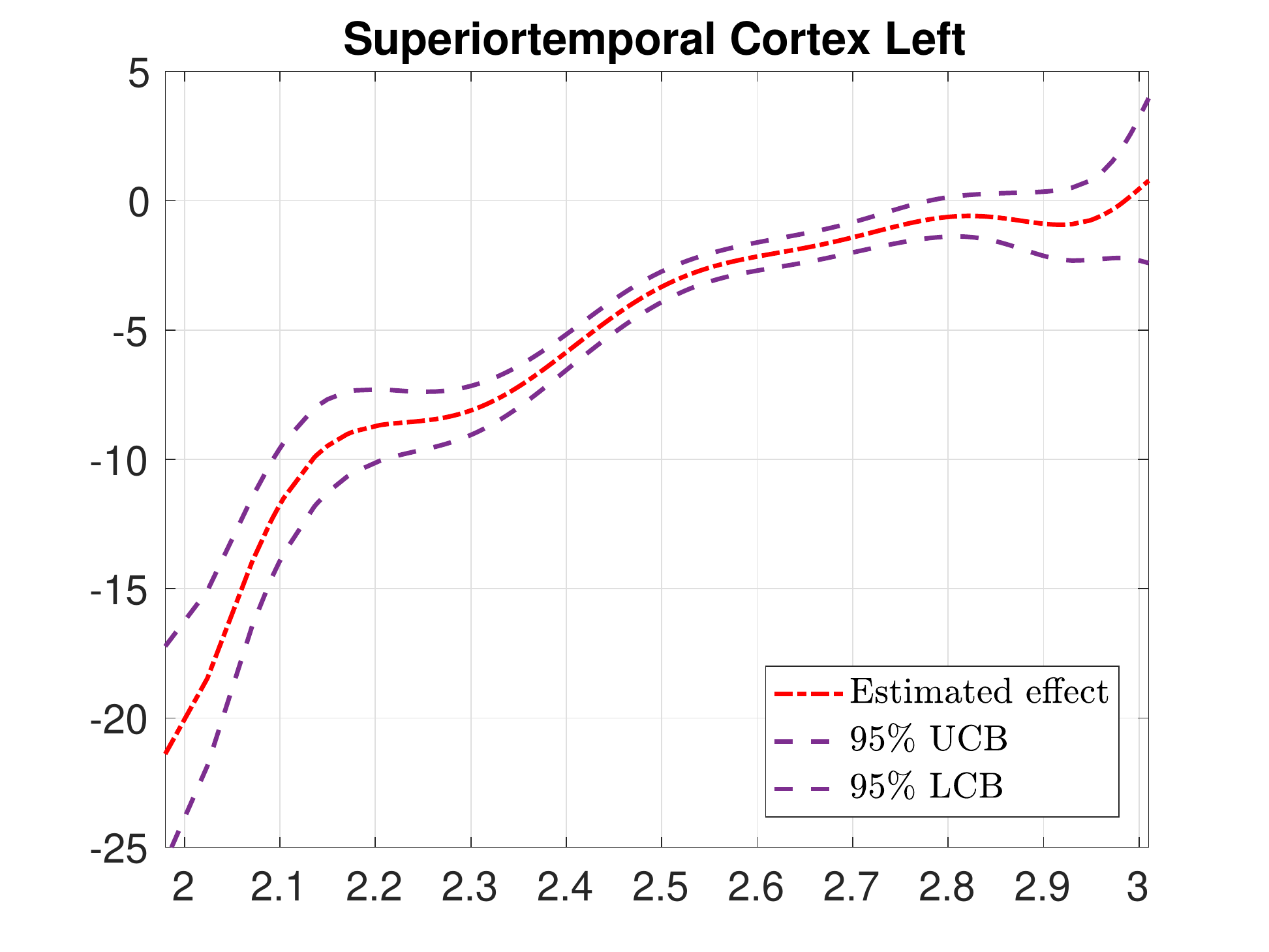} &
\includegraphics[width=0.32\textwidth,height=1.5in]{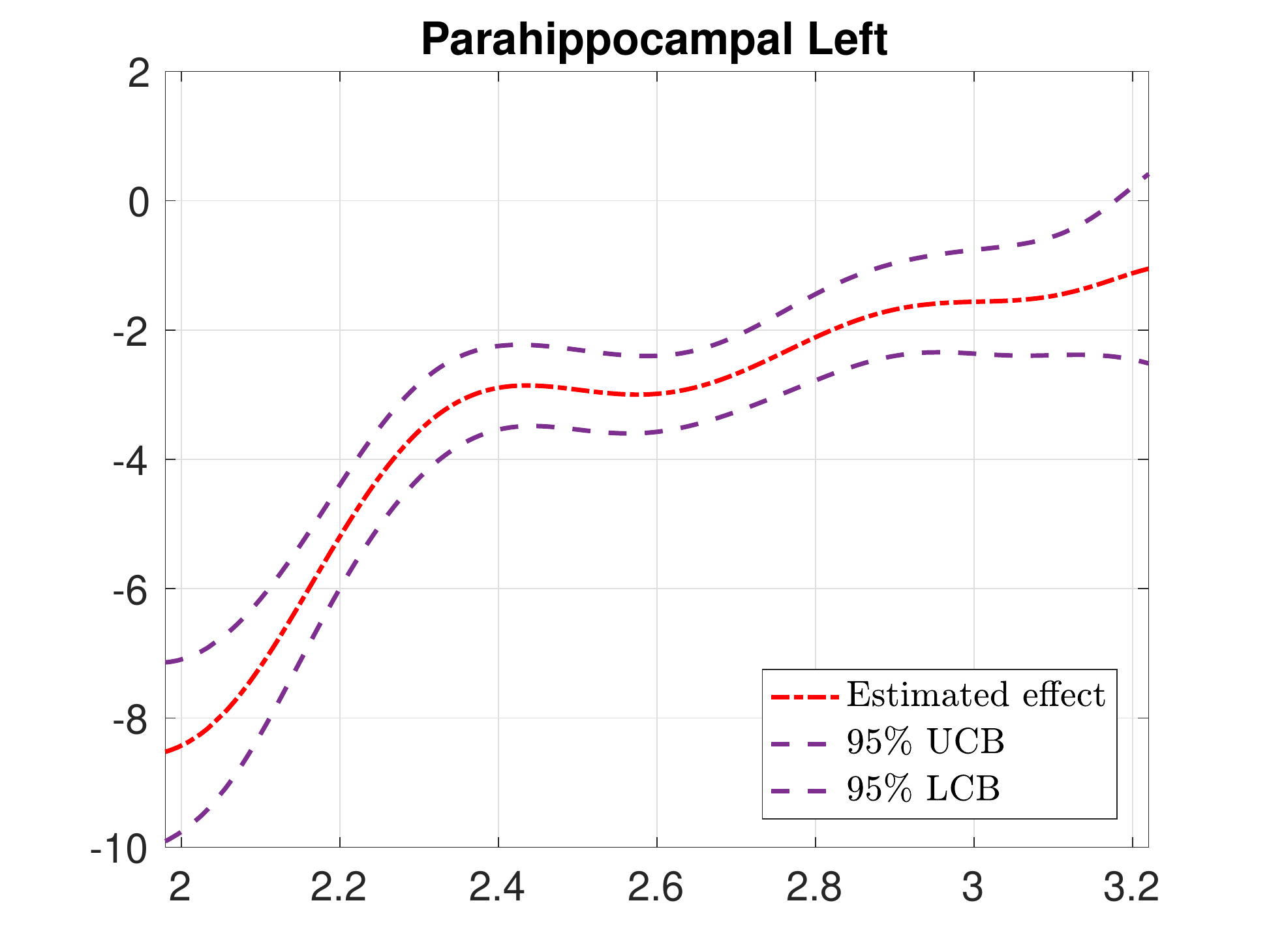} \\
\includegraphics[width=0.32\textwidth,height=1.5in]{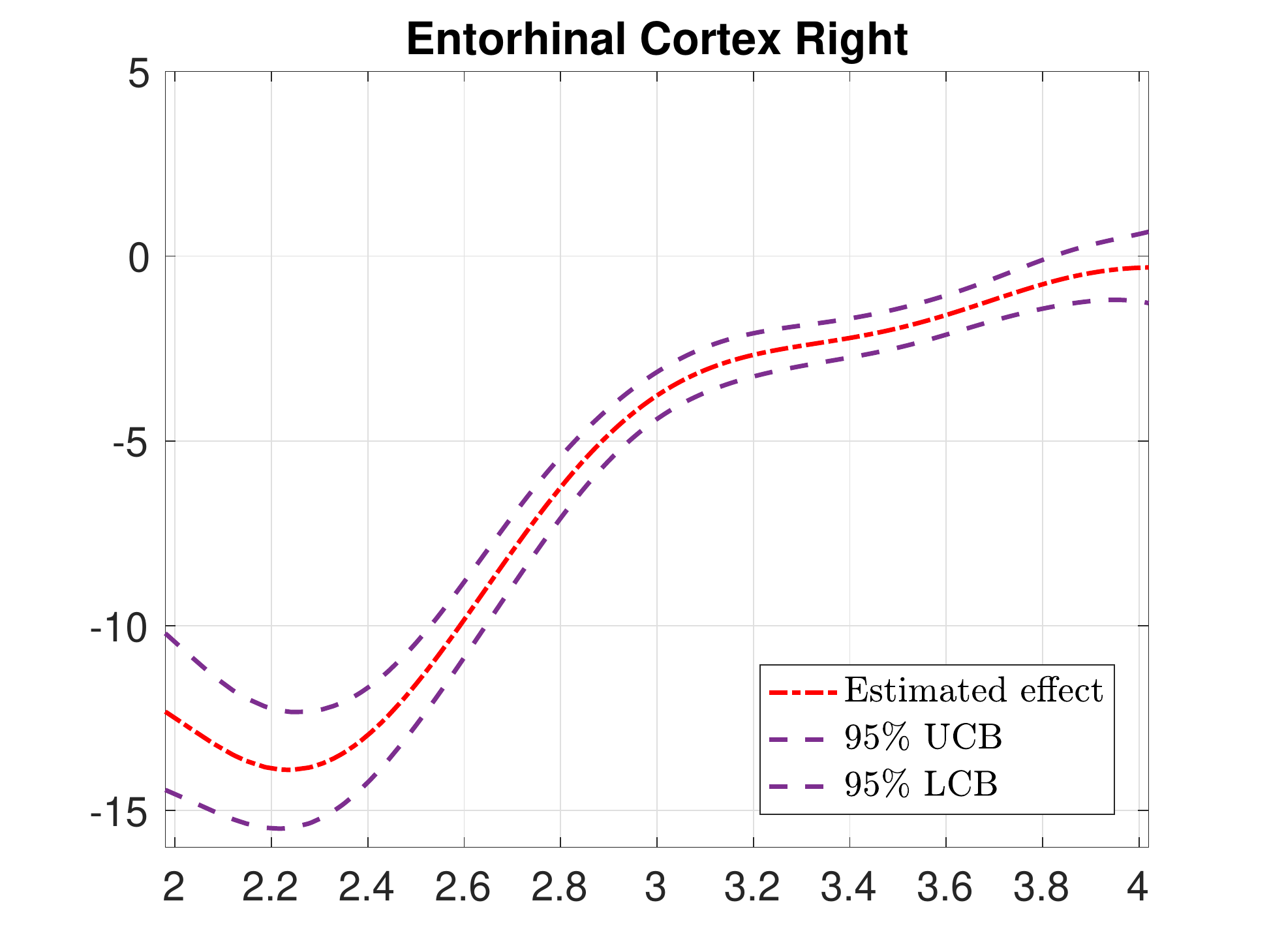} &
\includegraphics[width=0.32\textwidth,height=1.5in]{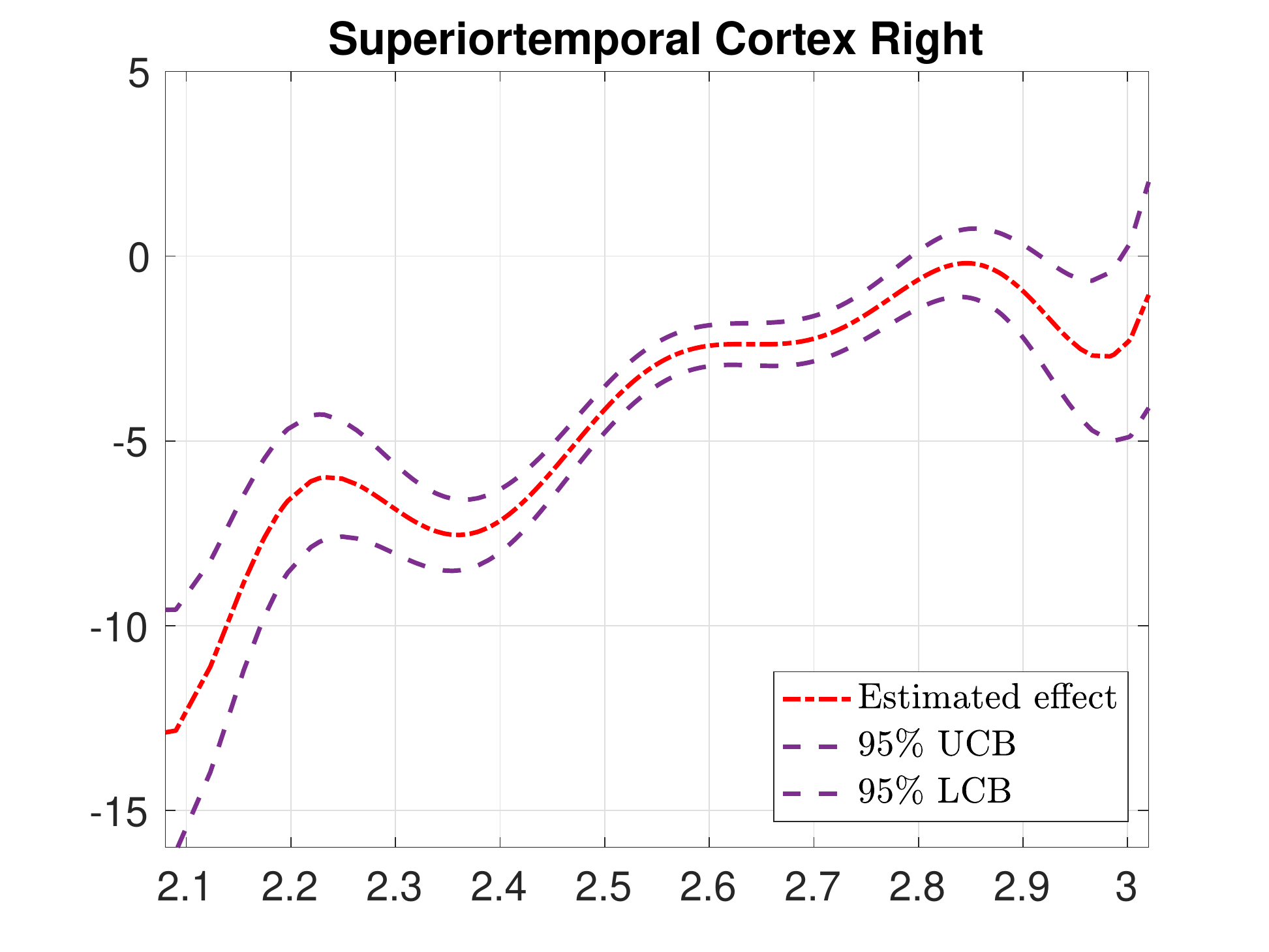} &
\includegraphics[width=0.32\textwidth,height=1.5in]{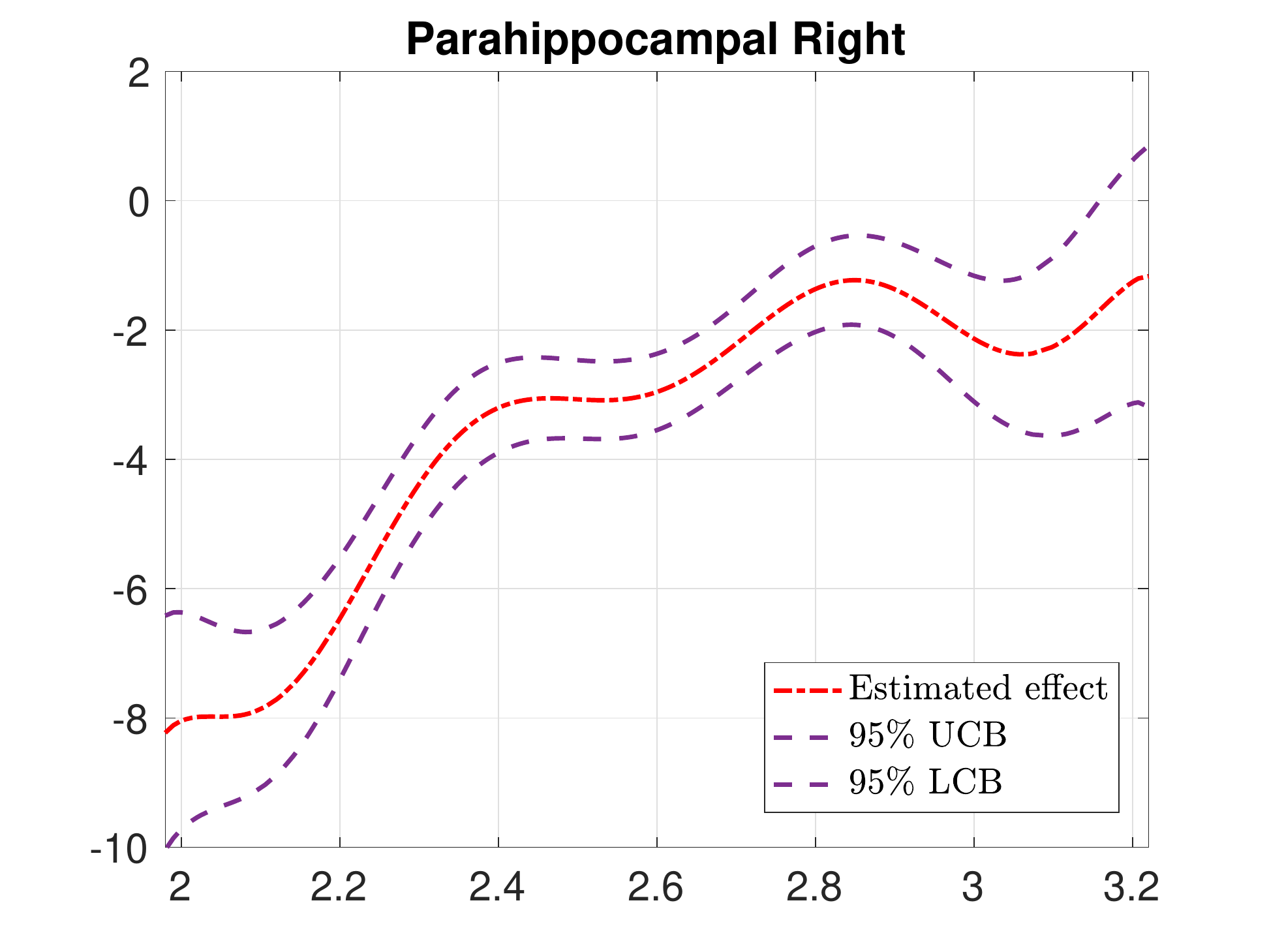} 
\end{tabular}
\caption{The estimated individual effect of the significant brain regions.}
\label{fig:app1} 
\end{figure}

For the first problem, we take the brain MRI cortical thickness as the primary modality, with $p = 68$, and take the PET tau deposition along with the demographic variables and the total amyloid-$\beta$ as the auxiliary modalities, resulting in $p' = 74$. We apply the proposed OKDML method to infer the effect of cortical thickness of individual brain regions on the cognitive outcome. We adopt a similar implementation as used in our first simulation example, and set $\eta(x, \theta_0) = \theta_0\trans x$. Table \ref{table:app1} reports the estimated effects of the brain regions where the cortical thickness is found to be significantly correlated with the cognitive outcome after controlling for amyloid-$\beta$, tau and other covariates, with the corresponding $p$-values under the FDR control at the $5\%$ level  \citep{benjamini1995controlling}. These findings agree well with the AD literature. Particularly, the entorhinal cortex is a brain area located in the medial temporal lobe, and functions as a hub in a widespread network for memory, navigation and the perception of time. Atrophy in the entorhinal cortex has been consistently reported in AD \citep{Pini2016}. The parahippocampal gyrus is a grey matter cortical region of the brain that surrounds the hippocampus, and plays an important role in memory encoding and retrieval. It is among the first to suffer damage from AD \citep{Jack2010}. The superior temporal gyrus locates in the temporal lobe, and contains the Wernicke's area responsible for processing of speech. Its connection with AD needs further verification. Moreover, Figure \ref{fig:app1} shows the confidence band for the estimated individual effect of each significant brain region. Besides, the $95\%$ confidence interval for the $R^2$ measure is $(0.402, 0.437)$, which supports the common belief that brain structural atrophy is closely related to the cognition outcome. 

For the second problem, \citet{Jack2010} suggested that tau deposition precedes structural atrophy in AD pathogenesis. To help verify this theory, we take the PET tau deposition as the primary modality, with $p=70$, then compare two model fits, one with the MRI cortical thickness as part of the auxiliary modalities, and the other without. In both models, we include age, gender, education and amyloid-$\beta$ as the auxiliary modalities. This yields $p' = 72$ when the cortical thickness is included, and $p' = 4$ if not. We obtain the $95\%$ confidence interval for the total effect of tau, which is $(-1.724, 0.702)$ when the cortical thickness is included, and $(-5.212, -3.945)$ when it is not. These results suggest that, not including structural atrophy as the auxiliary modality would result in a much larger effect of tau on cognition outcome, which in turn implies structural atrophy likely occurs after tau deposition, and thus lends some support to the existing theory.

%%%%%%%%%%%%%%%%%%%%%%%%%%%%%%%%%%%%%%%%%%%%%%%%%%%
\section{Discussion}
\label{sec:conclusion}

We conclude the paper by reiterating and further elaborating the innovation of our proposal and its difference from \citet{Chernozhukov2018}. We divide our discussion in two parts: the inference for the primary parameter $\theta_0$, and the inference for the primary function $f_0$. For each part, we first discuss why the question is important, what are the challenges, and why the existing solutions are not directly applicable. We then detail our methodological and theoretical contributions. 

\textbf{(A) Inference for $\theta_0$}: A key innovation of our proposal is that we allow an explicit and non-vanishing model error $\delta_0$ for the primary modality effect $f_0$ in \eqref{eqn:decomoff0}, whereas \citet{Chernozhukov2018} did not consider $\delta_0$. This difference has profound implications in model interpretation, estimation approach, and theoretical analysis, which in turn differentiates our proposal from the existing DML solutions such as \citet{Chernozhukov2018} and \citet{kozbur2020inference}. 
\begin{enumerate}[({A}1)]
\setlength\itemsep{0.1em}

\item In scientific studies such as multimodal analysis, it is crucial to balance model interpretability and model flexibility, which is also the main motivation for this article. In numerous applications, it is not uncommon for scientists to employ some relatively simple models, e.g., linear models, for the primary modality. Such models are easy to interpret, but may not be accurate, and can induce a non-negligible approximation error. In other applications, it is likely to employ more advanced and accurate but less interpretable models. It is thus pivotal to offer inferential robustness for both cases, and to achieve a balanced trade-off between model interpretability and model flexibility. 

\item \cite{Chernozhukov2018} focused on a low-dimensional primary parameter involving no additional error. \citet{kozbur2020inference} extended to a nonparametric primary function through basis expansion, but imposed that the error must be negligible, in that the squared approximation error is $o(N^{-1})$. However, this condition requires either the working model to be sufficiently close to the truth, or the number of basis functions to diverge to infinity with the sample size, which in effect excludes the use of simple yet inaccurate models in characterizing the effect of the primary modality. We also utilize basis expansion to approximate the primary modality effect, but we do not require a vanishing approximation error, nor a diverging number of basis functions, when we establish the asymptotic guarantees of the estimated $\theta_0$.

\item To decouple the primary parameter $\theta_0$ and the non-negligible model error $\delta_0$, we introduce the second form of orthogonality, the decomposition orthogonality, in addition to the Neyman orthogonality, into the framework of double/debiased machine learning. The new orthogonality is similar to the perpendicularity property in smoothing splines \citep{wahba1990}. We show in Proposition \ref{thm:orthogonalofphi} that, this decomposition orthogonality between the expanded basis functions and the model error ensures the identifiability of the primary parameter $\theta_0$. This is a new result, and is potentially useful for obtaining improved inferential robustness in other settings too when there exist non-negligible model error. 

\item Methodologically, the new decomposition orthogonality leads to the construction of a new RKHS, and a residual learning approach in our estimation algorithm, which helps decouple and remove the impact of the model error in parameter estimation. 

\item Theoretically, we successfully establish the $\sqrt{N}$-consistency and asymptotic normality of the estimated main parameter under model error. Compared to the existing semi-parametric inferential analysis, our proof relies on the score function that is Neyman orthogonal with respect to the model error $\delta_0$, and as such requires a weaker regularity condition (C3) than the Donsker conditions that are common but would often fail in multimodal analysis. Compared to the alternative multimodal  solutions, including uni-modality regression, debiased uni-modality regression, simple joint regression, and double/debiased machine learning without taking into account $\delta_0$, we show in Section \ref{sec:comparison} that our estimator is unbiased, but the alternative ones all suffer from a non-vanishing estimation bias when there is model error.  

\item We also show that our estimator is semi-parametric efficient, in that it achieves the highest possible efficiency, when the measurement error $U$ follows a normal distribution. This is also a new result, and its proof is based on constructing an oracle estimator from an ideal finite-dimensional parameter space that achieves the same asymptotic variance as our estimator from an infinite-dimensional parameter space. 
\end{enumerate}

\textbf{(B) Inference for $f_0$}: Another key innovation of our proposal is that we establish the confidence band for the nonparametric primary function $f_0$ in the presence of high-dimensional nonlinear nuisance function, whereas \cite{Chernozhukov2018} considered a low-dimensional primary parameter involving no nonparametric $f_0$.  
\begin{enumerate}[({B}1)]
\setlength\itemsep{0.1em}

\item The function $f_0$ captures the predicted effect of the primary modality, quantifies the amount of contribution of the primary modality in terms of the percentage of variation explained, and also has some causal interpretation under additional conditions. It is thus of great scientific interest to perform rigorous inference on $f_0$. 

\item The high-dimensional nonparametric inference of $f_0$ is challenging. Construction of confidence intervals in such a setting is often intertwined with penalized model estimation and selection, giving rise to post-regularization inference. There has been pioneering research on high-dimensional inference for parametric models such as linear and generalized linear models \citep[among others]{Zhang2014,Vandegeer2014,cai2017}. Early nonparametric inference usually focused on a fixed dimensionality \citep[e.g.,][]{Wahba1983, fan2005nonparametric}. More recently, \citet{lu2020kernel} and \citet{kozbur2020inference} studied high-dimensional inference for nonparametric models. However, as we point out after Theorem \ref{thm:infonpred}, \citet{lu2020kernel} required the variables to be only weakly correlated, which is unlikely to hold for multimodal data, whereas \citet{kozbur2020inference} required a fast vanishing approximation error, which sacrifices model interpretability. 

\item Our inference on $f_0$ is different from the existing literature, as it targets a high-dimensional nonparametric regression setting, allows the primary and auxiliary modalities to be strongly correlated, and also takes into account a non-negligible approximation error when modeling the primary modality effect.  

\item Technically, we extend the inferential framework of \citet{chernozhukov2014anti} to our system of models for multimodal data analysis. We construct the supremum of  high-dimensional empirical processes arising from our OKDML estimator, which enables us to control the supreme norm rate of our estimator, while allowing a diverging dimensionality. We then approximate the supremum with a Gaussian multiplier process to derive the corresponding quantiles and to obtain the asymptotically valid confidence band.
\end{enumerate}

In summary, our proposal integrates reproducing kernel learning \citep{wahba1990} with double/debiased machine learning \citep{Chernozhukov2018}. We believe it makes a useful addition to and also extends the scope of the general methodology and theory for multimodal data analysis, high-dimensional nonparametric inference, as well as double/debiased machine learning. Meanwhile, such an extension is far from simple and straightforward.

\baselineskip=17pt
\bibliographystyle{apa}
\bibliography{ref_orth}

\begin{thebibliography}{}

\bibitem[\protect\astroncite{Alam et~al.}{2018}]{WangYP2018}
Alam, M.~A., Lin, H.-Y., Deng, H.-W., Calhoun, V.~D., and Wang, Y.-P. (2018).
\newblock A kernel machine method for detecting higher order interactions in
  multimodal datasets: Application to schizophrenia.
\newblock {\em Journal of Neuroscience Methods}, 309:161--174.

\bibitem[\protect\astroncite{Baltrusaitis et~al.}{2019}]{Baltrusaitis2019}
Baltrusaitis, T., Ahuja, C., and Morency, L.-P. (2019).
\newblock Multimodal machine learning: A survey and taxonomy.
\newblock {\em IEEE Transactions on Pattern Analysis and Machine Intelligence},
  41(2):423--443.

\bibitem[\protect\astroncite{Benjamini and
  Hochberg}{1995}]{benjamini1995controlling}
Benjamini, Y. and Hochberg, Y. (1995).
\newblock Controlling the false discovery rate: a practical and powerful
  approach to multiple testing.
\newblock {\em Journal of the Royal Statistical Society, Series B.},
  57(1):289--300.

\bibitem[\protect\astroncite{Biau}{2012}]{biau2012analysis}
Biau, G. (2012).
\newblock Analysis of a random forests model.
\newblock {\em Journal of Machine Learning Research}, 13(1):1063--1095.

\bibitem[\protect\astroncite{Bickel et~al.}{1993}]{bickel1993efficient}
Bickel, P.~J., Klaassen, C. A.~J., Ritov, Y., and Wellner, J.~A. (1993).
\newblock {\em Efficient and Adaptive Estimation for Semiparametric Models}.
\newblock Johns Hopkins University Press, Baltimore, MD.

\bibitem[\protect\astroncite{Bickel et~al.}{2009}]{bickel2009simultaneous}
Bickel, P.~J., Ritov, Y., and Tsybakov, A.~B. (2009).
\newblock Simultaneous analysis of lasso and dantzig selector.
\newblock {\em The Annals of statistics}, 37(4):1705--1732.

\bibitem[\protect\astroncite{Breiman}{2001}]{breiman2001statistical}
Breiman, L. (2001).
\newblock Statistical modeling: The two cultures.
\newblock {\em Statistical Science}, 16(3):199--231.

\bibitem[\protect\astroncite{B{\"u}hlmann and van~de
  Geer}{2011}]{buhlmann2011statistics}
B{\"u}hlmann, P. and van~de Geer, S. (2011).
\newblock {\em Statistics for High-Dimensional Data: Methods, Theory and
  Applications}.
\newblock Springer Science \& Business Media.

\bibitem[\protect\astroncite{Buja et~al.}{1989}]{buja1989linear}
Buja, A., Hastie, T., and Tibshirani, R. (1989).
\newblock Linear smoothers and additive models.
\newblock {\em The Annals of Statistics}, 17(2):453--510.

\bibitem[\protect\astroncite{{Cai} et~al.}{2019}]{Cai2019}
{Cai}, Q., {Wang}, H., {Li}, Z., and {Liu}, X. (2019).
\newblock A survey on multimodal data-driven smart healthcare systems:
  Approaches and applications.
\newblock {\em IEEE Access}, 7:133583--133599.

\bibitem[\protect\astroncite{Cai and Guo}{2017}]{cai2017}
Cai, T.~T. and Guo, Z. (2017).
\newblock Confidence intervals for high-dimensional linear regression: Minimax
  rates and adaptivity.
\newblock {\em The Annals of Statistics}, 45(2):615--646.

\bibitem[\protect\astroncite{Chen and White}{1999}]{chen1999improved}
Chen, X. and White, H. (1999).
\newblock Improved rates and asymptotic normality for nonparametric neural
  network estimators.
\newblock {\em IEEE Transactions on Information Theory}, 45(2):682--691.

\bibitem[\protect\astroncite{Chernozhukov et~al.}{2018}]{Chernozhukov2018}
Chernozhukov, V., Chetverikov, D., Demirer, M., Duflo, E., Hansen, C., Newey,
  W., and Robins, J. (2018).
\newblock Double/debiased machine learning for treatment and structural
  parameters: Double/debiased machine learning.
\newblock {\em The Econometrics Journal}, 21:C1--C68.

\bibitem[\protect\astroncite{Chernozhukov et~al.}{2014}]{chernozhukov2014anti}
Chernozhukov, V., Chetverikov, D., and Kato, K. (2014).
\newblock Anti-concentration and honest, adaptive confidence bands.
\newblock {\em The Annals of Statistics}, 42(5):1787--1818.

\bibitem[\protect\astroncite{DeVore and Lorentz}{1993}]{devore1993constructive}
DeVore, R.~A. and Lorentz, G.~G. (1993).
\newblock {\em Constructive Approximation}, volume 303.
\newblock Springer Science \& Business Media.

\bibitem[\protect\astroncite{Fan and Jiang}{2005}]{fan2005nonparametric}
Fan, J. and Jiang, J. (2005).
\newblock Nonparametric inferences for additive models.
\newblock {\em Journal of the American Statistical Association},
  100(471):890--907.

\bibitem[\protect\astroncite{Fan and Lv}{2008}]{fan2008sure}
Fan, J. and Lv, J. (2008).
\newblock Sure independence screening for ultrahigh dimensional feature space.
\newblock {\em Journal of the Royal Statistical Society, Series B.},
  70(5):849--911.

\bibitem[\protect\astroncite{Friedman}{2001}]{friedman2001greedy}
Friedman, J.~H. (2001).
\newblock Greedy function approximation: a gradient boosting machine.
\newblock {\em The Annals of Statistics}, 29(5):1189--1232.

\bibitem[\protect\astroncite{Gin{\'e} and Nickl}{2009}]{gine2009exponential}
Gin{\'e}, E. and Nickl, R. (2009).
\newblock An exponential inequality for the distribution function of the kernel
  density estimator, with applications to adaptive estimation.
\newblock {\em Probability Theory and Related Fields}, 143(3-4):569--596.

\bibitem[\protect\astroncite{Hastie and
  Tibshirani}{1990}]{hastie1990generalized}
Hastie, T. and Tibshirani, R. (1990).
\newblock {\em Generalized Additive Models}.
\newblock CRC Press.

\bibitem[\protect\astroncite{Hinrichs et~al.}{2011}]{hinrichs2011predictive}
Hinrichs, C., Singh, V., Xu, G., Johnson, S.~C., and Initiative, A. D.~N.
  (2011).
\newblock Predictive markers for ad in a multi-modality framework: an analysis
  of mci progression in the adni population.
\newblock {\em Neuroimage}, 55(2):574--589.

\bibitem[\protect\astroncite{Huang et~al.}{2007}]{huang2007efficient}
Huang, J.~Z., Zhang, L., and Zhou, L. (2007).
\newblock Efficient estimation in marginal partially linear models for
  longitudinal/clustered data using splines.
\newblock {\em Scandinavian Journal of Statistics}, 34(3):451--477.

\bibitem[\protect\astroncite{Jack et~al.}{2010}]{Jack2010}
Jack, C.~R., Knopman, D.~S., Jagust, W.~J., Shaw, L.~M., Aisen, P.~S., Weiner,
  M.~W., Petersen, R.~C., and Trojanowski, J.~Q. (2010).
\newblock Hypothetical model of dynamic biomarkers of the alzheimer's
  pathological cascade.
\newblock {\em The Lancet Neurology}, 9(1):119 -- 128.

\bibitem[\protect\astroncite{Kosorok}{2007}]{kosorok2007introduction}
Kosorok, M.~R. (2007).
\newblock {\em Introduction to Empirical Processes and Semiparametric
  Inference}.
\newblock Springer Science \& Business Media, New York.

\bibitem[\protect\astroncite{Kozbur}{2020}]{kozbur2020inference}
Kozbur, D. (2020).
\newblock Inference in additively separable models with a high-dimensional set
  of conditioning variables.
\newblock {\em Journal of Business \& Economic Statistics}, pages 1--17.

\bibitem[\protect\astroncite{Li et~al.}{2019}]{LiChen2019}
Li, G., Liu, X., and Chen, K. (2019).
\newblock Integrative multi-view reduced-rank regression: Bridging group-sparse
  and low-rank models.
\newblock {\em Biometrics}, 75(2):593--602.

\bibitem[\protect\astroncite{Li and Li}{2021}]{LiLi2020factor}
Li, Q. and Li, L. (2021).
\newblock Integrative factor regression and its inference for multimodal data
  analysis.
\newblock {\em Journal of the American Statistical Association}, accepted.

\bibitem[\protect\astroncite{Lin and Zhang}{2006}]{LinZhang2006}
Lin, Y. and Zhang, H.~H. (2006).
\newblock Component selection and smoothing in multivariate nonparametric
  regression.
\newblock {\em The Annals of Statistics}, 34(5):2272--2297.

\bibitem[\protect\astroncite{Lock et~al.}{2013}]{Lock2013}
Lock, E.~F., Hoadley, K.~A., Marron, J.~S., and Nobel, A.~B. (2013).
\newblock Joint and individual variation explained (jive) for integrated
  analysis of multiple data types.
\newblock {\em The Annals of Applied Statistics}, 7(1):523.

\bibitem[\protect\astroncite{Lowe et~al.}{2017}]{Lowe2017}
Lowe, R., Wu, Y., Tamar, A., Harb, J., Abbeel, P., and Mordatch, I. (2017).
\newblock Multi-agent actor-critic for mixed cooperative-competitive
  environments.
\newblock In {\em Proceedings of the 31st International Conference on Neural
  Information Processing Systems}, pages 6382--6393. Curran Associates.

\bibitem[\protect\astroncite{Lu et~al.}{2020}]{lu2020kernel}
Lu, J., Kolar, M., and Liu, H. (2020).
\newblock Kernel meets sieve: Post-regularization confidence bands for sparse
  additive model.
\newblock {\em Journal of the American Statistical Association}, pages 1--16.

\bibitem[\protect\astroncite{Ma et~al.}{2015}]{shujie2015estimation}
Ma, S., Carroll, R.~J., Liang, H., and Xu, S. (2015).
\newblock Estimation and inference in generalized additive coefficient models
  for nonlinear interactions with high-dimensional covariates.
\newblock {\em Annals of Statistics}, 43(5):2102.

\bibitem[\protect\astroncite{Mai and Zhang}{2019}]{mai2019iterative}
Mai, Q. and Zhang, X. (2019).
\newblock An iterative penalized least squares approach to sparse canonical
  correlation analysis.
\newblock {\em Biometrics}, 75(3):734--744.

\bibitem[\protect\astroncite{Nathoo et~al.}{2019}]{Zhu2017review}
Nathoo, F.~S., Kong, L., Zhu, H., and for~the Alzheimer's Disease
  Neuroimaging~Initiative (2019).
\newblock A review of statistical methods in imaging genetics.
\newblock {\em Canadian Journal of Statistics}, 47(1):108--131.

\bibitem[\protect\astroncite{Newey}{1990}]{newey1990semiparametric}
Newey, W.~K. (1990).
\newblock Semiparametric efficiency bounds.
\newblock {\em Journal of Applied Econometrics}, 5(2):99--135.

\bibitem[\protect\astroncite{Newey}{1994}]{Newey1994asymptotic}
Newey, W.~K. (1994).
\newblock The asymptotic variance of semiparametric estimators.
\newblock {\em Econometrica}, pages 1349--1382.

\bibitem[\protect\astroncite{Newey and Robins}{2018}]{newey2018cross}
Newey, W.~K. and Robins, J.~R. (2018).
\newblock Cross-fitting and fast remainder rates for semiparametric estimation.
\newblock {\em arXiv preprint arXiv:1801.09138}.

\bibitem[\protect\astroncite{Neyman}{1959}]{Neyman1959}
Neyman, J. (1959).
\newblock Optimal asymptotic tests of composite statistical hypotheses.
\newblock {\em \text{In U. Grenander (Ed.),} Probability and Statistics}, pages
  416--444.

\bibitem[\protect\astroncite{Neyman}{1979}]{Neyman1979}
Neyman, J. (1979).
\newblock $c(\alpha)$ tests and their use.
\newblock {\em Sankhya}, pages 1--21.

\bibitem[\protect\astroncite{Pearl}{2009}]{pearl2009causality}
Pearl, J. (2009).
\newblock {\em Causality}.
\newblock Cambridge University Press, Cambridge.

\bibitem[\protect\astroncite{Pini et~al.}{2016}]{Pini2016}
Pini, L., Pievani, M., Bocchetta, M., Altomare, D., Bosco, P., Cavedo, E.,
  Galluzzi, S., Marizzoni, M., and Frisoni, G.~B. (2016).
\newblock Brain atrophy in alzheimer's disease and aging.
\newblock {\em Ageing Research Reviews}, 30:25--48.
\newblock Brain Imaging and Aging.

\bibitem[\protect\astroncite{Raskutti et~al.}{2011}]{Raskutti2011}
Raskutti, G., Wainwright, M.~J., and Yu, B. (2011).
\newblock Minimax rates of estimation for high-dimensional linear regression
  over $\ell_q$-balls.
\newblock {\em IEEE Transactions on Information Theory}, 57(10):6976--6994.

\bibitem[\protect\astroncite{Richardson et~al.}{2016}]{Richardson2016}
Richardson, S., Tseng, G.~C., and Sun, W. (2016).
\newblock Statistical methods in integrative genomics.
\newblock {\em Annual Reviews of Statistics and Its Applications}, 3:181--209.

\bibitem[\protect\astroncite{Robins and
  Rotnitzky}{1995}]{robins1995semiparametric}
Robins, J.~M. and Rotnitzky, A. (1995).
\newblock Semiparametric efficiency in multivariate regression models with
  missing data.
\newblock {\em Journal of the American Statistical Association},
  90(429):122--129.

\bibitem[\protect\astroncite{Shu et~al.}{2020}]{Shu2019dcca}
Shu, H., Wang, X., and Zhu, H. (2020).
\newblock D-cca: A decomposition-based canonical correlation analysis for
  high-dimensional datasets.
\newblock {\em Journal of the American Statistical Association},
  115(529):292--306.

\bibitem[\protect\astroncite{Sperling et~al.}{2019}]{Sperling2019}
Sperling, R.~A., Mormino, E.~C., Schultz, A.~P., et~al. (2019).
\newblock The impact of amyloid-beta and tau on prospective cognitive decline
  in older individuals.
\newblock {\em Annals of Neurology}, 85(2):181--193.

\bibitem[\protect\astroncite{Uluda{\u{g}} and Roebroeck}{2014}]{Uludaug2014}
Uluda{\u{g}}, K. and Roebroeck, A. (2014).
\newblock General overview on the merits of multimodal neuroimaging data
  fusion.
\newblock {\em Neuroimage}, 102:3--10.

\bibitem[\protect\astroncite{van~de Geer et~al.}{2014}]{Vandegeer2014}
van~de Geer, S., Bühlmann, P., Ritov, Y.~A., and Dezeure, R. (2014).
\newblock On asymptotically optimal confidence regions and tests for
  high-dimensional models.
\newblock {\em The Annals of Statistics}, 42(3):1166--1202.

\bibitem[\protect\astroncite{van~der Laan and Rubin}{2006}]{van2006targeted}
van~der Laan, M.~J. and Rubin, D. (2006).
\newblock Targeted maximum likelihood learning.
\newblock {\em The international journal of biostatistics}, 2(1).

\bibitem[\protect\astroncite{van~der Vaart}{1998}]{Vandevaart1998}
van~der Vaart, A.~W. (1998).
\newblock {\em Asymptotic Statistics. Cambridge Series in Statistical and
  Probabilistic Mathematics}.
\newblock Cambridge University Press, Cambridge.

\bibitem[\protect\astroncite{Wahba}{1983}]{Wahba1983}
Wahba, G. (1983).
\newblock Bayesian ``confidence intervals'' for the cross-validated smoothing
  spline.
\newblock {\em Journal of the Royal Statistical Society. Series B (Statistical
  Methodology)}, 45:133--150.

\bibitem[\protect\astroncite{Wahba}{1990}]{wahba1990}
Wahba, G. (1990).
\newblock {\em Spline Models for Observational Data}.
\newblock SIAM, Philadelphia.

\bibitem[\protect\astroncite{Wang et~al.}{2014}]{wang2014estimation}
Wang, L., Xue, L., Qu, A., and Liang, H. (2014).
\newblock Estimation and model selection in generalized additive partial linear
  models for correlated data with diverging number of covariates.
\newblock {\em The Annals of Statistics}, 42(2):592--624.

\bibitem[\protect\astroncite{Xue and Qu}{2020}]{XueQu2019}
Xue, F. and Qu, A. (2020).
\newblock Integrating multi-source block-wise missing data in model selection.
\newblock {\em Journal of the American Statistical Association}, accepted.

\bibitem[\protect\astroncite{Zhang and Zhang}{2014}]{Zhang2014}
Zhang, C. and Zhang, S. (2014).
\newblock Confidence intervals for low dimensional parameters in high
  dimensional linear models.
\newblock {\em Journal of the Royal Statistical Society. Series B.},
  76(1):217--242.

\bibitem[\protect\astroncite{Zhao and Hastie}{2021}]{zhao2019causal}
Zhao, Q. and Hastie, T. (2021).
\newblock Causal interpretations of black-box models.
\newblock {\em Journal of Business \& Economic Statistics}, 39(1):272--281.

\bibitem[\protect\astroncite{Zheng and van~der Laan}{2011}]{zheng2011cross}
Zheng, W. and van~der Laan, M.~J. (2011).
\newblock Cross-validated targeted minimum-loss-based estimation.
\newblock In {\em Targeted Learning}, pages 459--474. Springer.

\bibitem[\protect\astroncite{Zhu et~al.}{2014}]{ZhuHT2014}
Zhu, H., Khondker, Z., Lu, Z., and Ibrahim, J.~G. (2014).
\newblock {Bayesian Generalized Low Rank Regression Models for Neuroimaging
  Phenotypes and Genetic Markers}.
\newblock {\em Journal of the American Statistical Association},
  109(507):977--990.

\end{thebibliography}

\newpage
\appendix

\section{Proofs}

%%%%%%%%%%%%%%%%%%%%%%%%%%%%%%%%%%%%%%%%%%%%%%%%%%%
\subsection{Proof of Proposition \ref{thm:neymanscore}}

\begin{proof}
Since $g$ is infinite-dimensional, we apply the concentrated-out approach \citep{Newey1994asymptotic, Chernozhukov2018} to construct the Neyman orthogonal score. Consider the risk function, 
\begin{equation*}
L(\theta,\delta, g) = \E\Big\{\frac{1}{2}[Y-\eta(X,\theta)-\delta(X)-g(Z)]^2\Big\}.
\end{equation*} 
We have $(\theta_0,\delta_0,g_0) = \underset{\theta\in\R^d,\delta\in\Hcal_\delta,g\in\Hcal_g}{\arg\min}~L(\theta,\delta,g)$. For any $\theta\in\R^d$, let $g_\theta$ be the concentrated-out part of the model that is defined by $g_\theta(Z) = \underset{g\in\Hcal_g}{\arg\min}~L(\theta,\delta_0,g)$, for any $\theta\in\R^d$. Then $g_\theta(Z)$ has an explicit form
\begin{equation*}
g_{\theta}(Z) = \E\left\{[Y-\eta(X,\theta)-\delta_0(X)|Z] \right\}.
\end{equation*}
Denote a mapping $\Gcal(\theta):\R^d\to\Hcal_g$ with its true value $\Gcal_0$ given by $\Gcal_0(\theta)=g_\theta$, for any $\theta\in\R^d$. Consider the function,
\begin{equation*}
Q(\theta,t) = L\left\{ \theta,\delta_0+t(\delta-\delta_0),\Gcal_0(\theta)+t[\Gcal(\theta)-\Gcal_0(\theta)] \right\}, \quad \theta\in\R^d, \; t\in[0,1].
\end{equation*}
Then $\psi\{\theta,\delta_0+t(\delta-\delta_0),\Gcal_0(\theta)+t[\Gcal(\theta)-\Gcal_0(\theta)]\} = \partial_\theta Q(\theta,t)$. Therefore, 
\begin{align*}
& \partial_t\E\left( \psi\{\theta,\delta_0+t(\delta-\delta_0),\Gcal_0(\theta)+t[\Gcal(\theta)-\Gcal_0(\theta)]\} \right) 
= \partial_t\E[\partial_\theta Q(\theta,t)] 
= \partial_t\partial_\theta\E[Q(\theta,t)] \\
= \; & \partial_\theta\partial_t\E[Q(\theta,\delta,t)] 
= \partial_\theta\partial_t\E\left( L\{\theta,\delta_0+t(\delta-\delta_0),\Gcal_0(\theta)+t[\Gcal(\theta)-\Gcal_0(\theta)]\} \right).
\end{align*}
Because, 
\begin{equation*}
\partial_t\E( l\{\theta,\delta_0+t(\delta-\delta_0),\Gcal_0(\theta)+t[\Gcal(\theta)-\Gcal_0(\theta)]\}) \vert_{t=0} =0,\; \text{ for all }\theta\in\R^d,
\end{equation*}
we have that, 
\begin{equation*}
\partial_t\E(\psi\{\theta,\delta,\Gcal_0(\theta)+t[\Gcal(\theta)-\Gcal_0(\theta)]\}) \vert_{t=0}=0
\end{equation*}
Therefore, 
\begin{equation*}
\begin{aligned}
&\psi(\theta,\delta,g_\theta)\vert_{(\theta_0,\delta_0,g_0)} \\
= \; & \left\{ \partial_\theta \eta(X,\theta_0)-\E[\partial_\theta\eta(X,\theta_0)|Z]\right\} \left\{ Y-\delta_0(X)-\eta(X,\theta_0)-\E[Y-\delta_0(X)-\eta(X,\theta_0)|Z]\right \} \\
= \; & \left[\partial_\theta\eta(X,\theta_0)-r_0(Z)\right] \times \left[Y-\delta_0(X)-\eta(X,\theta_0)-g_0(Z)\right]
\end{aligned}
\end{equation*}
satisfies the Neyman orthogonality conditions. This completes the proof of Proposition \ref{thm:neymanscore}. 
\end{proof}

%%%%%%%%%%%%%%%%%%%%%%%%%%%%%%%%%%%%%%%%%%%%%%%%%%%
\subsection{Proof of Proposition \ref{thm:orthogonalofphi}}
\label{proof:proporthphi}

\begin{proof}
Since $\theta_0\in\R^d$, there exists a constant $\delta$ such that $\theta_0+ae_j\in\R^d$ for all $a\in[-\delta,\delta]$ and $j=1,\ldots,d$, where $e_j$ is the column vector of zeros except for a one at the $j$th position. Define the function, 
\begin{equation*}
Q_j(a) = \frac{\E_X\{ [f_0(X)-\eta(X,\theta_0)]^2\} - \E_X\{ [f_0(X)-\eta(X,\theta_0+ae_j)]^2\} }{2a}.
\end{equation*}
Since $\Phi(\cdot)$ is bounded on $\Xcal^p$, the dominated convergence theorem implies that $Q_j(a)$ has a limiting point at $a=0$:
\begin{equation*}
\lim_{a\to 0}Q_j(a) = e_j\trans\E_X\{ \Phi(X)[f_0(X)-\eta(X,\theta_0)] \} = e_j\trans\E_X[\Phi(X)\delta_0(X)],
\end{equation*}
where the last step is by model (\ref{eqn:decomoff0}). By the definition of $\theta_0$ in (\ref{eqn:lineartheta}) where $\eta(\cdot,\theta_0)$ is the unique projection, we have that $Q_j(a)\leq 0$ for any $a\in[0,\delta]$. Then taking the limit $a\to 0_+$, we have, 
\begin{equation*}
e_j\trans\E_X[\Phi(X)\delta_0(X)]\leq 0. 
\end{equation*} 
Moreover, for any $a\in[-\delta,0]$, $Q_i\geq 0$. Then taking the limit $a\to 0_-$, we have, 
\begin{equation*}
e_j\trans\E_X[\Phi(X)\delta_0(X)]\geq 0. 
\end{equation*} 
Repeat the above procedure for all $j=1,\ldots,d$, and we obtain that $\E_X[\Phi(X)\delta_0(X)] = 0$. Therefore, under models (\ref{eqn:decomoff0}) and (\ref{eqn:lineartheta}), $\theta_0$ is identifiable only if  $\Phi(X)$ and $\delta_0(X)$ satisfy the decomposition orthogonality in Definition \ref{defi:decomp-orth}. 

We next show the second part of this proposition. For any function that can be written as $\widehat{\delta}(x) = \sum_{i=1}^mc_iK_\delta(x,x_i)$ with $c_i\in\R, x_i\in\Xcal^p, m\geq 1$ and $K_\delta$ defined in Proposition \ref{thm:orthogonalofphi}, we have, 
\begin{equation*}
\begin{aligned}
\E_X\left[ \Phi(X)\widehat{\delta}(X) \right] &= \sum_{i=1}^mc_i\E_X[\Phi(X)K_\delta(X,x_i)] \\
& = \sum_{i=1}^mc_i\E_X[\Phi(X)K(X,x_i)]- \sum_{i=1}^mc_i \E_X[\Phi(X)] \E_{X'}[\Phi(X')\trans K(X,X')] \\
&\quad\quad \times\left(\E_{X'}\{\E_{X''}[\Phi(X'')K(X'',X')]\Phi(X')\trans\}\right)^{-1}\E_{X''}[\Phi(X'') K(X'',x_i)],
\end{aligned}
\end{equation*}
where $X, X'$ and $X''$ are i.i.d.\ copies of the primary modality. By Funibi's theorem, 
\begin{equation*}
\begin{aligned}
\E_X\left[ \Phi(X)\widehat{\delta}(X) \right] & =   \sum_{i=1}^mc_i\E_X[\Phi(X)K(X,x_i)]- \sum_{i=1}^mc_i\E_X\{\Phi(X)\E_{X'}[\Phi(X')\trans K(X,X')]\} \\
&\quad\quad\times\left(\E_{X''}\{\Phi(X'')\E_{X'}[\Phi(X')\trans K(X'',X')]\}\right)^{-1}\E_{X''}[\Phi(X'') K(X'',x_i)] \\
& = \sum_{i=1}^mc_i\E_X[\Phi(X)K(X,x_i)]- \sum_{i=1}^mc_i\E_{X''}[\Phi(X'') K(X'',x_i)] = 0.
\end{aligned}
\end{equation*}
Then, by definition, $\Phi(X)$ and $\delta(X)$ satisfy the decomposition orthogonality. This completes the proof of Proposition \ref{thm:orthogonalofphi}. 
\end{proof}

%%%%%%%%%%%%%%%%%%%%%%%%%%%%%%%%%%%%%%%%%%%%%%%%%%%
\subsection{Proof of Theorem \ref{thm:mainresulttri}}
\label{sec:pfthmparameter}

\begin{proof}
Rewrite the score in Proposition \ref{thm:neymanscore} as
\begin{equation*}
\psi(W;\theta,r,g,\delta) = [\Phi(X)-r(Z)]\Phi(X)\trans\theta+[r(Z)-\Phi(X)][Y-g(Z)-\delta(X)],
\end{equation*}
where $W=(X,Y,Z)$. Define the following quantities: 
\begin{align} \label{eqn:defofnotationsj0}
\begin{split}
& J_0 = \E\left\{[\Phi(X)-r_0(Z)]\Phi(X)\trans\right\} = \E(VV\trans), \quad 
\widehat{J}_0 = \frac{1}{Q}\sum_{q=1}^Q\frac{1}{n}\sum_{i\in I_q}[\Phi(X_i)-\widehat{r}_{q}(Z_i)]\Phi(X_i)\trans, \\ 
& R_{N,1} = \widehat{J}_0-J_0, \quad 
R_{N,2} =  \frac{1}{Q}\sum_{q=1}^Q \frac{1}{n}\sum_{i\in I_q}\psi(W_i;\theta_0,\widehat{r}_{q},\widehat{g}_{q},\widehat{\delta}_{q}) - \frac{1}{N}\sum_{i=1}^N\psi(W_i;\theta_0,r_0,g_0,\delta_0),
\end{split}
\end{align}
where $W_i = (X_i,Y_i,Z_i)$. We divide the proof of this theorem into four steps.

\bigskip
\noindent
\textbf{Step 1: Bounding $R_{N,1}$}. We aim to show that 
\begin{equation} \label{eqn:bdonRn1}
\|R_{N,1}\|_{\ell_2} = o_p(N^{-1/4}).
\end{equation}
For any $q\in[Q]$, by the triangle inequality, we have $\|R_{N,1}\|_{\ell_2} \leq Q^{-1} \sum_{q=1}^Q(\mathcal I_{1,q}+ \mathcal I_{2,q})$, where 
\begin{equation*}
\begin{aligned}
\mathcal I_{1,q} &= \bigg\| \frac{1}{n}\sum_{i\in I_q}[\Phi(X_i)-\widehat{r}_{q}(Z_i)]\Phi(X_i)\trans - \E\left\{ [\Phi(X)-\widehat{r}_{q}(Z)] \Phi(X)\trans | (Z_i,X_i)_{i\in I_q^c} \right\} \bigg\|_2,\\
\mathcal I_{2,q} &= \left\| \E\left\{ [\Phi(X)-\widehat{r}_{q}(Z)] \Phi(X)\trans | (Z_i,X_i)_{i\in I_q^c} \right\} - \E(VV\trans) \right\|_2, 
\end{aligned}
\end{equation*}
where $\|\cdot\|_2$ denotes the matrix $2$-norm. We next bound $\mathcal I_{1,q}$ and $\mathcal I_{2,q}$, respectively. 

To bound $\mathcal I_{1,q}$, we have that, 
\begin{equation*}
\begin{aligned}
\E\left[\mathcal I_{1,q}^2| (Z_i,X_i)_{i\in I_q^c}\right] &\leq n^{-1}\E\left\{ \|[\Phi(X)-\widehat{r}_{q}(Z)]\Phi(X)\trans \|_2^2 \ \big| \ (Z_i,X_i)_{i\in I_q^c} \right\} \\
& = n^{-1}\E\left\{ \|[\Phi(X)-r_{0}(Z)]\Phi(X)\trans \|_2^2 \right\} + o(n^{-1}) = O(n^{-1}),
\end{aligned}
\end{equation*}
where the second step is due to conditions (C1) and (C3), together with the fact that $\widehat{r}_q$ in (\ref{eqn:estimateofr}) only uses the subset of data indexed by $I_q^c$, and the last step is due to condition (C2). Therefore, $\mathcal I_{1,q} = O_p(n^{-1/2})=O_p(N^{-1/2})$ for a finite $Q$.

To bound $\mathcal I_{2,q}$, we have that,
\begin{equation*}
\begin{aligned}
\mathcal I_{2,q} &= \left\| \E\left\{ [\Phi(X)-r_0(Z)+r_0(Z)-\widehat{r}_{q}(Z)] \Phi(X)\trans | (Z_i,X_i)_{i\in I_q^c}] \right\} - \E(VV\trans) \right\|_2 \\
&\leq \left( \E\left\{ [r_0(Z)-\widehat{r}_q(Z)]^2 \right\} \E\left[ \|\Phi(X)\|^2_{\ell_2} \right] \right)^{1/2} = o(N^{-1/4}),
\end{aligned}
\end{equation*}
where second step is due to Cauchy-Schwarz inequality and condition (C1), and the last step is due to conditions (C1) and (C3).

Combining the bounds for $\mathcal I_{1,q}$ and $\mathcal I_{2,q}$ yields (\ref{eqn:bdonRn1}).

\bigskip
\noindent
\textbf{Step 2: Bounding $R_{N,2}$}. We aim to show that
\begin{equation} \label{eqn:eqn:bdonRn2}
\|R_{N,2}\|_{\ell_2} = o_p(N^{-1/2}).
\end{equation}
For any $q\in[Q]$, by the triangle inequality, we have, 
\begin{equation*}
\bigg| \frac{1}{n}\sum_{i\in I_q}\psi(W_i;\theta_0,\widehat{r}_{q},\widehat{g}_{q},\widehat{\delta}_{q})-\frac{1}{n}\sum_{i\in I_q}\psi(W_i;\theta_0,r_0,g_0,\delta_0) \bigg| \leq n^{-1/2}(\mathcal I_{3,q}+\mathcal I_{4,q}),
\end{equation*}
where 
\begin{equation*}
\begin{aligned}
\mathcal I_{3,q} &= \Big\|\frac{1}{\sqrt{n}}\sum_{i\in I_q}\{\psi(W_i;\theta_0,\widehat{r}_{q},\widehat{g}_{q},\widehat{\delta}_{q})-\E[\psi(W;\theta_0,\widehat{r}_{q},\widehat{g}_{q},\widehat{\delta}_{q})|(W_i)_{i\in I_q^c}]\} \\
&\quad\quad\quad\;-\frac{1}{\sqrt{n}}\sum_{i\in I_q}\left\{\psi(W_i;\theta_0,r_0,g_{0},\delta_{0})-\E[\psi(W;\theta_0,r_0,g_{0},\delta_{0})] \right\}\Big\|_{\ell_2},\\
\mathcal I_{4,q} &= \sqrt{n} \Big\|\E[\psi(W;\theta_0,\widehat{r}_{q},\widehat{g}_{q},\widehat{\delta}_{q})|(W_i)_{i\in I_q^c}]-\E[\psi(W;\theta_0,r_0,g_{0},\delta_{0})] \Big\|_{\ell_2}.
\end{aligned}
\end{equation*}
We next bound $\mathcal I_{3,q}$ and $\mathcal I_{4,q}$, respectively. 

To bound $\mathcal I_{3,q}$, we have that, 
\begin{equation*}
\begin{aligned}
& \E\left[ \mathcal I_{3,q}^2|(W_i)_{i\in I_q^c} \right] \\
\leq \; & \E\left[ \|\psi(W;\theta_0,\widehat{r}_{q},\widehat{g}_{q},\widehat{\delta}_{q})-\psi(W;\theta_0,r_{0},g_{0},\delta_{0})\|_{\ell_2}^2 \ \big| \ (W_i)_{i\in I_q^c} \right] \\
= \; & O\left[ \E\left[ \|\widehat{r}_q(Z)-r_0(Z)\|^2_{\ell_2} \right] \left( \E\left\{ \left[ \widehat{g}_q(Z)-g_0(Z) \right]^2 \right\} + \E\left\{ \left[ \widehat{\delta}_q(X)-\delta_0(X) \right]^2 \right\} \right) \right] \\
& \quad\quad + O\left( \E\left[ \|\widehat{r}_q(Z)-r_0(Z)\|_{\ell_2}^2 \right] + \E\left\{ \left[\widehat{g}_q(Z)-g_0(Z)\right]^2 \right\} + \E\left\{ \left[ \widehat{\delta}_q(X)-\delta_0(X) \right]^2 \right\} \right)\\
= \; & o(N^{-1/2}),
\end{aligned}
\end{equation*}
where the second step is by Cauchy-Schwarz inequality and condition (C2), and the last step is due to condition (C3). Therefore, $\mathcal I_{3,q}=o_p(N^{-1/4})$. 

To bound $I_{4,q}$,  we apply the Taylor expansion and obtain that, 
\begin{equation*}
\begin{aligned}
n^{-1/2}\mathcal I_{4,q} &= \left\| \E\left[ \psi(W;\theta_0,\widehat{r}_{q},\widehat{g}_{q},\widehat{\delta}_{q})|(W_i)_{i\in I_q^c} \right] - \E\left[ \psi(W;\theta_0,r_0,g_0,\delta_0) \right] \right\|_{\ell_2}\\
&= \left\| \E\left[ \psi(W;\theta_0,r_0,g_0,\delta_0)|(W_i)_{i\in I_q^c} \right] - \E\left[ \psi(W;\theta_0,r_0,g_0,\delta_0) \right] \right\|_{\ell_2}\\
& \quad\quad + O\left( \E\left[ \left\| \widehat{r}_q(Z)-r_0(Z) \right\|_{\ell_2} \left| \widehat{g}_q(Z)-g_0(Z)+\widehat{\delta}_q(X)-\delta_0(X) \right| \right] \right)\\
& = o(N^{-1/2}).
\end{aligned}
\end{equation*}
Therefore, $\mathcal I_{4,q} = o(1)$. 

Combining the bounds for $\mathcal I_{3,q}$ and $\mathcal I_{4,q}$ yields (\ref{eqn:eqn:bdonRn2}).

\bigskip
\noindent
\textbf{Step 3: Bounding $\psi(W_i)$}. We aim to show that 
\begin{equation}
\label{eqn:bdonmeanpsi}
\bigg\| N^{-1/2}\sum_{i=1}^N\psi(W_i;\theta_0,r_0,g_0,\delta_0) \bigg\|_{\ell_2} = O_p(1).
\end{equation}
Since $W_i$'s are independent, we have that, 
\begin{equation*}
\E\left[ \bigg\| N^{-1/2}\sum_{i=1}^N\psi(W_i;\theta_0,r_0,g_0,\delta_0) \bigg\|_{\ell_2}^2 \right] = \E\left[ \|\psi(W;\theta_0,r_0,g_0,\delta_0)\|_{\ell_2}^2 \right] 
= \E(U^2) \E(V\trans V) = O(1),
\end{equation*}
where the first step is due to $\E[\psi(W_i;\theta_0,r_0,g_0,\delta_0)]=0$ for all $i$, and the last step is due to $\E[U^2]<\infty$ and condition (C2). Then by the Markov's inequality, we obtain (\ref{eqn:bdonmeanpsi}).

\bigskip
\noindent
\textbf{Step 4: Deriving $\widehat{\theta}-\theta_0$}. By condition (C2), $J_0$ is positive definite. Together with (\ref{eqn:bdonRn1}), all singular values of $\widehat{J}_0$ are bounded below from zero. The estimator in (\ref{eqn:TDE}) can be rewritten as
\begin{equation*}
\widehat{\theta} = - \widehat{J}_0^{-1}\frac{1}{Q}\sum_{q=1}^Q\frac{1}{n}\sum_{i\in I_q} \left[ Y_i-\widehat{g}_{q}(Z_i)-\widehat{\delta}_{q}(X_i) \right] \left[ \widehat{r}_{q}(Z_i)-\Phi(X_i) \right].
\end{equation*}
By the definition of $R_{N,1}$ and $R_{N,2}$, we have that, 
\begin{equation*}
\begin{aligned}
\widehat{\theta}-\theta_0 & = -\widehat{J}_0^{-1}\frac{1}{Q}\sum_{q=1}^Q\frac{1}{n}\sum_{i\in I_q}\psi(W_i;\theta_0,\widehat{r}_{q},\widehat{g}_{q},\widehat{\delta}_{q})\\
&=-(J_0+R_{N,1})^{-1} \left[ \frac{1}{N}\sum_{i=1}^N\psi(W_i,\theta_0,r_0,g_0,\delta_0)+R_{N,2} \right] \\
&=-\left\{ \left[ (J_0+R_{N,1})^{-1}-J_0^{-1} \right] + J_0^{-1} \right\} \left[ \frac{1}{N}\sum_{i=1}^N\psi(W_i,\theta_0,r_0,g_0,\delta_0)+R_{N,2} \right].
\end{aligned}
\end{equation*}
Again by (\ref{eqn:bdonRn1}) and condition (C2), we obtain that, 
\begin{equation} \label{eqn:bdonjandr}
\begin{aligned}
\left\| (J_0+R_{N,1})^{-1}-J_0^{-1} \right\|_2 & = \left\| (J_0+R_{N,1})^{-1}R_{N,1}J_0^{-1} \right\|_2 \\
& \leq \left\| (J_0+R_{N,1})^{-1} \right\|_2 \times \|R_{N,1}\|_2 \times \left\| J_0^{-1} \right\|_2 = o_p(N^{-1/4}).
\end{aligned}
\end{equation}
Then by (\ref{eqn:eqn:bdonRn2}), (\ref{eqn:bdonmeanpsi}) and condition (C2),  we obtain that, 
\begin{equation*}
\begin{aligned}
\widehat{\theta}-\theta_0  = - J_0^{-1}\left[ \frac{1}{N}\sum_{i=1}^N\psi(W_i,\theta_0,r_0,g_0,\delta_0) \right] + o_p(N^{-1/2}).
\end{aligned}
\end{equation*}
This completes the proof of Theorem \ref{thm:mainresulttri}.
\end{proof}

%%%%%%%%%%%%%%%%%%%%%%%%%%%%%%%%%%%%%%%%%%%%%%%%%%%
\subsection{Proof of Corollary \ref{thm:covarestimation}}
\label{sec:proofofcovarianceestimation}
\begin{proof}
Recall the orthogonal score function in Proposition \ref{thm:neymanscore}:
\begin{equation*}
\psi(W;\theta,r,g,\delta) =[r(Z)-\Phi(X)][Y-\Phi(X)\trans\theta-g(Z)-\delta(X)].
\end{equation*}
Note that $\psi$ is a $d$-dimensional vector. Let $\psi_l$ denote its $l$th component, $l \in [d]$.

For any $q\in [Q]$ and $l_1,l_2\in[d]$, define
\begin{equation*}
\mathcal I_{ql_1l_2} = \bigg \vert\frac{1}{n}\sum_{i\in I_q}\psi_{l_1}(W_i;\widehat{\theta},\widehat{r}_{q},\widehat{g}_{q},\widehat{\delta}_{q})\psi_{l_2}(W_i;\widehat{\theta},\widehat{r}_{q},\widehat{g}_{q},\widehat{\delta}_{q}) - \E(U^2V_{l_1}V_{l_2}) \bigg\vert
\end{equation*}
By the triangle inequality, we have $\mathcal I_{ql_1l_2}\leq \mathcal I_{ql_1l_2,1} + \mathcal I_{ql_1l_2,2}$, where 
\begin{equation*}
\begin{aligned}
\mathcal I_{ql_1l_2,1}  &= \bigg\vert\frac{1}{n}\sum_{i\in I_q}\psi_{l_1}(W_i;\theta_0,r_0,g_0,\delta_0)\psi_{l_2}(W_i;\theta_0,r_0,g_0,\delta_0) - \E(U^2V_{l_1}V_{l_2})\bigg\vert, \\
\mathcal I_{ql_1l_2,2}  &= \bigg\vert\frac{1}{n}\sum_{i\in I_q}\psi_{l_1}(W_i;\widehat{\theta},\widehat{r}_{q},\widehat{g}_{q},\widehat{\delta}_{q})\psi_{l_2}(W_i;\widehat{\theta},\widehat{r}_{q},,\widehat{g}_{q},\widehat{\delta}_{q})\\
  &\quad\quad\quad\quad-\frac{1}{n}\sum_{i\in I_q}\psi_{l_1}(W_i;\theta_0,r_0,g_0,\delta_0)\psi_{l_2}(W_i;\theta_0,r_0,g_0,\delta_0)\bigg\vert.
\end{aligned}
\end{equation*}
We divide the proof of this corollary into three steps.

\bigskip
\noindent
\textbf{Step 1: Bounding $ \mathcal I_{ql_1l_2,1}$}. We aim to show that
\begin{equation} \label{eqn:bdonikl1}
\mathcal I_{ql_1l_2,1}=O_p(N^{-1/2})
\end{equation}
Note that
\begin{equation*}
\begin{aligned}
\E( \mathcal I_{ql_1l_2,1}^2 )  \leq n^{-1}\E\left[ \psi_{l_1}(W;\theta_0,r_0,g_0,\delta_0)^2\psi_{l_2}(W;\theta_0,r_0,g_0,\delta_0)^2 \right] 
= n^{-1}\E(U^4) \E(V^2_{l_1}V^2_{l_2}) = O(n^{-1}).
\end{aligned}
\end{equation*}
where the last step is due to the assumption that $U$ and the entries of $V$ have bounded fourth moment. Since $Q$ is finite, $\E( \mathcal I_{ql_1l_2,1}^2 ) = O(n^{-1})= O(N^{-1})$. Therefore, (\ref{eqn:bdonikl1}) holds.

\bigskip
\noindent
\textbf{Step 2:  Bounding $\mathcal I_{ql_1l_2,2}$}. We aim to show that
\begin{equation} \label{eqn:bdonikl2}
\mathcal I_{ql_1l_2,2}=o_p(N^{-1/4})
\end{equation}
To simplify the notation, write $\psi_{l}(W) = \psi_{l}(W;\theta_0,r_0,g_0,\delta_0)$, and $\widehat{\psi}_l(W) = \psi_{l}(W;\widehat{\theta},\widehat{r}_{q},\widehat{g}_{q},\widehat{\delta}_{q})$ for $l\in[d]$. Let $a\vee b = \max\{a,b\}$. Note that,
\begin{equation} \label{eqn:bdoniq2}
\begin{aligned}
\mathcal I_{ql_1l_2,2} & \leq \frac{1}{n}\sum_{i\in I_q}\Big\vert\widehat{\psi}_{l_1}(W_i)\widehat{\psi}_{l_2}(W_i) - \psi_{l_1}(W_i)\psi_{l_2}(W_i)\Big\vert\\
 &\leq \frac{2}{n}\sum_{i\in I_q} \left( |\widehat{\psi}_{l_1}(W_i)-\psi_{l_1}(W_i)|\vee|\widehat{\psi}_{l_2}(W_i)-\psi_{l_2}(W_i)| \right)\\
 &\quad\quad\quad\times \left( |\psi_{l_1}(W_i)|\vee |\psi_{l_2}(W_i)|+|\widehat{\psi}_{l_1}(W_i)-\psi_{l_1}(W_i)|\vee |\widehat{\psi}_{l_2}(W_i)-\psi_{l_2}(W_i)| \right)\\
 & \leq \left( \frac{2}{n}\sum_{i\in I_q}|\widehat{\psi}_{l_1}(W_i)-\psi_{l_1}(W_i)|^2\vee|\widehat{\psi}_{l_2}(W_i)-\psi_{l_2}(W_i)|^2 \right)^{1/2}\\
 &\quad\quad\quad\times \bigg[ \bigg(\frac{4}{n}\sum_{i\in I_q}|\psi_{l_1}(W_i)|^2\vee |\psi_{l_2}(W_i)|^2 \bigg)^{1/2}\\
 &\quad\quad\quad\quad\;+\bigg(\frac{4}{n}\sum_{i\in I_q}|\widehat{\psi}_{l_1}(W_i)-\psi_{l_1}(W_i)|^2\vee |\widehat{\psi}_{l_2}(W_i)-\psi_{l_2}(W_i)^2|\bigg)^{1/2} \bigg].
\end{aligned}
\end{equation}
By condition (C2), we have that $\E[|\psi(W;\theta_0,r_0,g_0,\delta_0)|^2] = \E(U^2) \E(V\trans V) = O(1)$. Therefore, 
\begin{equation} \label{eqn:bdonpsi12}
\left( \frac{1}{n}\sum_{i\in I_q}|\psi_{l_1}(W_i)|^2\vee |\psi_{l_2}(W_i)|^2 \right)^{1/2} = O_p(1).
\end{equation}
Note that
\begin{equation} \label{eqn:decomppsinhat0}
\begin{aligned}
& \frac{1}{n}\sum_{i\in I_q}\Big\|\psi(W_i;\widehat{\theta},\widehat{r}_{q},\widehat{g}_{q},\widehat{\delta}^{\infty)}_{q})-\psi(W;\theta_0,r_0,g_0,\delta_0)\Big\|_{\ell_2}^2\\
\leq \; & \frac{2}{n}\sum_{i\in I_q} \Big\| [\Phi(X_i)-\widehat{r}_{q}(Z_i)]\Phi(X_i)\trans(\widehat{\theta}-\theta_0) \Big\|_{\ell_2}^2\\
& + \frac{2}{n}\sum_{i\in I_q} \Big\| \psi(W_i;\theta_0,\widehat{r}_{q},\widehat{g}_{q},\widehat{\delta}_{q})-\psi(W_i;\theta_0,r_0,g_0,\delta_0) \Big\|_{\ell_2}^2.
\end{aligned}
\end{equation}
We next bound the two terms on the right-hand-side of (\ref{eqn:decomppsinhat0}) separately. 

For the first term, we have that, 
\begin{equation} \label{eqn:bdonterm1phi}
\begin{aligned}
&\frac{2}{n}\sum_{i\in I_q} \Big\| [\Phi(X_i)-\widehat{r}_{q}(Z_i)]\Phi(X_i)\trans(\widehat{\theta}-\theta_0) \Big\|_{\ell_2}^2 \\
\leq \; & \bigg[ \frac{2}{n}\sum_{i\in I_q}\|[\Phi(X_i)-\widehat{r}_{q}(Z_i)]\Phi(X_i)\trans\|_{2}^2 \bigg]\|\widehat{\theta}-\theta_0\|_{\ell_2}^2\\
\leq \; & \bigg( \frac{2}{n}\sum_{i\in I_q} \left\{ \|VV\trans\|_{2}^2 + \|[r_0(Z_i)-\widehat{r}_q(Z_i)] \Phi(X_i)\trans\|_2^2 \right\} \bigg) \|\widehat{\theta}-\theta_0\|_{\ell_2}^2 \\
= \; & O_p(\|\widehat{\theta}-\theta_0\|_{\ell_2}^2)  = O_p(N^{-1}).
\end{aligned}
\end{equation}
where the third step is due to conditions (C1) to (C3), and the last step is by Theorem \ref{thm:mainresulttri}. 

For the second term, we apply the Taylor expansion and obtain that, 
\begin{equation} \label{eqn:bdonterm2phi}
\begin{aligned}
& \frac{2}{n}\sum_{i\in I_q} \Big\| \psi(W_i;\theta_0,\widehat{r}_{q},\widehat{g}_{q},\widehat{\delta}_{q})-\psi(W_i;\theta_0,r_0,g_0,\delta_0) \Big\|_{\ell_2}^2\\
\leq \; & O_p\left\{\E(U^2) \|\widehat{r}_q-r_0\|_{\ell_2}^2+\E(V\trans V) \E\left[(\widehat{g}_q-g_0)^2+(\widehat{\delta}_q-\delta_0)^2 \right] \right\} = o_p(N^{-1/2}),
\end{aligned}
\end{equation}
where the last step is due to conditions (C2) and (C3).

Combining (\ref{eqn:bdoniq2}) to (\ref{eqn:bdonterm2phi}), we obtain that $\mathcal I_{ql_1l_2,2}  = o_p(N^{-1/4})$.

\bigskip
\noindent
\textbf{Step 3: Establishing the consistency}. By (\ref{eqn:bdonikl1}) and (\ref{eqn:bdonikl2}), we have, 
\begin{equation*}
\bigg\vert\frac{1}{n}\sum_{i\in I_q}\psi_{l_1}(W_i;\widehat{\theta},\widehat{r}_{q},\widehat{g}_{q},\widehat{\delta}_{q})\psi_{l_2}(W_i;\widehat{\theta},\widehat{r}_{q},\widehat{g}_{q},\widehat{\delta}_{q}) - \sigma^2\E(V_{l_1}V_{l_2}) \bigg\vert = o_p(N^{-1/4}).
\end{equation*} 
Note that
\begin{equation*}
\widehat{J}_0 = \frac{1}{Q}\sum_{q=1}^Q\frac{1}{n}\sum_{i\in I_q}[\Phi(X_i)-\widehat{r}_{q}(Z_i)]\Phi(X_i)\trans\overset{p}{\to} \E(VV\trans).
\end{equation*}
Then applying the continuous mapping theorem completes the proof of Corollary \ref{thm:covarestimation}.
\end{proof}

%%%%%%%%%%%%%%%%%%%%%%%%%%%%%%%%%%%%%%%%%%%%%%%%%%%
\subsection{Proof of Theorem \ref{thm:semvariance}}
\label{sec:pfofsemi-parametric}

We first begin with a quick review of estimation efficiency for semi-parametric problems in Section \ref{sec:rwofefficiency}. We then provide the proof of Theorem \ref{thm:semvariance} in Section \ref{pf:thm:semvariance}, which is built on the concepts discussed in Section \ref{sec:rwofefficiency}.

\subsubsection{Review of semi-parametric efficiency}
\label{sec:rwofefficiency}

In statistics, a parametric model is generally referred to as the one whose parameter space is finite-dimensional. A nonparametric model is the one whose parameter space is infinite-dimensional. Different from parametric or nonparametric models, a semi-parametric model involves a more complicated definition \citep{bickel1993efficient, Vandevaart1998, kosorok2007introduction}. To put in simple terms, a semi-parametric model is the one that has an infinite-dimensional parameter space, but whose parameter of interest is only finite-dimensional. 

The estimation problem in a semi-parametric model is described as follows.  Let $\Hcal$ denote an infinite-dimensional parameter space. Let $\xi_0\in\Hcal$ denote the true function. Let $\widetilde{\theta}$ be an estimator for the parameter of interest $\theta_0$ under the space $\Hcal$ using $N$ independent samples. Suppose that $\widetilde{\theta}$ satisfies the asymptotic normality, such that $\sqrt{N}(\widetilde{\theta}-\theta_0)$ follows a normal distribution when $N$ tends to infinity. Let $\Hcal_0$ denote a finite-dimensional subspace of $\Hcal$. Suppose that $\Hcal_0$ contains the true function $\xi_0$. Now we compare the estimation problems with the same observational data, but different parameter spaces: $\Hcal_0$ and $\Hcal$. Let $\widetilde{\theta}^{\Hcal_0}$ denote the maximum likelihood estimator of $\theta_0$ under the space $\Hcal_0$. Since the construction of $\widetilde{\theta}^{\Hcal_0}$ uses more information than $\widetilde{\theta}$, the asymptotic variance of $\widetilde{\theta}^{\Hcal_0}$ should be smaller than or equal to that of $\widetilde{\theta}$. Moreover, we reiterate the definition of the \emph{semi-parametric efficiency} as follows and refer to \citet{bickel1993efficient} and \citet{kosorok2007introduction} for details.

\begin{defi}
An estimator $\widetilde{\theta}$ is said to be semi-parametric efficient, if there exists a finite-dimensional space $\Hcal_0$, such that $\widetilde{\theta}^{\Hcal_0}$ has the same asymptotic variance as $\widetilde{\theta}$.
\end{defi}

Back to the estimation problem we target under the system of models (\ref{eqn:regeqn}) to (\ref{eqn:confandmediator}), we construct the finite-dimensional subspace $\Hcal_0$ by letting $\delta_0 = 0$, and consider the following $d$-dimensional parametric model indexed by the parameter $\gamma\in\R^d$: 
\begin{equation}
\label{eqn:semieff}
\xi_\gamma(v,z) = \xi_0(v,z)+v\trans\gamma,
\end{equation}
where $\delta_0$ is as defined in (\ref{eqn:decomoff0}), the function $\xi_0$ in (\ref{eqn:semieff}) is defined as $\xi_0(v,z) = [r_0(z)+v]\trans\theta_0+g_0(z)$, and the variables $v\in\R^d$ and $z\in\R^{p'}$. Regarding (\ref{eqn:regeqn}), the true value of $\gamma$ in (\ref{eqn:semieff}) is $\gamma_0=0$.
The observational data in the system of models (\ref{eqn:regeqn}) to (\ref{eqn:confandmediator}) can be rewritten as $\{(V_i,Z_i,Y_i): i=1,\ldots,N\}$, where each sample is an independent copy of $(V,Z,Y)$ following
\begin{equation}
\label{eqn:transdata}
Y = \xi_0(V,Z) + U.
\end{equation}
Then (\ref{eqn:semieff}) and (\ref{eqn:transdata}) form a linear regression model with parameter of interest $\gamma\in\R^d$.  Suppose the measurement error $U$ in (\ref{eqn:regeqn}) follows $\Ncal(0,\sigma^2)$. The maximum likelihood estimator  with the observational data $\{(V_i,Z_i,Y_i): i=1,\ldots,N\}$ is
\begin{equation*}
\widetilde{\gamma}_N = \underset{\gamma\in\R^d}{\arg\min} \left\{ \frac{1}{N}\sum_{i=1}^N \left[ Y_i-\xi_0(V_i,Z_i)- V_i\trans\gamma \right]^2 \right\}.
\end{equation*}
Then $\widetilde{\gamma}_N$ has the asymptotic expression, 
\begin{equation} \label{eqn:defofhatgamman}
\begin{aligned}
\widetilde{\gamma}_N & = \left[\E(VV\trans)\right]^{-1} \frac{1}{N} \sum_{i=1}^NV_iU_i+ o_p(N^{-1/2}).
\end{aligned}
\end{equation}
Given (\ref{eqn:semieff}), a natural estimator for $\theta_0$ is,
\begin{equation} \label{eqn:defofthetah0}
\widetilde{\theta}^{\Hcal_0} = \underset{\theta\in\R^d}{\arg\min} \ \E\left[ \xi_{\widetilde{\gamma}_N}(V,Z) - V\trans\theta \right]^2.
\end{equation}

\subsubsection{Proof of Theorem \ref{thm:semvariance}}
\label{pf:thm:semvariance}

\begin{proof}
Following the definition of the semi-parametric efficiency in Section \ref{sec:rwofefficiency}, it suffices to show that $\widetilde{\theta}^{\Hcal_0}$ defined in (\ref{eqn:defofthetah0}) has the same asymptotic variance as the estimator $\widehat{\theta}$ in (\ref{eqn:TDE}).

Toward that end, the asymptotic variance of $\widetilde{\theta}^{\Hcal_0} $ can be obtained by the delta method. Define
\begin{equation}
\label{eqn:defofthetat}
\theta(\gamma) =  \underset{\theta\in\R^d}{\arg\min} \ \E[\xi_\gamma(V,Z) -V\trans\theta]^2,
\end{equation}
for each $\gamma$ near $0$. Here if $\gamma=0$, then $\theta(0) = \theta_0$. Let 
\begin{equation*}
\begin{aligned}
\Psi(\theta,\gamma) &= \partial_\theta\E\left[ \xi_\gamma(V,Z)-V\trans\theta \right]^2\\
& = \partial_\theta\E\left[\xi_0(V,Z)+V\trans \gamma-V\trans\theta \right]^2,
\end{aligned}
\end{equation*}
where the second step is by (\ref{eqn:semieff}). Then (\ref{eqn:defofthetat}) implies that $\Psi[\theta(\gamma),\gamma]=0$ for all $\gamma$ near $0$. By the implicit function theorem, 
\begin{equation} \label{eqn:derofthetat}
\begin{aligned}
\left.\frac{\partial\theta(\gamma)}{\partial \gamma}\right\vert_{\gamma=0} & = -\left[\partial_{\theta\trans}\Psi(\theta,0)\vert_{\theta=\theta_0}\right]^{-1}\partial_{\gamma}\Psi(\theta_0,0)\\
& = -\left( \E\left\{ \partial^2_{\theta\theta\trans}[\xi_0(V,Z)-V\trans\theta]^2 \right\} \big\vert_{\theta=\theta_0} \right)^{-1}\times 2\E(VV\trans) \\
& = -1.
\end{aligned}
\end{equation}
By the delta method, we have, 
\begin{equation*}
\widetilde{\theta}^{\Hcal_0}  - \theta_0 = \theta(\widetilde{\gamma}_N) - \theta(0) =\left.\frac{\partial\theta(\gamma)}{\partial \gamma}\right\vert_{\gamma=0}\widehat{\gamma}_n + o_p(N^{-1/2}).
\end{equation*}
Together with (\ref{eqn:defofhatgamman}) and (\ref{eqn:derofthetat}), we have, 
\begin{equation*}
\begin{aligned}
\widetilde{\theta}^{\Hcal_0}  -\theta_0 & = -\left[\E(VV\trans)\right]^{-1} \frac{1}{N} \sum_{i=1}^NU_iV_i+ o_p(N^{-1/2}),
\end{aligned}
\end{equation*}
which implies the asymptotic normality:
\begin{equation*}
\sqrt{N}(\widetilde{\theta}^{\Hcal_0}-\theta_0)\overset{d}{\to} {\mathcal N}\left( 0,\sigma^2[\E(VV\trans)]^{-1} \right).
\end{equation*}
Compared to (\ref{eqn:asympnormalityoftheta}), we see that $\widetilde{\theta}^{\Hcal_0}$ achieves the same asymptotic variance as the estimator $\widehat{\theta}$ in (\ref{eqn:TDE}). Then by definition, the estimator $\widehat{\theta}$ is semi-parametric efficient. This completes the proof of Theorem \ref{thm:semvariance}.
\end{proof}

%%%%%%%%%%%%%%%%%%%%%%%%%%%%%%%%%%%%%%%%%%%%%%%%%%%
\subsection{Proof of Theorem \ref{thm:infonpred}}
\label{sec:pfofasympthonestci}

\begin{proof}
Define the empirical process,
\begin{equation*}
\widetilde{Z}_N(x) = \sqrt{N}\left[ \Phi(x)\trans\widehat{\theta}-f_0(x) \right],\quad\forall x\in\Xcal^p.
\end{equation*}
Then by definition of $\widehat{\theta}$ in (\ref{eqn:TDE}), it is equivalent to write
\begin{equation*}
\begin{aligned}
\widetilde{Z}_N(x)  & =  \sqrt{N}\Phi(x)\left\{ \frac{1}{Q}\sum_{q=1}^Q\frac{1}{n}\sum_{i\in I_q}\left[ \Phi(X_i)-\widehat{r}_{q}(Z_i) \right] \Phi(X_i)\trans \right\}^{-1}\\
&\quad\times \frac{1}{Q}\sum_{q=1}^Q\frac{1}{n}\sum_{i\in I_q}\left[ \Phi(X_i)-\widehat{r}_{q}(Z_i) \right] \left[ Y_i-\widehat{g}_{q}(Z_i)-\widehat{\delta}_{q}(X_i) \right]-\sqrt{N}f_0(x).
\end{aligned}
\end{equation*}
Define $\widetilde{V}^Z = \sup_{x\in\Xcal^p}\widetilde{Z}_N(x)$. We divide the proof of this theorem into four steps.

\bigskip
\noindent
\textbf{Step 1}. We aim to prove the following statement: There exists a Gaussian process $\widetilde{\mathbb H}_N(x)$, such that  $\E[\sup_{x\in\Xcal^p}\widetilde{\mathbb H}_N(x) ]\leq C\sqrt{\log N}$, for some constant $C>0$, and  a sequence of random variables $W_N^0$, such that $W_N^0 =\sup_{x\in\Xcal^p}\widetilde{\mathbb H}_N(x)$ and $\P\left( |W_N^0-\widetilde{V}^Z|>\epsilon_{1N} \right) < \delta_{1N}$, for some $(\epsilon_{1N},\delta_{1N})\to 0$ as $N\to\infty$.

We construct the Gaussian process $\widetilde{\mathbb H}_N(x) $ as
\begin{equation*}
\begin{aligned}
\widetilde{\mathbb H}_N(x) & = \sqrt{N}\Phi(x)\trans\left\{\frac{1}{Q}\sum_{q=1}^Q\frac{1}{n}\sum_{i\in I_q}\left[ \Phi(X_i)-\widehat{r}_{q}(Z_i) \right]\Phi(X_i)\trans\right\}^{-1}\\
&\quad\quad\times \frac{1}{Q}\sum_{q=1}^Q\frac{1}{n}\sum_{i\in I_q}\left[\Phi(X_i)-\widehat{r}_{q}(Z_i)\right] U_{i},
\end{aligned}
\end{equation*}
where $\{U_i:i=1,\ldots,N\}$ are independent copies of the error term $U$ in (\ref{eqn:regeqn}).
Then $\widetilde{\mathbb H}_N(x)$ is a Gaussian variable conditional on $\{(X_i,Z_i)\}_{i=1}^N$. By Jensen's inequality, there exists some constant $C>0$, such that
\begin{equation*}
\exp\left[ t\E(V_N^0) \right] \leq \E\exp\left( tV_N^0 \right) = \E\left\{ \sup_{x\in\Xcal^p}\exp\left[t\widetilde{\mathbb H}_N(x)\right] \right\} \leq N\exp\left( Ct^2 \right), 
\end{equation*}
where the last inequality follows from the definition of the Gaussian moment generating function. Rewriting this inequality, we have $\E(V_N^0) \leq \log N/t + Ct$. Setting $t = \sqrt{\log N/C}$, we obtain, 
\begin{equation*}
\E\left[ \sup_{x\in\Xcal^p}\widetilde{\mathbb H}_N(x) \right] \leq C\sqrt{\log N}.
\end{equation*}
Note that
\vspace{-0.05in}
\begin{equation*}
\begin{aligned}
& \widetilde{Z}_N(x) - \widetilde{\mathbb H}_N(x) \\
= \; & \sqrt{N}[\Phi(x)\trans\theta_0-f_0(x)] + \sqrt{N}\Phi(x)\trans(J_0+R_{N,1})^{-1} \\
&\quad \times \frac{1}{Q}\sum_{q=1}^Q\frac{1}{n}\sum_{i\in I_q}\left[ \Phi(X_i)-\widehat{r}_{q}(Z_i) \right] \left[ \delta_0(X_i)-\widehat{\delta}_{q}(X_i)+g_0(Z_i)-\widehat{g}_{q}(Z_i) \right], 
\end{aligned}
\end{equation*}
where the quantities $J_0$ and $R_{N,1}$ are defined in (\ref{eqn:defofnotationsj0}). Then by (\ref{eqn:bdonRn1}), condition (C2), and Cauchy-Schwarz inequality, we have that, 
\begin{equation*}
\begin{aligned}
& \widetilde{Z}_N(x) - \widetilde{\mathbb H}_N(x)  \leq \sqrt{N} \left[ \Phi(x)\trans\theta_0-f_0(x) \right] \\
&\quad + \sqrt{N}\Phi(x)\trans J_0^{-1}  O_p\left( \E\left\{[\widehat{g}_q(Z)-g_0(Z)]^2 \right\} + \E\left\{ \left[\widehat{\delta}_q(X)-\delta_0(X)\right]^2 \right\} \right) + \sqrt{N}\Phi(x)\trans J_0^{-1}\\
&\quad \times O_p\left( \E\left[ \|\widehat{r}_q(Z)-r_0(Z)\|_{\ell_2} \right] \left[\left( \E\left\{ [\widehat{g}_q(Z)-g_0(Z)]^2 \right\} \right)^{1/2} + \left(\E\left\{ [\widehat{\delta}_q(X)-\delta_0(X)]^2 \right\} \right)^{1/2} \right] \right).
\end{aligned}
\end{equation*}
By conditions (C1) and (C3$'$), we have, 
\begin{equation} \label{eqn:difofhnzn}
\begin{aligned}
\widetilde{Z}_N(x) - \widetilde{\mathbb H}_N(x)  \leq O\left( \sqrt{N}\{\E[\delta^2_0(X)]\}^{1/2} \right) + O_p(N^{-c_{\min}}),
\end{aligned}
\end{equation}
where $c = \min\{c_r,c_g,c_\delta\}>0$. Under condition (C4'), the approximation error $\delta_0$ can be bounded as $\E[\delta_0^2(X)] \leq O(s^{-2k})$ \citep{devore1993constructive}. Therefore, by the condition that $s=\ceil{N^{(1+2c)/2k}}\geq N^{(1+2c)/2k}$ for some $c\in(0,c_{\min}]$, we have, 
\begin{equation*}
\E[\delta_0^2(X)] \leq O\left( N^{-(1+2c)} \right).
\end{equation*}
Define $V_N^0 = \sup_{x\in\Xcal^p}\widetilde{\mathbb H}_N(x)$. Recall that $\widetilde{V}^Z = \sup_{x\in\Xcal^p}\widetilde{Z}_N(x)$. Then by (\ref{eqn:difofhnzn}), there exists some constant $C>0$, such that 
\begin{equation} \label{eqn:bdonvnz}
\P\left( \left| V_N^0-\widetilde{V}^Z \right| > CN^{-c} \right) \leq \P\left(\sup_{x\in\Xcal^p} \left| \widetilde{\mathbb H}_N(x)-\widetilde{Z}_N(x) \right| > CN^{-c}\right)\leq N^{-1}.
\end{equation}
Letting $\epsilon_{1N} = CN^{-c}$, $\delta_{1N} = N^{-1}$ and $W_N^0 \overset{d}{=} V_N^0$ completes the proof of Step 1.

\bigskip
\noindent
\textbf{Step 2}. We aim to prove the following anti-concentration inequality for any $\epsilon>0$,
\begin{equation*}
\sup_{t\in\R}\P\left[ \left\vert\sup_{x\in\Xcal^p} \left| \widetilde{\mathbb H}_N(x) \right|-t\right\vert \leq \epsilon \right] \leq C\epsilon\sqrt{\log N}.
\end{equation*}
This is true due to the result of Step 1 and Corollary 2.1 of \citet{chernozhukov2014anti}.

\bigskip
\noindent
\textbf{Step 3}.We aim to prove the following statement: Let $c_N(\alpha)$ and $\widehat{c}_N(\alpha)$ be the $(1-\alpha)$-quantiles of $\widetilde{V}^Z$ and  $V_N^0$, respectively. Then there exist $\tau_N,\epsilon_{2N},\delta_{2N}>0$, such that
\begin{equation*}
\P\big[ \widehat{c}_N(\alpha)<c_N(\alpha+\tau_N)-\epsilon_{2N} \big] \leq \delta_{2N}, \quad \P\big[ \widehat{c}_N(\alpha)>c_N(\alpha-\tau_N)+\epsilon_{2N} \big] \leq \delta_{2N},
\end{equation*}
and $(\tau_N, \epsilon_{2N},\delta_{2N})\to 0$ as $N\to\infty$. 

Recall that the Gaussian multiplier process $\widehat{\mathbb H}_N(x)$ in Section \ref{sec:uqoff0} is defined as
\begin{equation*}
\begin{aligned}
\widehat{\mathbb H}_N(x) & = \sqrt{N}\Phi(x)\trans\left\{\frac{1}{Q}\sum_{q=1}^Q\frac{1}{n}\sum_{i\in I_q}\left[ \Phi(X_i)-\widehat{r}_{q}(Z_i) \right]\Phi(X_i)\trans\right\}^{-1}\\
&\quad\quad\quad\quad\quad\quad \times \frac{1}{Q}\sum_{q=1}^Q\frac{1}{n}\sum_{i\in I_q} \left[ \Phi(X_i)-\widehat{r}_{q}(Z_i) \right] \widehat{\sigma}(\widehat{\theta})\xi_{i},
\end{aligned}
\end{equation*}
where $\xi = (\xi_1,\ldots,\xi_N)\trans$ consists of independent standard normal variables. We consider the following process:
\begin{equation*}
\begin{aligned}
\widehat{\mathbb H}^{(1)}_N(x) & = \sqrt{N}\Phi(x)\trans\left\{\frac{1}{Q}\sum_{q=1}^Q\frac{1}{n}\sum_{i\in I_q}[\Phi(X_i)-\widehat{r}_{q}(Z_i)]\Phi(X_i)\trans\right\}^{-1}\\
&\quad\quad\quad\quad\quad\quad\quad \times \frac{1}{Q}\sum_{q=1}^Q\frac{1}{n}\sum_{i\in I_q}[\Phi(X_i)-\widehat{r}_{q}(Z_i)]\sigma\xi_{i},
\end{aligned}
\end{equation*}
Let $\widehat{V}_N = \sup_{x_0\in\Xcal^p}\widehat{\mathbb H}_N(x)$, and $\widehat{V}^{(1)}_N = \sup_{x\in\Xcal^p}\widehat{\mathbb H}_N^{(1)}(x)$. Denote $\Delta \mathbb H^{(1)}(x) = \widehat{\mathbb H}_N^{(1)}(x) - \widehat{\mathbb H}_N(x)$. By the triangle inequality,
\begin{equation}
\label{eqn:bdonsupdeltah}
\begin{aligned}
\sup_{x\in\Xcal^p} \big| \Delta \mathbb H^{(1)}(x) \big| & \leq \left[ \sup_{x_0\in\Xcal^p}\mathcal I_1(x)+\sup_{x\in\Xcal^p}\mathcal I_2(x) \right] \big| \sigma-\widehat{\sigma}(\widehat{\theta}) \big|,
\end{aligned}
\end{equation}
where 
\begin{equation*}
\begin{aligned}
\mathcal I^{\mathbb H}_1(x) &= \|\Phi(x)\|_{\ell_2}\Big\|\sqrt{N}\left( \widehat{J}_0^{-1}-J_0^{-1} \right) \times \frac{1}{Q}\sum_{q=1}^Q\frac{1}{n}\sum_{i\in I_q}[\Phi(X_i)-\widehat{r}_{q}(Z_i)]\xi_i\Big\|_{\ell_2},\\
\mathcal I^{\mathbb H}_2(x) &=  \|\Phi(x)\|_{\ell_2}  \Big\|\sqrt{N}J_0^{-1} \times \frac{1}{Q}\sum_{q=1}^Q\frac{1}{n}\sum_{i\in I_q}[\Phi(X_i)-\widehat{r}_{q}(Z_i)]\xi_i\Big\|_{\ell_2}.
\end{aligned} 
\end{equation*}
and $J_0,\widehat{J}_0$ are as defined in (\ref{eqn:defofnotationsj0}). By (\ref{eqn:bdonjandr}) and condition (C1), we have that, 
\begin{equation}
\label{eqn:bdonsupi1h}
\begin{aligned}
& \sup_{x\in\Xcal^p}\mathcal I^{\mathbb H}_1(x) \leq \sqrt{N}\times o_p(N^{-1/4}) \\
& \quad \times \frac{1}{Q}\sum_{q=1}^Q\frac{1}{\sqrt{n}}O_p\left( \left[ \E \left( \|V\xi\|_{\ell_2}^2 \right) \right]^{1/2} + \left\{ \E\left[ \|\widehat{r}_q(Z)-r_0(Z)\|^2_{\ell_2} \right] \E(\xi^2) \right\}^{1/2} \right) = o_p(N^{-1/4}).
\end{aligned}
\end{equation}
Moreover, by conditions (C1) and (C2), we have, 
\begin{equation} \label{eqn:bdonsupi2h}
\begin{aligned}
\sup_{x\in\Xcal^p}\mathcal I^{\mathbb H}_2(x) &\leq \sqrt{N} \times \frac{1}{Q}\sum_{q=1}^Q\frac{1}{\sqrt{n}}O_p\left( \left[\E\left(\|V\xi\|_{\ell_2}^2 \right) \right]^{1/2} + \left\{\E\left[ \|\widehat{r}_q(Z)-r_0(Z)\|^2_{\ell_2} \right] \E(\xi^2) \right\}^{1/2} \right)\\
&= O_p(1).
\end{aligned}
\end{equation}
We next bound $|\sigma-\widehat{\sigma}(\widehat{\theta})|$. Note that
\begin{equation}
\label{eqn:bdonsigma2iq}
\begin{aligned}
\big| \widehat{\sigma}^2(\widehat{\theta}) -\sigma^2 \big| & \leq \frac{1}{Q}\sum_{q=1}^Q\bigg| \frac{1}{n}\sum_{i\in I_q} \left[ Y_i-\Phi(X_i)\widehat{\theta}-\widehat{\delta}_q(X_i)-\widehat{g}_q(Z_i) \right]^2-\sigma^2 \bigg|\\
&\leq \frac{1}{Q}\sum_{q=1}^Q[I_{q,1}+I_{q,2}],
\end{aligned}
\end{equation}
where 
\begin{equation*}
\begin{aligned}
I_{q,1} & =  \bigg|\frac{1}{n}\sum_{i\in I_q}\left[ Y_i-\Phi(X_i)\theta_0-\delta_0(X_i)-g_0(Z_i) \right]^2-\sigma^2\bigg| = \bigg|\frac{1}{n}\sum_{i\in I_q}U_i^2-\sigma^2\bigg|,\\
I_{q,2} & =  \bigg|\frac{1}{n}\sum_{i\in I_q}\left[Y_i-\Phi(X_i)\widehat{\theta}-\widehat{\delta}_q(X_i)-\widehat{g}_q(Z_i) \right]^2-\frac{1}{n}\sum_{i\in I_q}[Y_i-\Phi(X_i)\theta_0-\delta_0(X_i)-g_0(Z_i)]^2\bigg|.
\end{aligned}
\end{equation*}
To bound $I_{q,1}$, we have that, 
\begin{equation*}
\E(I^2_{q,1}) \leq n^{-1} \E(U^4) = O(n^{-1}),
\end{equation*}
where the last step is due to that $U$ is a normal random variable and hence $U$ has a bounded fourth moment. Since $Q$ is finite, we have $\E(I_{q,1}^2)= O(N^{-1})$ and 
\begin{equation}
\label{eqn:bdiq2q}
I_{q,1}=O_p(N^{-1/2}).
\end{equation}
To bound $I_{q,2}$, we have that, 
\begin{equation*}
\begin{aligned}
I_{q,2} &\leq \frac{1}{n}\sum_{i\in I_q}\Big| \left[ Y_i-\Phi(X_i)\widehat{\theta}-\widehat{\delta}_q(X_i)-\widehat{g}_q(Z_i) \right]^2-U_i^2\Big|\\
&\leq \frac{1}{n}\sum_{i\in I_q}\Big|Y_i-\Phi(X_i)\widehat{\theta}-\widehat{\delta}_q(X_i)-\widehat{g}_q(Z_i)-U_i\Big|\\
&\quad\quad\quad\times \left( |U_i| + \Big|Y_i-\Phi(X_i)\widehat{\theta}-\widehat{\delta}_q(X_i)-\widehat{g}_q(Z_i)-U_i\Big| \right)\\
& \leq \left( \frac{1}{n}\sum_{i\in I_q} \left[ Y_i-\Phi(X_i)\widehat{\theta}-\widehat{\delta}_q(X_i)-\widehat{g}_q(Z_i)-U_i \right]^2 \right)^{1/2}\\
&\quad\quad\quad\times\left[\left(\frac{2}{n}\sum_{i\in I_q}U_i^2\right)^{1/2}+\left\{\frac{2}{n}\sum_{i\in I_q}\left[Y_i-\Phi(X_i)\widehat{\theta}-\widehat{\delta}_q(X_i)-\widehat{g}_q(Z_i)-U_i \right]^2\right\}^{1/2}\right]
\end{aligned}
\end{equation*}
Since $\E(U^2)=\sigma^2<\infty$, we have that, 
\begin{equation*}
\left( \frac{2}{n}\sum_{i\in I_q}U_i^2 \right)^{1/2}= O_p(1).
\end{equation*}
Note that
\begin{equation*}
\begin{aligned}
&\frac{1}{n}\sum_{i\in I_q}\left[ Y_i-\Phi(X_i)\widehat{\theta}-\widehat{\delta}_q(X_i)-\widehat{g}_q(Z_i)-U_i \right]^2\\
\leq \; & \frac{2}{n}\sum_{i\in I_q}\left[ \Phi(X_i)\trans(\widehat{\theta}-\theta_0) \right]^2+\frac{2}{n}\sum_{i\in I_q}\left[ Y_i-\Phi(X_i)\theta_0-\widehat{\delta}_q(X_i)-\widehat{g}_q(Z_i)-U_i \right]^2\\
\leq \; & O_p\left( \|\widehat{\theta}-\theta_0\|_{\ell_2}^2 \right) + O_p\left( \E\left[(\widehat{g}_q-g_0)^2+(\widehat{\delta}_q-\delta_0)^2 \right] \right)\\
\leq \; & O_p(N^{-1}s^p) + O_p(N^{-1/2-c})
\end{aligned}
\end{equation*}
where the second step is by condition (C1), and the last step is by Theorem \ref{thm:mainresulttri}, condition (C3$'$), and the condition that $0<c\leq \frac{k-p}{2(k+p)}< \frac{1}{2}$. 
By condition (C4) that $k>p$, and $c\leq \frac{k-p}{2(k+p)}$,  there exists constant $C>0$ such that
\begin{equation*}
N^{-1}s^p\leq CN^{-1+p/2k+cp/k}\leq CN^{-1/2-c}.
\end{equation*} 
Therefore,
\begin{equation}
\label{eqn:bdiq214}
I_{q,2}\leq O_p(N^{-1/4-c/2}).
\end{equation}
Combining (\ref{eqn:bdonsigma2iq}) to (\ref{eqn:bdiq214}), we have that, 
\begin{equation*}
\begin{aligned}
\widehat{\sigma}(\widehat{\theta}) -\sigma& = O\left( |\widehat{\sigma}^2(\widehat{\theta}) -\sigma^2| \right) \leq O_p(N^{-1/4-c/2}).
\end{aligned}
\end{equation*}
By (\ref{eqn:bdonsupdeltah}) to (\ref{eqn:bdonsupi2h}), we have that, 
\begin{equation*}
\sup_{x\in\Xcal^p}|\Delta \mathbb H^{(1)}(x)| \leq O_p(N^{-1/4-c/2})
\end{equation*}
Then there exists a constant $C>0$, such that
\begin{equation}
\label{eqn:bdonvnhatvn1}
\P\left( \big| \widehat{V}_N-\widehat{V}_N^{(1)} \big|>CN^{-1/4-c/2} \right) \leq \P\left( \sup_{x\in\Xcal^p}|\Delta \mathbb H^{(1)}(x)|>CN^{-1/4-c/2} \right)\leq N^{-1}.
\end{equation}
Since $\sigma\xi\overset{d}{=}(U_1,\ldots,U_N)\trans$, we have $\sup_{x\in\Xcal^p}\widehat{\mathbb H}_N^{(1)}(x)\overset{d}{=}\sup_{x\in\Xcal^p}\widetilde{\mathbb H}_N(x)$. That is, $\widehat{V}^{(1)}_N\overset{d}{=} V_N^0$. Combining (\ref{eqn:bdonvnz}) with (\ref{eqn:bdonvnhatvn1}), we have that,
\begin{equation*}
\P\left(\big| \widehat{V}_N-\widetilde{V}_N^Z \big|>CN^{-c}\right)\leq N^{-1}.
\end{equation*}
Therefore, by the definition of $\widehat{c}_N(\alpha)$,
\begin{equation*}
\begin{aligned}
\P\left(\widetilde{V}_N^Z\leq \widehat{c}_N(\alpha)+CN^{-c}\right) \geq \P\left(\widehat{V}_N\leq \widehat{c}_N(\alpha)\right) - \P\left(\big| \widehat{V}_N-\widetilde{V}_N^Z \big| > CN^{-c}\right)
\geq 1-\alpha-N^{-1},
\end{aligned}
\end{equation*}
which implies that the estimated quantile is lower bounded as
\begin{equation*}
\widehat{c}_N(\alpha)\geq c_N(\alpha+N^{-1})-CN^{-c},\quad\text{for some }c\in(0,c_{\min}].
\end{equation*}
Similarly, we also have $\widehat{c}_N(\alpha)\leq c_N(\alpha-N^{-1})+CN^{-c}$. Setting $\tau_N=N^{-1}$, $\epsilon_{2N} = CN^{-c}$ and $\delta_{2N}=N^{-1}$ completes the proof of Step 3.

\bigskip
\noindent
\textbf{Step 4}. By verifying the statements in Steps 1 to 3, we now apply Corollary 3.1 of \citet{chernozhukov2014anti} and obtain that, 
\begin{equation*}
\P\big[ f_0(x)\in\Ccal_N(x),\text{ for all }x\in\Xcal^p \big] \geq 1-\alpha-CN^{-c}, \; \textrm{ for any } \; 0 < \alpha < 1.
\end{equation*}
Therefore, the confidence band $\Ccal_N$ in \eqref{eqn:cioff0} is asymptotically valid.

%\bigskip
%\noindent
%\textbf{Step 5}. Finally, we establish the $\sqrt{N}$-consistency of $\widehat{f}$, i.e., $\E_X\{[\widehat{f}(X)-f_0(X)]^2\}=O(N^{-1})$. This is true from the decomposition that
%\begin{equation*}
%\begin{aligned}
%\E\left\{ \left[\widehat{f}(X)-f_0(X) \right]^2 \right\} &\leq 2\E\left\{ \left[\widehat{f}(X)-\Phi(X)\trans\theta_0\right]^2 \right\} + 2\E\left\{ \left[\Phi(X)\trans\theta_0-f_0(X)\right]^2 \right\} \\
%& \leq 2\E\left[ \|\Phi(X)\|_{\ell_2}^2 \right] \E\left[(\widehat{\theta}-\theta_0)^2\right] + 2\E[\delta^2(X)]\\
%& = O(N^{-1}) + O(N^{-1-2c}) = O(N^{-1}),
%\end{aligned}
%\end{equation*}
%where the third step is by condition (C1) and Theorem \ref{thm:mainresulttri}.

This completes the proof of Theorem \ref{thm:infonpred}.
\end{proof}

%%%%%%%%%%%%%%%%%%%%%%%%%%%%%%%%%%%%%%%%%%%%%%%%%%%
\subsection{Proof of Proposition \ref{prop:unimodelregression}}
\label{sec:pfofunimodal}

\begin{proof}
Define
\begin{equation} \label{eqn:defofrn1tilde}
\widetilde{R}_{N,1} = \frac{1}{N}\sum_{i=1}^N\Phi(X_i)\Phi(X_i)\trans - \E[\Phi(X)\Phi(X)\trans] .
\end{equation}
Since $\E(\|\widetilde{R}_{N,1}\|_2^2) \leq N^{-1}\E[\|\Phi(X)\Phi(X)\trans\|_2^2] = O(N^{-1})$,
we have
\begin{equation}
\label{eqn:bdrn1}
\|\widetilde{R}_{N,1}\|_2 = O_p(N^{-1/2}), 
\end{equation}
where $\|\cdot\|_2$ denotes the matrix $2$-norm. Note that
\begin{equation*}
\begin{aligned}
\widehat{\theta}_{\text{UR}} - \theta_0 = \left\{\E[\Phi(X)\Phi(X)\trans]+\widetilde{R}_{N,1}\right\}^{-1} \left\{ \frac{1}{N}\sum_{i=1}^N\Phi(X_i)[\delta_0(X_i)+g_0(Z_i)+U_i]\right\}.
\end{aligned}
\end{equation*}
By (\ref{eqn:bdrn1}), we have that, 
\begin{equation}
\label{eqn:boundrb1}
\begin{aligned}
& \Big\|\left\{\E[\Phi(X)\Phi(X)\trans]+\widetilde{R}_{N,1}\right\}^{-1} - \left\{\E[\Phi(X)\Phi(X)\trans]\right\}^{-1}\Big\|_2 \\
= \; & \Big\|\left\{\E[\Phi(X)\Phi(X)\trans]+\widetilde{R}_{N,1}\right\}^{-1} \widetilde{R}_{N,1} \left\{\E[\Phi(X)\Phi(X)\trans]\right\}^{-1}\Big\|_2 \\
\leq \; & \Big\|\left\{\E[\Phi(X)\Phi(X)\trans]+\widetilde{R}_{N,1}\right\}^{-1}\Big\|_2 \times \|\widetilde{R}_{N,1} \|_2\times \|\left\{\E[\Phi(X)\Phi(X)\trans]\right\}^{-1}\|_2 \\
= \; & o_p(1).
\end{aligned}
\end{equation}
Therefore, 
\begin{equation*}
\begin{aligned}
\widehat{\theta}_{\text{UR}} - \theta_0  =  \left\{ \E[\Phi(X)\Phi(X)\trans] \right\}^{-1} \left\{ \frac{1}{N}\sum_{i=1}^N\Phi(X_i)[\delta_0(X_i)+g_0(Z_i)+U_i] \right\} + o_p(N^{-1/2}).
\end{aligned}
\end{equation*}
This completes the proof of Proposition \ref{prop:unimodelregression}.
\end{proof}

%%%%%%%%%%%%%%%%%%%%%%%%%%%%%%%%%%%%%%%%%%%%%%%%%%%
\subsection{Proof of  Proposition \ref{thm:twostep}}
\label{sec:pfofpropsingledebiased}

\begin{proof}
By the definition of $\widehat{\theta}_{\text{DUR}}$, we have that, 
\begin{equation*}
\widehat{\theta}_{\text{DUR}} -\theta_0 =\left\{\E[\Phi(X)\Phi(X)\trans]+\widetilde{R}_{N,1}\right\}^{-1} \left\{\frac{1}{N}\sum_{i=1}^N\Phi(X_i)\left[ \delta_0(X_i)-\widehat{\delta}_{\text{DUR}}(X_i)+g_0(Z_i)+U_i \right] \right\},
\end{equation*}
where $\widetilde{R}_{N,1}$ is as defined in (\ref{eqn:defofrn1tilde}). Let 
\begin{equation*}
\widetilde{R}^{\text{DUR}}_{N,2} = \frac{1}{N}\sum_{i=1}^N\Phi(X_i)\left[ \delta_0(X_i)-\widehat{\delta}_{\text{DUR}}(X_i) \right].
\end{equation*}
Then by triangle inequality, we have,
\begin{equation*}
\big\| \widetilde{R}^{\text{DUR}}_{N,2} \big\|_{\ell_2} \leq N^{-1/2}\left( \Ical_1^{\text{DUR}} + \Ical_2^{\text{DUR}} \right),
\end{equation*}
where
\vspace{-0.05in}
\begin{equation*}
\begin{aligned}
\mathcal I^{\text{DUR}}_1 & = \bigg\|\frac{1}{\sqrt{N}}\sum_{i=1}^N \left\{ \Phi(X_i)\widehat{\delta}_{\text{DUR}}(X_i) - \E\left[ \Phi(X)\widehat{\delta}_{\text{DUR}}(X) \right] \right\}\\
&\quad\quad\quad - \frac{1}{\sqrt{N}}\sum_{i=1}^N\{\Phi(X_i)\delta_0(X_i) - \E[\Phi(X)\delta_0(X)]\} \bigg\|_{\ell_2}, \\
\mathcal I^{\text{DUR}}_2 & = \sqrt{N}\bigg\| \E\left[ \Phi(X)\widehat{\delta}_{\text{DUR}}(X) \right] - \E\left[ \Phi(X)\delta_0(X) \right] \bigg\|_{\ell_2}.
\end{aligned}
\end{equation*}
To bound $\mathcal I^{\text{DUR}}_1$, note that,
\begin{equation*}
\E\left[ (\mathcal I^{\text{DUR}}_1)^2 \right]  \leq \frac{1}{N} \E\left\{ \bigg\|\sum_{i=1}^N\Phi(X_i)\left[ \widehat{\delta}_{\text{DUR}}(X_i)-\delta_0(X_i) \right]\bigg\|_{\ell_2}^2 \right\}.
\end{equation*}
Then under condition (C1),
\begin{equation}
\label{eqn:bdoni1udr}
\mathcal I^{\text{DUR}}_1 = O_p\left[ N^{1/2}\left( \E\left\{  [\widehat{\delta}_{\text{DUR}}(X)-\delta_0(X)]^2 \right\} \right)^{1/2} \right].
\end{equation}
To bound $\mathcal I^{\text{DUR}}_2$, note that, 
\begin{equation}
\label{eqn:bdoni2udr}
\begin{aligned}
\mathcal I^{\text{DUR}}_2 & = \sqrt{N}\bigg\| \E\left\{ \Phi(X)\left[ \widehat{\delta}_{\text{DUR}}(X)-\delta_0(X) \right] \right\} \bigg\|_{\ell_2} \\
&= O\left[ N^{1/2}\left(\E\left\{ \left[\widehat{\delta}_{\text{DUR}}(X)-\delta_0(X) \right]^2 \right\} \right)^{1/2} \right].
\end{aligned}
\end{equation}
Combining (\ref{eqn:bdoni1udr}) and (\ref{eqn:bdoni2udr}), we have that, 
\begin{equation*}
\widetilde{R}^{\text{DUR}}_{N,2} = O_p\left[ \left(\E\left\{ \left[\widehat{\delta}_{\text{DUR}}(X)-\delta_0(X) \right]^2 \right\}\right)^{1/2} \right].
\end{equation*}
Together with the derived bound of $\widetilde{R}_{N,1}$ in (\ref{eqn:boundrb1}) and that $\E[\Phi(X)\Phi(X)\trans]$ is invertible, we have, 
\begin{equation*}
\begin{aligned}
 \widehat{\theta}_{\text{DUR}} - \theta_0 &=  \left\{ \E[\Phi(X)\Phi(X)\trans] \right\}^{-1} \left\{ \frac{1}{N}\sum_{i=1}^N\Phi(X_i)[g_0(Z_i)+U_i] \right\} \\
 &\quad\; +  O_p\left[ \left( \E\left\{ \left[\widehat{\delta}_{\text{DUR}}(X)-\delta_0(X) \right]^2 \right\} \right)^{1/2} \right] + o_p(N^{-1/2}).
 \end{aligned}
 \end{equation*}
This completes the proof of Proposition \ref{thm:twostep}.
\end{proof}

%%%%%%%%%%%%%%%%%%%%%%%%%%%%%%%%%%%%%%%%%%%%%%%%%%%
\subsection{Proof of  Proposition \ref{thm:sjr}}
\label{sec:pfofsjr}

\begin{proof}
By the definition of $\widehat{\theta}_{\text{SJR}}$, we have that, 
\begin{equation*}
\widehat{\theta}_{\text{SJR}} -\theta_0 =\left\{\E[\Phi(X)\Phi(X)\trans]+\widetilde{R}_{N,1}\right\}^{-1}\left\{\frac{1}{N}\sum_{i=1}^N\Phi(X_i)\left[ g_0(Z_i) - \widehat{g}_{\text{SJR}}(Z_i)+\delta_0(X_i)+U_i \right]\right\},
\end{equation*}
where $\widetilde{R}_{N,1}$ is as defined in (\ref{eqn:defofrn1tilde}). Let 
\begin{equation*}
\widetilde{R}^{\text{SJR}}_{N,2} = \frac{1}{N}\sum_{i=1}^N\Phi(X_i)[g_0(Z_i)-\widehat{g}_{\text{SJR}}(Z_i)].
\end{equation*}
Then by triangle inequality,
\begin{equation*}
\big\| \widetilde{R}^{\text{SJR}}_{N,2} \big\|_{\ell_2} \leq N^{-1/2}\left( \Ical_1^{\text{SJR}} + \Ical_2^{\text{SJR}} \right),
\end{equation*}
where
\begin{equation*}
\begin{aligned}
\mathcal I^{\text{SJR}}_1 & = \bigg\|\frac{1}{\sqrt{N}}\sum_{i=1}^N \left\{\Phi(X_i)\widehat{g}_{\text{SJR}}(Z_i) - \E\left[\Phi(X)\widehat{g}_{\text{SJR}}(Z) \right] \right\}\\
&\quad\quad\quad - \frac{1}{\sqrt{N}}\sum_{i=1}^N\left\{ \Phi(X_i)g_0(Z_i) - \E[\Phi(X)g_0(Z)] \right\} \bigg\|_{\ell_2}, \\
\mathcal I^{\text{SJR}}_2 & = \sqrt{N}\big\| \E[\Phi(X)\widehat{g}_{\text{SJR}}(Z)]-\E[\Phi(X)g_0(Z)] \big\|_{\ell_2}.
\end{aligned}
\end{equation*}
To bound $\mathcal I^{\text{SJR}}_1$, note that, 
\begin{equation*}
\E\left[ (\mathcal I^{\text{SJR}}_1)^2 \right] \leq \frac{1}{N} \E\left\{ \bigg\| \sum_{i=1}^Nr_0(Z_i)[\widehat{g}_{\text{SJR}}(Z_i)-g_0(Z_i)] \bigg\|_{\ell_2}^2 \right\} + \E\left\{\big\| V[\widehat{g}_{\text{SJR}}(Z)-g_0(Z)] \big\|_{\ell_2}^2\right\}.
\end{equation*}
Then under condition (C1), 
\begin{equation}
\label{eqn:bdoni1sjr}
\mathcal I^{\text{SJR}}_1 = O_p\left[N^{1/2}\left(\E\left\{ \left[ \widehat{g}_{\text{SJR}}(Z)-g_0(Z) \right]^2 \right\} \right)^{1/2} \right].
\end{equation}
To bound $\mathcal I^{\text{SJR}}_2$, note that, 
\begin{equation}
\label{eqn:bdoni2sjr}
\begin{aligned}
\mathcal I^{\text{SJR}}_2 & = \sqrt{N}\|\E\left\{ \Phi(X)\left[\widehat{g}_{\text{SJR}}(Z)-g_0(Z)\right] \right\}\|_{\ell_2} \\
&= O\left[ N^{1/2}\left(\E\left\{ \left[\widehat{g}_{\text{SJR}}(Z)-g_0(Z)\right]^2 \right\} \right)^{1/2} \right].
\end{aligned}
\end{equation}
Combining (\ref{eqn:bdoni1sjr}) and (\ref{eqn:bdoni2sjr}), we have that, 
\begin{equation*}
\widetilde{R}^{\text{SJR}}_{N,2} = O_p\left[ \left( \E\left\{ \left[ \widehat{g}_{\text{SJR}}(Z)-g_0(Z) \right]^2 \right\} \right)^{1/2} \right].
\end{equation*}
Together with the derived bound of $\widetilde{R}_{N,1}$ in (\ref{eqn:boundrb1}) and that $\E[\Phi(X)\Phi(X)\trans]$ is invertible, we have, 
\begin{equation*}
\begin{aligned}
 \widehat{\theta}_{\text{SJR}} - \theta_0 & =  \left\{ \E[\Phi(X)\Phi(X)\trans] \right\}^{-1} \left\{ \frac{1}{N}\sum_{i=1}^N\Phi(X_i)\left[ \delta_0(X_i)+U_i \right] \right \} \\
 &\quad+  O_p\left[\left( \E\left\{ \left[ \widehat{g}_{\text{SJR}}(Z)-g_0(Z) \right]^2 \right\} \right)^{1/2} \right] + o_p(N^{-1/2}).
 \end{aligned}
 \end{equation*}
This completes the proof of Proposition \ref{thm:sjr}.
\end{proof}

%%%%%%%%%%%%%%%%%%%%%%%%%%%%%%%%%%%%%%%%%%%%%%%%%%%
\subsection{Proof of Proposition \ref{prop:propertyofDML}}
\label{sec:pfofdml}

\begin{proof}
By the definition of $\widehat{\theta}_{\text{DML}}$, we have that, 
\begin{equation*}
\widehat{\theta}_{\text{DML}} -\theta_0 = \left\{J_0+R_{N,1}\right\}^{-1} \left\{ \frac{1}{Q}\sum_{q=1}^Q\frac{1}{n}\sum_{i\in I_q}\left[ \Phi(X_i)-\widehat{r}_{q}(Z_i) \right]\left[ g_0(Z_i) - \widehat{g}_{\text{DML},q}(Z_i)+\delta_0(X_i)+U_i \right] \right\},
\end{equation*}
where $R_{N,1}$ is as defined in (\ref{eqn:defofnotationsj0}).  

Let
\begin{equation*}
\widetilde{R}^{\text{DML}}_{N,2} = \frac{1}{Q}\sum_{q=1}^Q\frac{1}{n}\sum_{i\in I_q}[\Phi(X_i)-\widehat{r}_{q}(Z_i)][g_0(Z_i)-\widehat{g}_{\text{DML},q}(Z_i)+U_i] - \frac{1}{N}\sum_{i=1}^NV_iU_i.
\end{equation*}
Then by triangle inequality,
\begin{equation*}
\big\| \widetilde{R}^{\text{DML}}_{N,2} \big\|_{\ell_2} \leq  \frac{1}{Q}\sum_{q=1}^Q\frac{1}{\sqrt{n}}\left( \Ical_{1,q}^{\text{DML}} + \Ical_{2,q}^{\text{DML}} \right),
\end{equation*}
where
\begin{equation*}
\begin{aligned}
\mathcal I^{\text{DML}}_{1,q} & = \bigg\| \frac{1}{\sqrt{n}}\sum_{i\in I_q} \Big( [\Phi(X_i)-\widehat{r}_{q}(Z_i)][g_0(Z_i)-\widehat{g}_{\text{DML},q}(Z_i)+U_i]\\
&\quad - \E\left\{ [\Phi(X)-\widehat{r}_{q}(Z)] [ g_0(Z)-\widehat{g}_{\text{DML},q}(Z)+U ] | (W_i)_{i\in I_q^c} \right\} \Big)
- \frac{1}{\sqrt{n}}\sum_{i\in I_q} [V_iU_i - \E(VU)] \bigg\|_{\ell_2}, \\
\mathcal I^{\text{DML}}_{2,q} & = \sqrt{n} \big\| \E\left\{ [\Phi(X)-\widehat{r}_{q}(Z)] [g_0(Z)-\widehat{g}_{\text{DML},q}(Z)+U]|(W_i)_{i\in I_q^c} \right\} - \E(VU) \big\|_{\ell_2}.
\end{aligned}
\end{equation*}

To bound $\mathcal I^{\text{DML}}_{1,q}$, note that, 
\begin{equation*}
\begin{aligned}
\E\left[ (\mathcal I^{\text{DML}}_{1,q})^2|(W_i)_{i\in I_q^c} \right] 
&\leq \E\left\{ \big\|[\Phi(X)-\widehat{r}_q(Z)][g_0(Z)-\widehat{g}_{\text{DML},q}(Z)+U]-VU \big\|^2_{\ell_2}|(W_i)_{i\in I_q^c} \right\} \\
&=O\left( \E\left[ \|\widehat{r}_q(Z)-r_0(Z)\|^2_{\ell_2} \right] \E\left\{ [\widehat{g}_q(Z)-g_0(Z)]^2 \right\} \right) \\
&\quad\quad + O\left( \E\left[ \|\widehat{r}_q(Z)-r_0(Z)\|_{\ell_2}^2 \right] + \E\left\{ [\widehat{g}_q(Z)-g_0(Z)]^2 \right\} \right)\\
&=o(N^{-1/2}),
\end{aligned}
\end{equation*}
where the second step is by Cauchy-Schwarz inequality and condition (C2), and the last step is due to condition (C3). Therefore, $\mathcal I_{1,q}^{\text{DML}}=o_p(N^{-1/4})$. 

To bound $\mathcal I^{\text{DML}}_{2,q}$, we apply the Taylor expansion and obtain that, 
\begin{equation*}
\begin{aligned}
n^{-1/2}\mathcal I_{2,q}^{\text{DML}} & = \big\| \E\left\{ [\Phi(X)-\widehat{r}_{q}(Z)] [g_0(Z)-\widehat{g}_{\text{DML},q}(Z)+U] | (W_i)_{i\in I_q^c} \right\}-\E(VU) \big\|_{\ell_2} \\
& = \big\| \E\left\{ [\Phi(X)-r_0(Z)] [g_0(Z)-g_0(Z)+U] | (W_i)_{i\in I_q^c} \right\} - \E(VU) \big\|_{\ell_2} \\
& \quad\quad + O\left( \E\left[ \|\widehat{r}_q(Z)-r_0(Z)\|_{\ell_2}|\widehat{g}_q(Z)-g_0(Z)| \right] \right) \\
& = O\left[ \E\left[ \|\widehat{r}_q(Z)-r_0(Z)\|_{\ell_2} \right] (\E\{ [\widehat{g}_q(Z)-g_0(Z)]^2 \} )^{1/2} \right]\\
& = o(N^{-1/2}),
\end{aligned}
\end{equation*}
where the  last step is due to condition (C3). Since $Q$ is finite, we have that $\mathcal I_{2,q}^{\text{DML}} = o(1)$. 

Combining the derived bounds for $\mathcal I_{1,q}^{\text{DML}}$ and $\mathcal I_{2,q}^{\text{DML}}$, we obtain that, 
\begin{equation}
\label{eqn:bdrdmln2}
\big\| \widetilde{R}^{\text{DML}}_{N,2} \big\|_{\ell_2} \leq o_p(N^{-1/2}).
\end{equation}

Next, let  
\begin{equation*}
\widetilde{R}^{\text{DML}}_{N,3} = \frac{1}{Q}\sum_{q=1}^Q\frac{1}{n}\sum_{i\in I_q}[\Phi(X_i)-\widehat{r}_{q}(Z_i)]\delta_0(X_i).
\end{equation*}
Then by triangle inequality,
\begin{equation*}
\big\| \widetilde{R}^{\text{DML}}_{N,3} \big\|_{\ell_2}\leq \frac{1}{Q}\sum_{q=1}^Q\frac{1}{\sqrt{n}}\left( \Ical_{3,q}^{\text{DML}}+\Ical_{4,q}^{\text{DML}} \right),
\end{equation*}
where 
\begin{equation*}
\begin{aligned}
\Ical_{3,q}^{\text{DML}} &= \bigg\| \frac{1}{\sqrt{n}}\sum_{i\in I_q} \left( [\Phi(X_i)-\widehat{r}_{q}(Z_i)]\delta_0(X_i) - \E\left\{ [\Phi(X)-\widehat{r}_q(Z)] \delta_0(X) | (W_i)_{i\in I_q^c} \right\} \right) \bigg\|_{\ell_2},\\
\mathcal I^{\text{DML}}_{4,q} & = \sqrt{n} \Big\| \E\left\{ [\Phi(X)-\widehat{r}_{q}(Z)] \delta_0(X) | (W_i)_{i\in I_q^c} \right\} \Big \|_{\ell_2}.
\end{aligned}
\end{equation*}

To bound $\Ical_{3,q}^{\text{DML}}$, note that, 
\begin{equation*}
\begin{aligned}
&\E\left[ (\mathcal I^{\text{DML}}_{3,q})^2|(W_i)_{i\in I_q^c} \right] \leq \E\left\{ \|[\Phi(X)-\widehat{r}_q(Z)]\delta_0(X)\|^2_{\ell_2}|(W_i)_{i\in I_q^c} \right\} \\
= \; & O\left( \E\left[ \|\widehat{r}_q(Z)-r_0(Z)\|_{\ell_2}^2 \right] \E[\delta^2_0(X)] \right) + O\left(\E\left[\|V\|_{\ell_2}^2\right] \E[\delta^2_0(X)] \right) 
=O(\E[\delta^2_0(X)]),
\end{aligned}
\end{equation*}
where the second step is by Cauchy-Schwarz inequality, and the last step is by conditions (C2) and (C3). Therefore, $\mathcal I_{3,q}^{\text{DML}}=O_p(\{\E[\delta^2_0(X)]\}^{1/2})$.  

To bound $\mathcal I^{\text{DML}}_{4,q}$, we apply the Taylor expansion and obtain that, 
\begin{equation*}
\begin{aligned}
n^{-1/2}\mathcal I_{4,q}^{\text{DML}} & = \big\| \E\left\{  [\Phi(X)-r_0(Z)] \delta_0(X)|(W_i)_{i\in I_q^c} \right\} \big\|_{\ell_2} + O\left( \E[\|\widehat{r}_q(Z)-r_0(Z)\|_{\ell_2}] \right) \\
&\leq \left( \E\left\{ \|[\Phi(X)-r_0(Z)]\delta_0(X)\|^2_{\ell_2}|(W_i)_{i\in I_q^c} \right\} \right)^{1/2}+ O\left( \E[\|\widehat{r}_q(Z)-r_0(Z)\|_{\ell_2}] \right) \\
& = O\left( \{\E[\|V\|_{\ell_2}^2]\}^{1/2}\{\E[\delta^2_0(X)]\}^{1/2}+\E[\|\widehat{r}_q(Z)-r_0(Z)\|_{\ell_2}] \right) \\
& = O\left( \{\E[\delta^2_0(X)]\}^{1/2} \right),
\end{aligned}
\end{equation*}
where the third step is by Cauchy-Schwarz inequality, and the last step is by conditions (C2) and (C3).

Combining the derived bounds for $\mathcal I_{3,q}^{\text{DML}}$ and $\mathcal I_{4,q}^{\text{DML}}$, we obtain that, 
\begin{equation}
\label{eqn:bdrdmln3}
\big\| \widetilde{R}^{\text{DML}}_{N,3} \big\|_{\ell_2} \leq O_p\left( \{\E[\delta^2_0(X)]\}^{1/2} \right).
\end{equation}

Finally, combining the derived bounds for $\widetilde{R}_{N,1},\widetilde{R}^{\text{DML}}_{N,2}, \widetilde{R}^{\text{DML}}_{N,4}$ in (\ref{eqn:boundrb1}), (\ref{eqn:bdrdmln2}), and (\ref{eqn:bdrdmln3}), respectively, together with (\ref{eqn:bdonmeanpsi}), we obtain that, 
\begin{equation*}
\begin{aligned}
 \widehat{\theta}_{\text{DML}} - \theta_0 &=  \{\E[VV\trans]\}^{-1} \left\{\frac{1}{N}\sum_{i=1}^NV_iU_i\right\} +O_p\left( \{\E[\delta^2_0(X)]\}^{1/2} \right) +o_p(N^{-1/2}).
 \end{aligned}
 \end{equation*}
This completes the proof of Proposition \ref{prop:propertyofDML}.
\end{proof}

%%%%%%%%%%%%%%%%%%%%%%%%%%%%%%%%%%%%%%%%%%%%%%%%%%%
\subsection{Sensitivity of nuisance function modeling for inference on $\theta_0$}
\label{sec:sim-sensitivity}

We study the sensitivity of using different machine learning methods for nuisance function estimation when inferring $\theta_0$. We compare with SJR and DML, but exclude UR and DUR since they do not involve any nuisance function estimation. We consider a complex nonlinear model with interactions, $Y_i = f_0(X_i) + h_{01}(Z_{i1})g_{02}(Z_{i2})+ h_{02}(Z_{i1})g_{03}(Z_{i3})+ h_{03}(Z_{i3})g_{04}(Z_{i4}) + U_i$, where 
\begin{equation*}
\begin{aligned}
f_0(x)& = -2\sin(2\pi x)+5(1-e^x)^2; \\
h_{01}(z_1)&=\frac{15}{8} \left[ 1-(4z_1-1)^2 \right]^2,\quad h_{02}(z_1)=3\cos(2\pi z_1),\quad  h_{03}(z_1)=4;\\
 g_{02}(z_z) & = z_z^2-\frac{1}{3},\quad  g_{03}(z_3)=z_3-\frac{1}{2},\quad g_{04}(z_4)=e^{z_4}+e^{-1}-1.
\end{aligned}
\end{equation*}
A similar model has been considered in \citet{lu2020kernel}. We generate random variables $E_1,\ldots,E_6$ independently from Uniform$[0,1]$, and set the primary and auxiliary modalities as $X = (E_1+\rho E_6)/(1+\rho)$, and $Z_j=(E_{j+1}+\rho E_6)/(1+\rho)$, for $\rho=1$ and $j=1,\ldots,4$. We generate i.i.d.\ copies $(X_i,Z_{i1},Z_{i2},Z_{i3},Z_{i4})$ of $(X,Z_{1},Z_{2},Z_{3},Z_{4})$, and generate the error $U_i$ from $\Ncal(0,\sigma^2)$ with $\sigma=1$. We set the sample size  $N\in\{400, 500,600\}$. We set $\eta(x,\theta_0) = \theta_0\sin(2\pi x)$. 

We apply numerous nonlinear machine learning methods to estimate the nuisance functions $\{r_0,g_0\}$, including random forests, boosted trees, and neural networks. We tune the parameters by ten-fold cross-validation. For neural networks, we use five hidden layers with ten neuron at each hidden layer, and choose the learning rate of $0.02$ and a linear activation function.

\begin{figure}[t!]
\centering
\includegraphics[width=\textwidth]{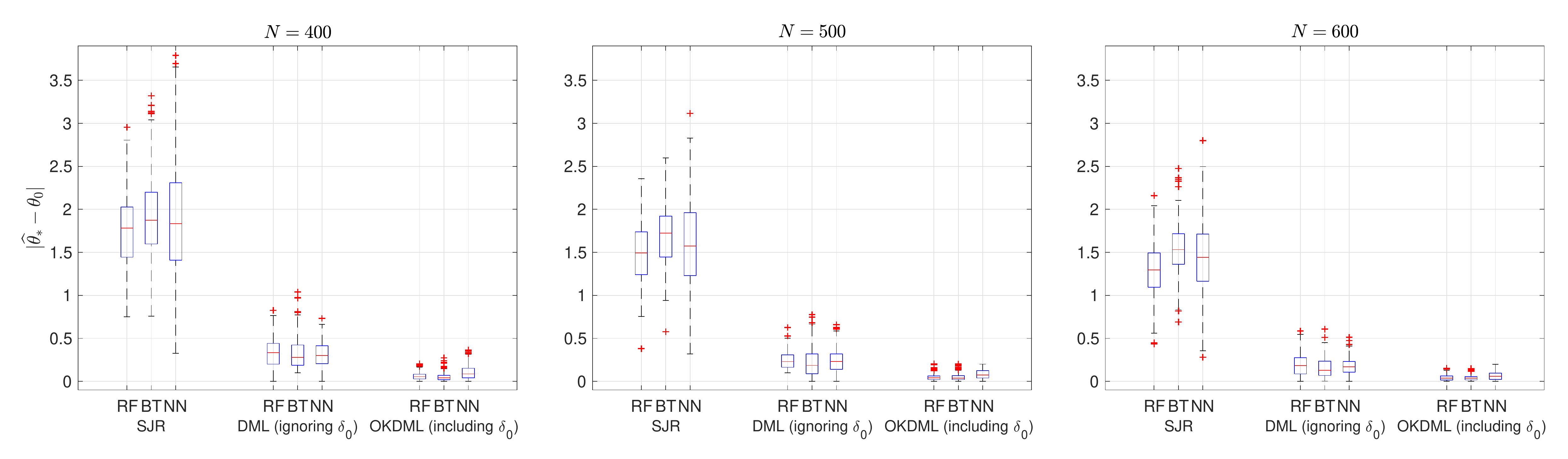}
\caption{Absolute error of the estimation of $\theta_0$, with varying sample size and different machine learning methods, including random forests (RF),   boosted trees (BT), and neural networks (NN).}
\label{fig:eg2} 
\end{figure}

Figure \ref{fig:eg2} reports the absolute error of estimating $\theta_0$ under various combinations of the sample size $N$ and the nonlinear modeling methods, based on 500 data replications. It is seen that our OKDML estimator achieves the smallest bias and standard deviation, and the results are relatively stable across different choices of the nonlinear modeling methods for the nuisance functions.

\end{document}